\documentclass[11pt]{article}
\usepackage[a4paper,margin=0.6in]{geometry}
\usepackage[titletoc,title]{appendix}
\usepackage{xcolor}
\usepackage{amsmath,amssymb,amsfonts}
\usepackage[bookmarks=false]{hyperref}
\hypersetup{pdfstartview={XYZ null null 1.00}}
\usepackage{authblk}

\usepackage{graphicx}
\usepackage{subcaption}

\DeclareMathOperator{\erf}{erf}
\DeclareMathOperator{\erfc}{erfc}

\newcommand{\sfrac}[2]{{\textstyle\frac{#1}{#2}}}

\title{{\sc Complex singularities for Burgers’ equation with piecewise-continuous initial conditions}}

\author[1]{Jacob C. Gentner}
\author[1]{Michael C. Dallaston}
\author[1]{Scott~W. McCue\thanks{scott.mccue@qut.edu.au}}
\affil[1]{\footnotesize{School of Mathematical Sciences, Queensland University of Technology, Brisbane QLD 4001, Australia}}

\begin{document}

\maketitle
\begin{abstract}
There is a body of research devoted to understanding how complex singularities of solutions of nonlinear partial differential equations (pdes) spontaneously emerge at $t=0^+$ and propagate for $t>0$, and how their behaviour affects the solution on the real axis. Despite the importance of the small-time limit in these studies, there is still a lack of understanding of how complex singularities are born at $t=0^+$, including for initial conditions that are not analytic functions of the spatial variable. In this paper, we use Burgers' equation as a prototype nonlinear pde and study the complex-plane singularities for initial conditions that are piecewise smooth. Using matched asymptotic expansions, we show how infinitely many singularities emerge from points of discontinuity in a pattern that can be described using branches of the Lambert-$W$ function. For various initial conditions, we observe how these singularities rearrange themselves to align with the appropriate exactly-described long-time behaviour, including sigmoid-shaped travelling waves, constant-area (triangular wave) similarity solutions and $N$-wave solutions. In terms of Burgers' equation, our small-time asymptotic analysis of the singularity propagation for piecewise-continuous initial conditions illustrates the types of generic behaviours that arise for inner regions when diffusion dominates advection.  More generally, this work is a step towards understanding complex-plane behaviour of solutions of nonlinear partial differential equations with non-analytic initial conditions.
\end{abstract}


\newpage
\section{Introduction}\label{sec:intro}

While the study of the formation and motion of complex singularities of solutions of nonlinear partial differential equations (pdes) is not commonplace, it does have a rich history that involves a variety of applications in fluid mechanics and the physical sciences.  Some examples of this research relate to Burgers' equation
\begin{equation}
\frac{\partial u}{\partial t}+u\frac{\partial u}{\partial x}=\mu\frac{\partial^2u}{\partial x^2},\quad \mu>0
\label{eq:burgers0}
\end{equation}
\cite{bessis1984pole,bessis1990complex,caflisch2015,deconinck2007,derevianko2026,senouf1997dynamics,senouf1997dynamicsII,sulem1983,trefethen2023,weideman2022dynamics},
which is the topic of the present study.  Other examples are concerned with similar types of advection and/or diffusion equations, including the Sivashinsky equation as a model for winkled flame fronts \cite{thual1985}, analogue models for vortex sheets \cite{baker1996analytic,matsuno1991,weideman2003computing} and a one-dimensional model for vorticity \cite{constantin1985,schochet1986}.  Further studies come from more complicated models such as those with applications to Hele-Shaw flows \cite{tanveer93heleshaw}, water waves \cite{baker2011,tanveer91waterwaves} and Prandtl's equations \cite{gargano2009}, or third-order pdes such as the KdV equation \cite{bona2009, mccue2026} or the Harry-Dym equation \cite{costin2004,costin2006}.  The kinds of research questions that are pertinent in this body of work include: how many complex singularities are there; how is their initial motion affected by the initial condition; what is the trajectory of each of the singularities over time; do any of the complex singularities intersect the real axis in finite time; and, if not, what is the long-time behaviour of each singularity?  Furthermore, an important challenge is to determine how all of these issues affect the solution on the real line.

In this context, the choice of Burgers' equation is advantageous for a number of reasons, the most obvious being that there is an exact solution (via the Cole-Hopf transform) that can be used to plot solutions in the complex plane and to compare with various numerical and analytical predictions.  Another is that each complex singularity of a solution to Burgers' equation, $x=s(t)$, must be a simple pole with the local behaviour \cite{Joshi1996}
\begin{equation}
u\sim -\frac{2\mu}{x-s(t)}+\frac{\mathrm{d}s}{\mathrm{d}t}+\mathcal{O}(x-s(t))
\quad\mbox{as}\quad x\rightarrow s(t);
\label{eq:burgerpoleequation}
\end{equation}
therefore, any such study of Burgers' equation will not involve the various complications that arise with branch point singularities (such as \cite{fasondini2024}).  Further, Burgers' equation is an excellent prototype model for studying the formation and propagation of smoothed shocks \cite{chapman2007} in a second-order model without the complications of the highly oscillatory waves that arise in the small-dispersion limit of third-order models such as the KdV equation \cite{claeys2009}.  More generally, Burgers' equation is worth studying as it has been used extensively as a toy model for one-dimensional viscous flow and turbulence \cite{matharu2025,protas2024} and, in its complex form, has been used recently as a model for Marangoni-driven flow with insoluble surfactant on a bubble boundary \cite{crowdy2021,crowdy2024,temprano2024}.  Another reason for studying Burgers' equation is that it is a relatively simple nonlinear pde that has exact travelling-wave solutions and similarity solutions, both of which will prove useful for us here.

We are particularly interested in the small-time limit of solutions of Burgers' equation.  Our approach is to consider the analytic continuation of the solutions into the complex-$x$ plane, following the recent study of Burgers' equation by VandenHeuvel et al.~\cite{vandenheuvel2023burgers} and other important studies such as Tanveer's work on complex singularities in Rayleigh-Taylor flow \cite{tanveer93rayleigh}, Cowley et al.'s study of curvature singularities in vortex sheets \cite{cowley1999} and Fasondini et al.'s asymptotic analysis of a semi-linear heat equation \cite{fasondini2024}.  For these and other similar studies where analytic initial conditions are used \cite{costin2004,mccue2026}, a generic feature is that a regular power series expansion in $t$ of the type
\begin{equation}
u\sim \sum_{n=0}^\infty t^n u_n(x)
\quad\mbox{as}\quad t\rightarrow 0^+
\label{eq:outergeneric}
\end{equation}
will break down (become disordered) because higher-order terms that are smaller than previous terms (even exponentially smaller) on the real line become larger (more singular) in the complex plane either a) near a singularity of the initial condition, b) in the far field (e.g., due to an essential singularity at infinity), and/or c) between certain anti-Stokes lines.  Subsequent analysis therefore requires a rescaling of dependent and independent variables using matched asymptotic expansions and Stokes phenomenon, so that the birth and initial dynamics of the singularities can be described.  In all of these cases, the singularities are born off the real axis, either at a singularity of the initial condition or at infinity.
In the present study, we are motivated to study initial conditions that are {\em not} analytic and revisit this general approach given that there cannot be a regular power series of the form (\ref{eq:outergeneric}) and that there will be complex singularities spontaneously emerging from points on the real line.

As a first step in studying Burgers' equation with non-analytic initial conditions, in section~\ref{Heaviside_initial_conditions} we consider the simplest piecewise-continuous initial conditions; that is, we study
\begin{equation}
\frac{\partial u}{\partial t}+u\frac{\partial u}{\partial x}=\mu\frac{\partial^2u}{\partial x^2},
\quad u(x,0)=u_0(x), \quad x \in \mathbb{R},
\label{eq:burgers}
\end{equation}
with
\begin{equation}
u(x,0)=u_L+(u_R-u_L)H(x),
\label{eq:heaviside}
\end{equation}
where $H(x)$ is the Heaviside step function defined to be $H(x)=0$ for $x<0$ and $H(x)=1$ for $x>0$.
By employing matched asymptotic expansions, we explain how infinitely many simple poles are born at $x=0$ at $t=0^+$.   We describe their early-time spatial structure (that turns out to be related to the Lambert-$W$ function) and how the closest poles to the real axis move with a speed that scales like $t^{-1/2}\ln^{1/2}(1/t)$ as $t\rightarrow 0^+$, while the speed of the furthest poles scales like $t^{-1/2}$.  In order to resolve the asymptotic details, we need to work on incredibly short time-scales for which $\ln(1/t)\gg 1$ (e.g., we use $t=10^{-24}$ in some figures to illustrate extremely early-time behaviour).  For moderate and large times, we use exact solutions to track the singularities and show how they reorganise themselves to produce known late-time limits.  For example, for the case $u_L>u_R$, there are two arrays of singularities in the upper half plane for moderate times (extending out at angles $\pi/4$ and $3\pi/4$) that eventually merge in a complicated way in order to produce a travelling wave solution with a very simple singularity structure (a vertical array of equally-spaced poles).

In section~\ref{sec:other}, we cover two further examples, involving a rectangular initial condition (top-hat function) that evolves to a well-known constant-area similarity solution in the long-time limit, and a saw-tooth initial condition that evolves to the well-known $N$-wave solution.  The solutions with these initial conditions demonstrate qualitatively similar features to the simpler Heaviside examples considered in section~\ref{Heaviside_initial_conditions} in the small-time limit, especially near points of discontinuity of the initial condition, as well as some new behaviours caused by poles originating from different discontinuities crossing paths at small times.  We close the main part of the paper with a discussion in section~\ref{sec:discussion}, where, together with Appendix~\ref{appendix:inner}, we argue that the small-time asymptotic analysis for the Heaviside examples in section~\ref{Heaviside_initial_conditions} contains many generic structures for initial conditions that are piecewise continuous or involve diffusion-dominated asymptotic regimes.  We provide a recipe for attacking such problems.  Finally, we include in Appendix~\ref{appendix:discderiv} an example for which the initial condition is continuous but the derivative is discontinuous, demonstrating that, while some details of the matched asymptotic expansions are more complicated, the overall analysis follows our general approach.

\section{Heaviside initial conditions} \label{Heaviside_initial_conditions}

\subsection{Preamble}

In this section we consider the example (\ref{eq:burgers})-(\ref{eq:heaviside}).  The initial condition
(\ref{eq:heaviside}) is simply $u(x,0)=u_L$ for $x<0$ and $u(x,0)=u_R$ for $x>0$.
For $u_L>u_R$, the solution behaves like a smoothed shock and is well known to evolve to an exact travelling wave solution as $t\rightarrow\infty$ with speed $(u_L+u_R)/2$, while for $u_L<u_R$ the solution behaves like a smoothed rarefaction fan as $t\rightarrow\infty$. In both cases, the initial condition has a single discontinuity at $x=0$. Our study of Burgers’ equation with (\ref{eq:heaviside}) will demonstrate how complex singularities emerge from this point of discontinuity and how they propagate for small time. Further, depending on the sign of $u_L-u_R$, we shall also observe how the singularities rearrange themselves to approach the
known large-time behaviour (i.e., travelling wave or smoothed rarefaction fan).


It is well known that Burgers' equation \eqref{eq:burgers} has an exact solution found by applying the Cole-Hopf transform \cite{whitham1974linear}
\begin{equation}
    u = -2\mu\frac{\partial}{\partial x}\ln{\phi} = -2\mu\frac{\phi_x}{\phi}.
    \label{eq:colehopf1}
\end{equation}
The change of variables \eqref{eq:colehopf1} linearises \eqref{eq:burgers} since $\phi$ satisfies the linear heat equation with the solution
\begin{equation}
    \phi\left(x, t\right) = \frac{1}{2\sqrt{\mu\pi t}}\int^\infty_{-\infty}\phi\left(s, 0\right)
    \mathrm{e}^{-(x-s)^2/4\mu t}\mathrm{d}s.
    \label{eq:colehopf2}
\end{equation}
Given the initial condition $u_0(x)$ in \eqref{eq:burgers}, from \eqref{eq:colehopf1} we now have
\begin{equation}
    \phi\left(x, 0\right) = \exp{\left(-\frac{1}{2\mu }\int^x_{0}u\left(s, 0\right)\mathrm{d} s \right)}.
    \label{eq:colehopf3}
\end{equation}
Thus, for a given initial condition $u_0(x)$, we can first calculate $\phi(x, 0)$ using \eqref{eq:colehopf3}, then generate $\phi(x, t)$ from \eqref{eq:colehopf2}, and finally recover $u(x, t)$ in \eqref{eq:colehopf1}.  Note that the Gaussian kernel in (\ref{eq:colehopf2}) allows for a broad class of initial conditions and, for example, does not require that $u(x,0)$ decays to zero as either $x\rightarrow\pm\infty$.

\subsection{Step-down initial condition: exact solution}

For very simple initial conditions such as \eqref{eq:heaviside}, we can calculate all of (\ref{eq:colehopf1})-(\ref{eq:colehopf3}) explicitly. For example, the step-down initial condition with $u_L = 1$ and $u_R = 0$, namely
\begin{equation}
u\left(x, 0\right) = \begin{cases}
      1 & x < 0 \\
      0 & x > 0,
   \end{cases}
   \label{eq:stepdownIC}
\end{equation}
gives rise to the exact solution
\begin{equation}
\begin{aligned}
    u\left(x, t\right) &= \frac{\mathrm{erf}\! \left(\dfrac{t-x}{2 \sqrt{\mu t}}\right)+1}{\mathrm{erf}\! \left(\dfrac{t-x}{2 \sqrt{\mu t}}\right)+1+\exp{\left({-\dfrac{t-2 x}{4 \mu}}\right)} \left(\mathrm{erf}\! \left(\dfrac{x}{2 \sqrt{\mu t}}\right)+1\right)},
\end{aligned}
\label{eq:stepdownsolution}
\end{equation}
where $\mathrm{erf}(z)$ is the error function \cite{nistDLMFAbout}.

\begin{figure}[!h]
        \centering
        \begin{subfigure}[t]{0.4\textwidth}
            \includegraphics[width=\linewidth]{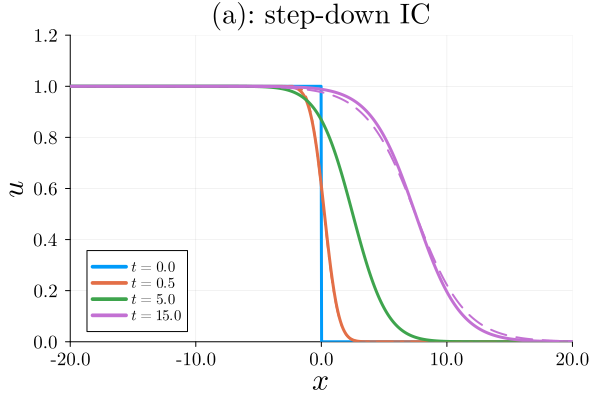}
        \end{subfigure}
        \begin{subfigure}[t]{0.35\textwidth}
            \includegraphics[width=\linewidth]{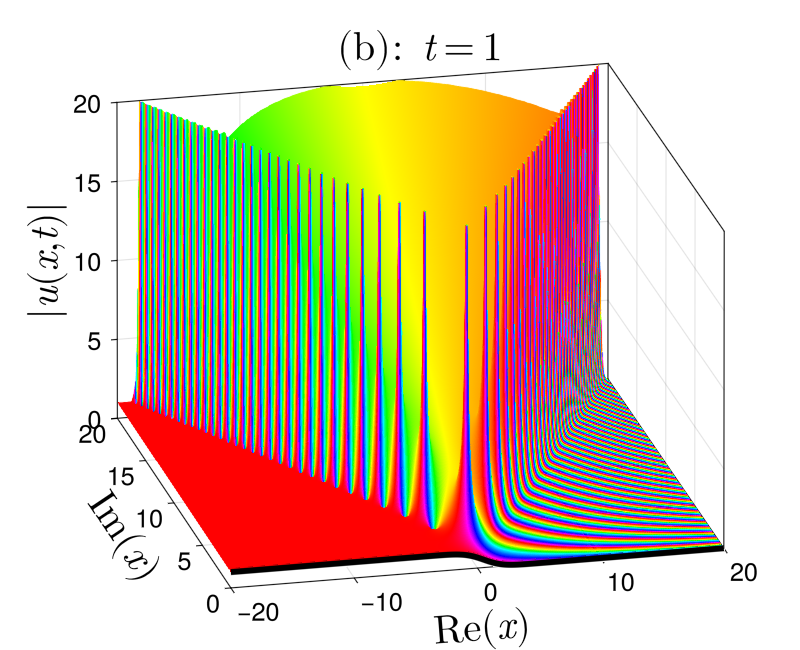}
        \end{subfigure}

        \begin{subfigure}[t]{0.32\textwidth}
            \includegraphics[width=\linewidth]{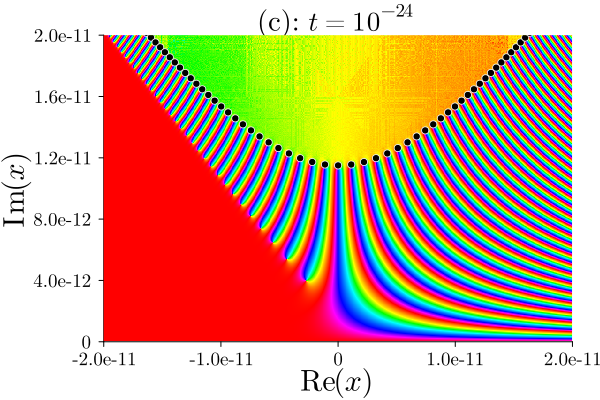}
        \end{subfigure}
        \begin{subfigure}[t]{0.32\textwidth}
            \includegraphics[width=\linewidth]{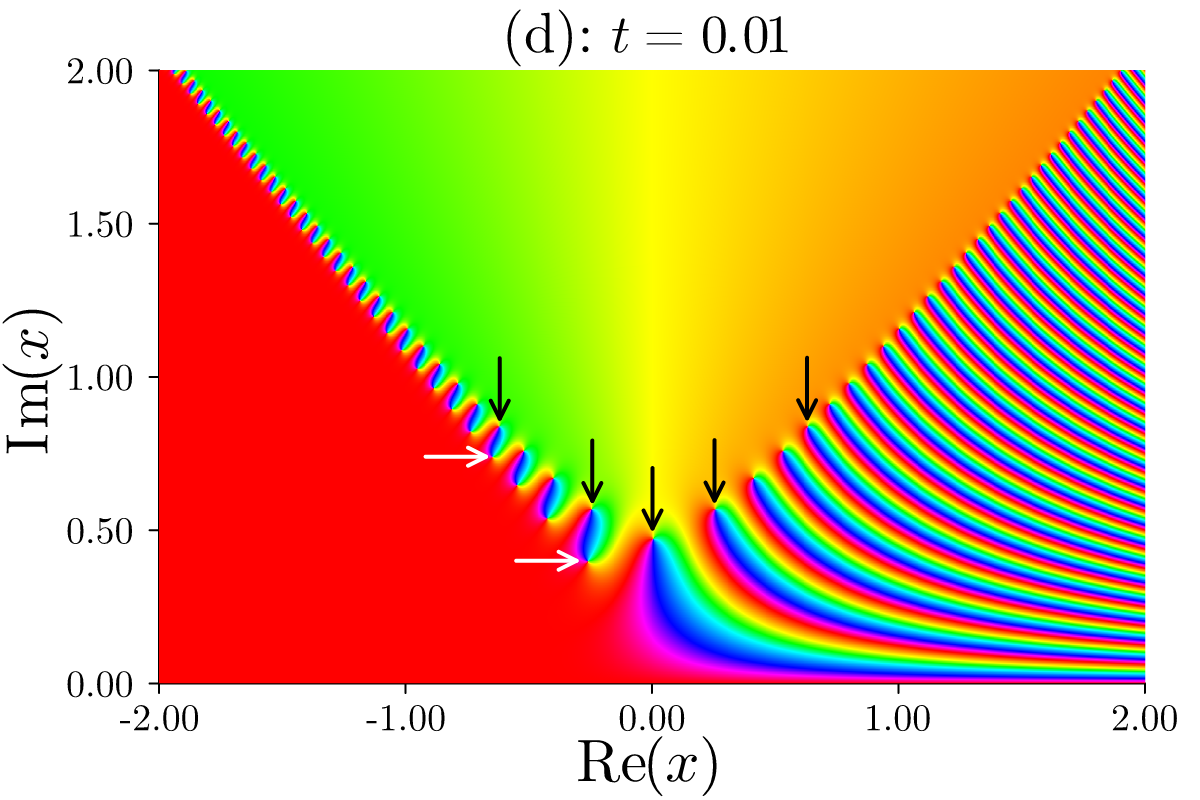}
        \end{subfigure}
        \begin{subfigure}[t]{0.32\textwidth}
            \includegraphics[width=\linewidth]{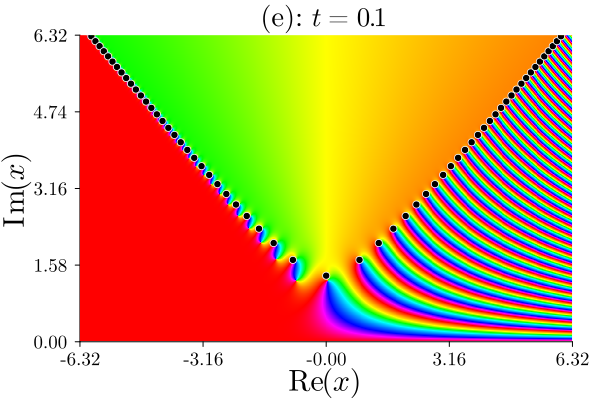}
        \end{subfigure}
        \begin{subfigure}[t]{0.32\textwidth}
            \includegraphics[width=\linewidth]{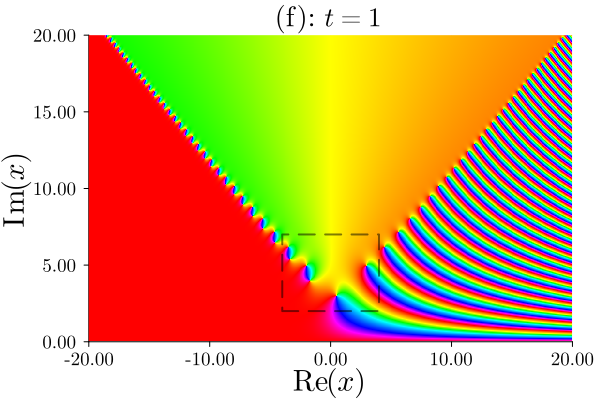}
        \end{subfigure}
        \begin{subfigure}[t]{0.32\textwidth}
            \includegraphics[width=\linewidth]{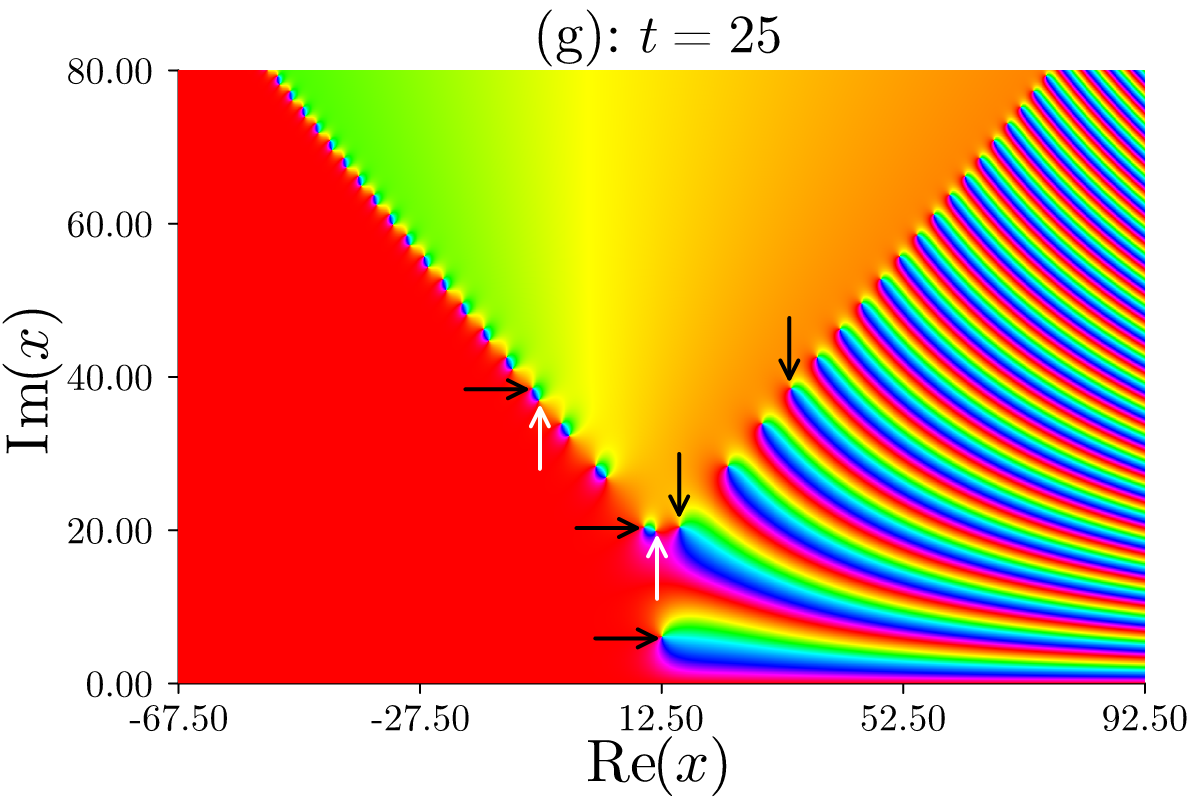}
        \end{subfigure}
        \begin{subfigure}[t]{0.32\textwidth}
            \includegraphics[width=\linewidth]{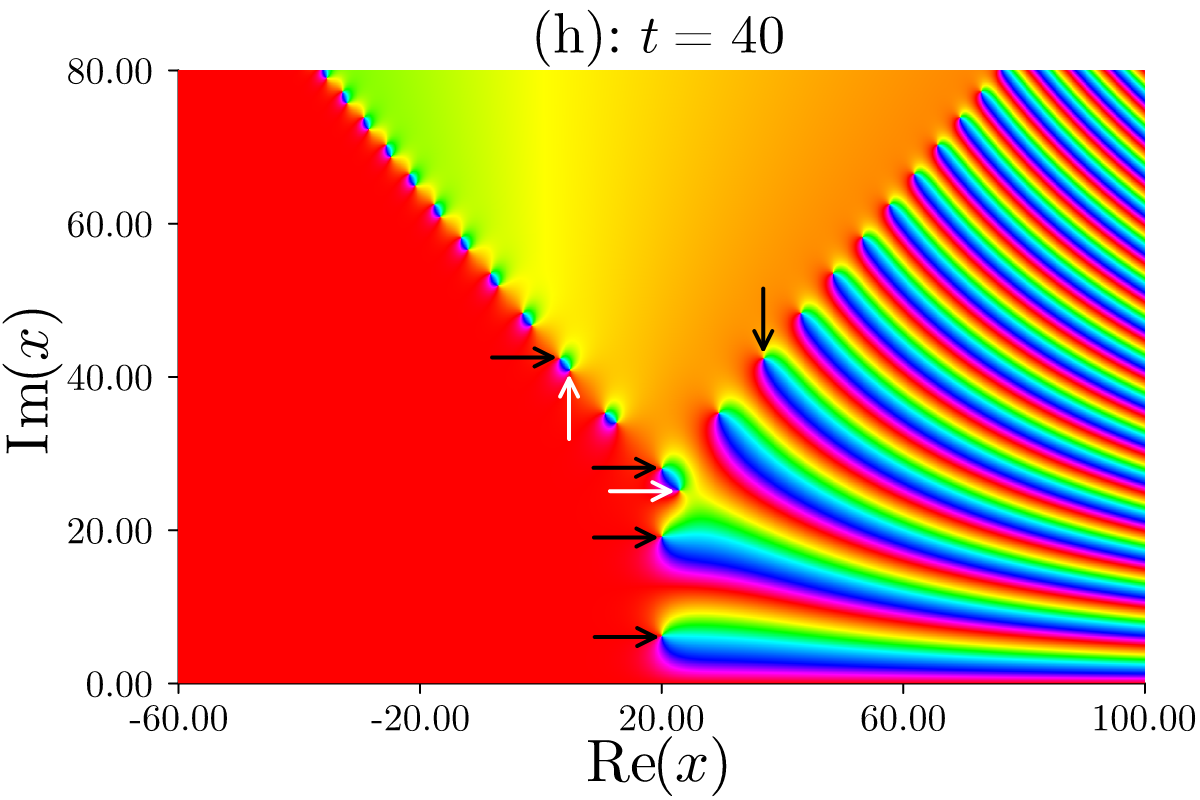}
        \end{subfigure}
        \begin{subfigure}[t]{0.32\textwidth}
            \includegraphics[width=\linewidth]{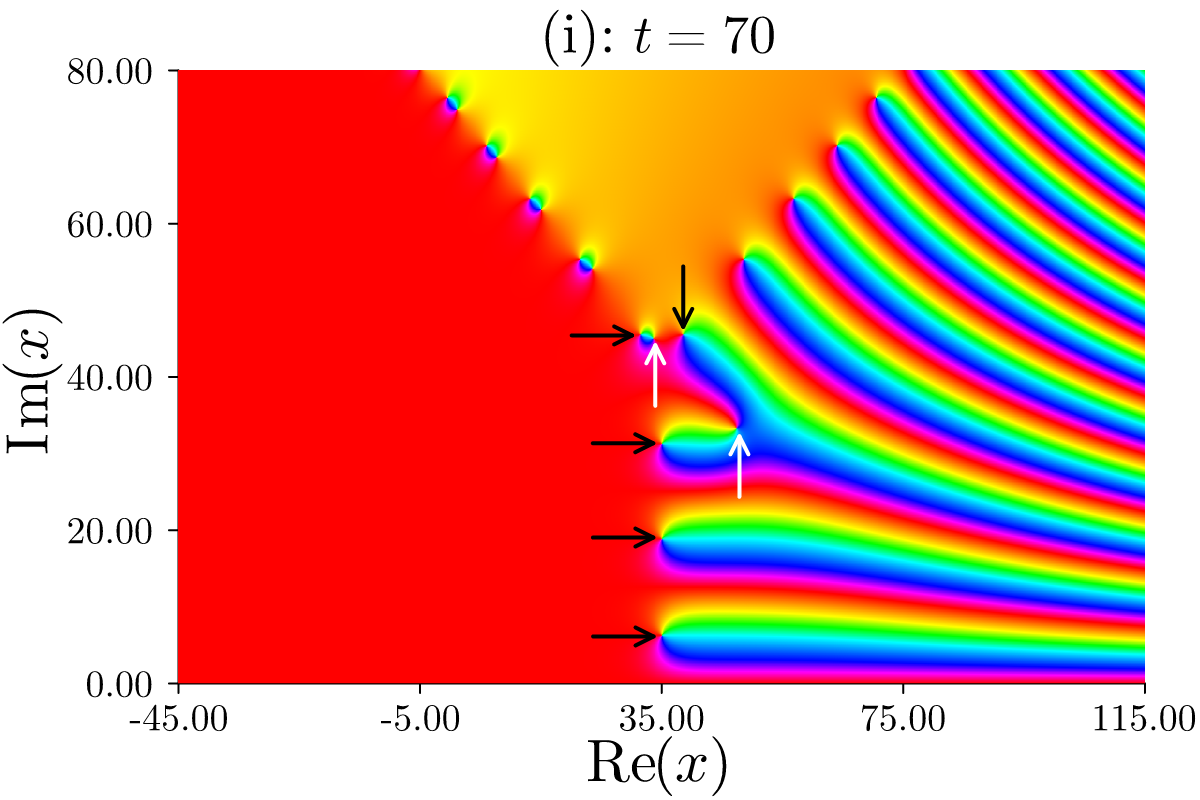}
        \end{subfigure}
        \begin{subfigure}[t]{0.32\textwidth}
            \includegraphics[width=\linewidth]{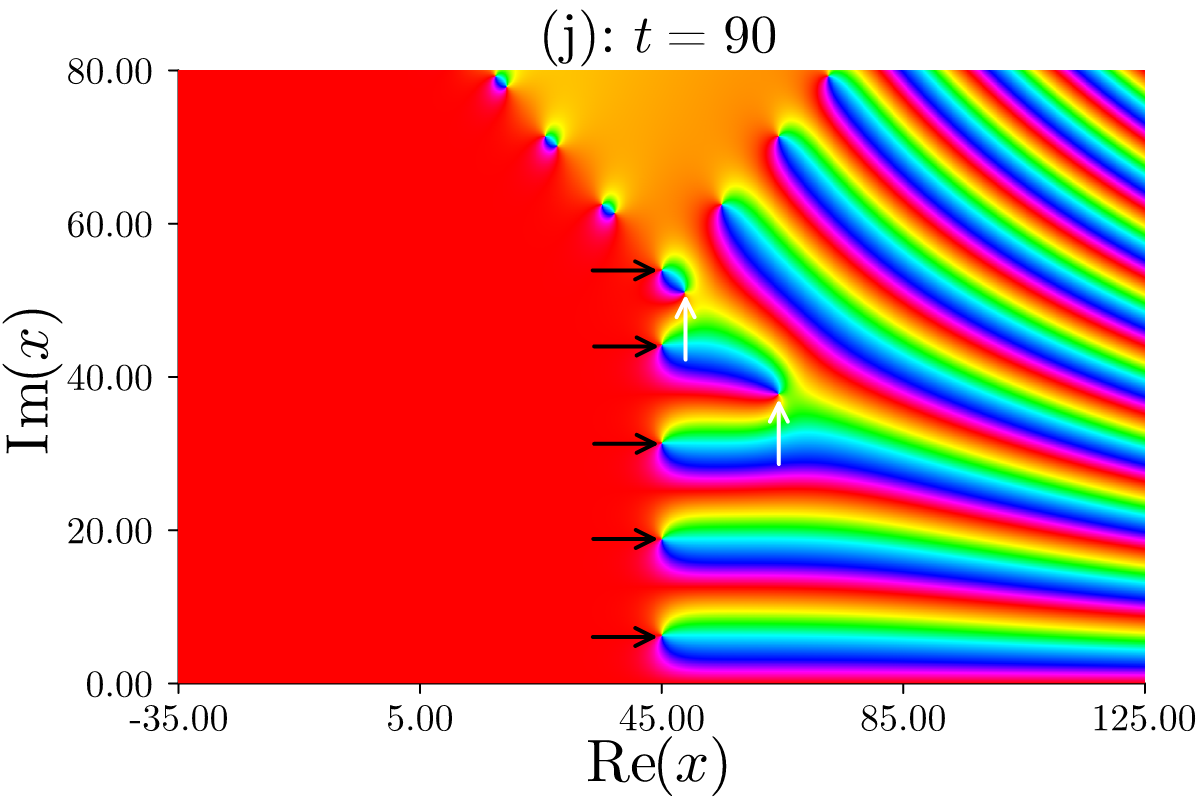}
        \end{subfigure}
        \begin{subfigure}[t]{0.32\textwidth}
            \includegraphics[width=\linewidth]{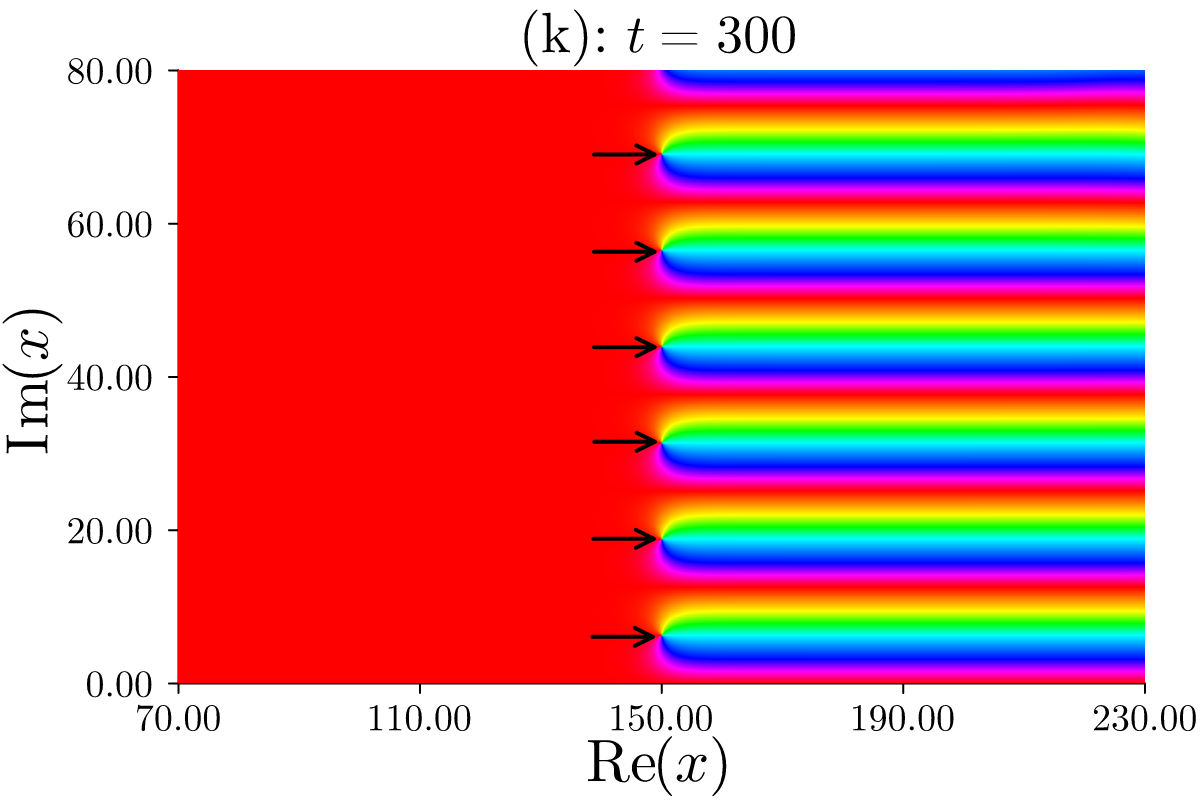}
        \end{subfigure}
        \caption{(a) Real-valued solution profiles of Burgers' equation \eqref{eq:burgers} with $\mu = 1$ at $t=0$, $0.5$, $5$ and $15$ using step-down initial condition \eqref{eq:stepdownIC} (solid lines), together with the travelling wave solution \eqref{eq:travwavesolution2} computed with $x_0=0$ (dashed line) and the small-time approximation $f\sim f_0$ with \eqref{eq:asymptoticsolf0} at $t=0.5$ (dot-dashed line, difficult to observe on this scale).
        (b) Analytic landscape plot of the solution at $t=1$ with the black curve along $\mathrm{Im}(x)=0$ indicating the real-valued solution on the real line.
        (c)-(k)
        Phase portraits plotted at various times illustrating early-time behaviour, with poles emerging from $x=0$ along a convex-shaped curve, as well as late-time behaviour where
        poles and zeros collide and ultimately
        approach the travelling wave solution \eqref{eq:travwavesolution}.
        The black dots in (c) and (e) indicate the approximate location of the simple poles according to (\ref{eq:innerextended}).  Equivalent approximations on this scale could be drawn using the Lambert-$W$ function (\ref{eq:lambertscaled}).  The black arrows in (d), (g)-(k) point to poles, while white arrows point to zeros.
        While this last image in panel (k) is drawn for a finite time $t=300$, it is visually undistinguishable from the travelling wave solution \eqref{eq:travwavesolution} on this scale.}
        \label{fig:stepdowntotravwave}
    \end{figure}

\begin{figure}[!h]
        \centering
        \begin{subfigure}[t]{0.23\textwidth}
        \includegraphics[width=\linewidth]{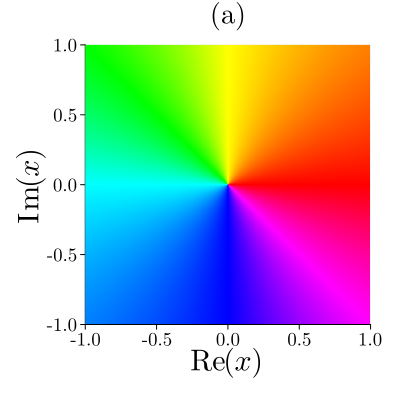}
        \end{subfigure}
        \hspace{5ex}
        \begin{subfigure}[t]{0.23\textwidth}
        \includegraphics[width=\linewidth]{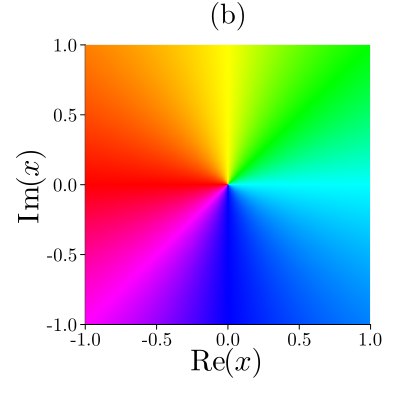}
        \end{subfigure}

        \begin{subfigure}[t]{0.32\textwidth}
        \includegraphics[width=\linewidth]{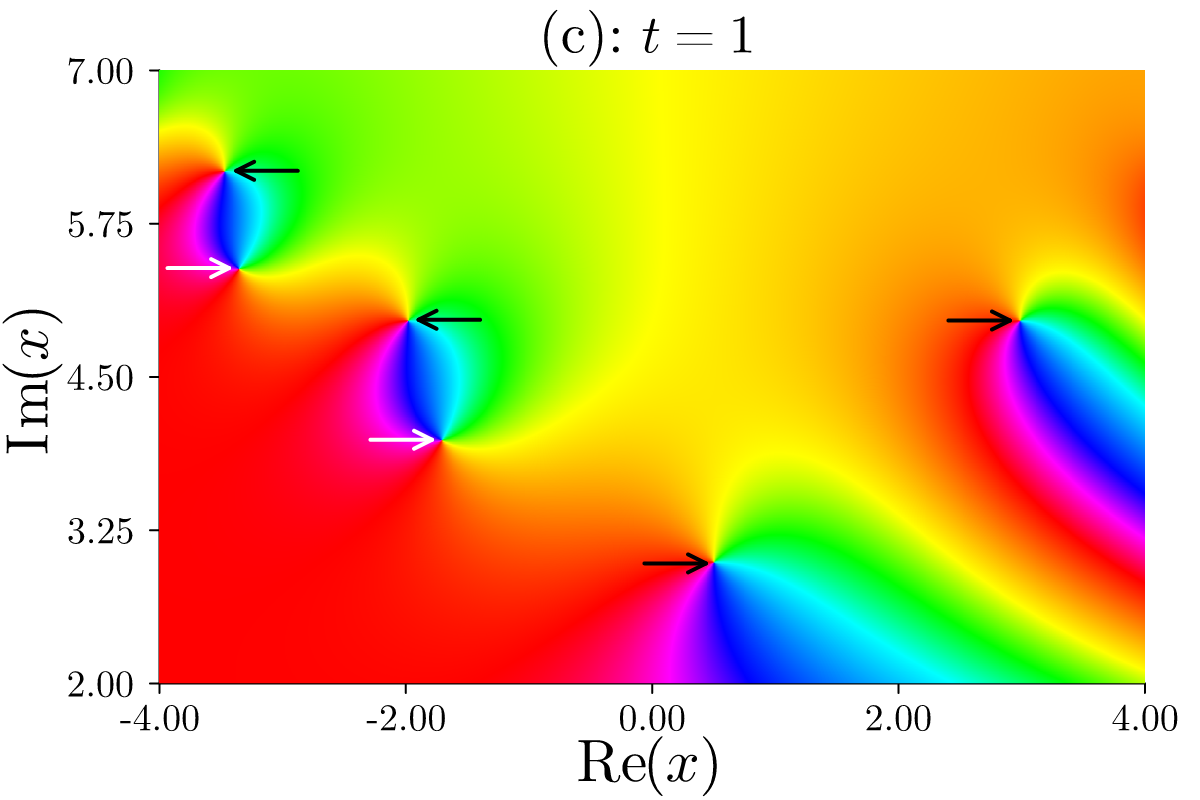}
        \end{subfigure}
        \begin{subfigure}[t]{0.32\textwidth}
        \includegraphics[width=\linewidth]{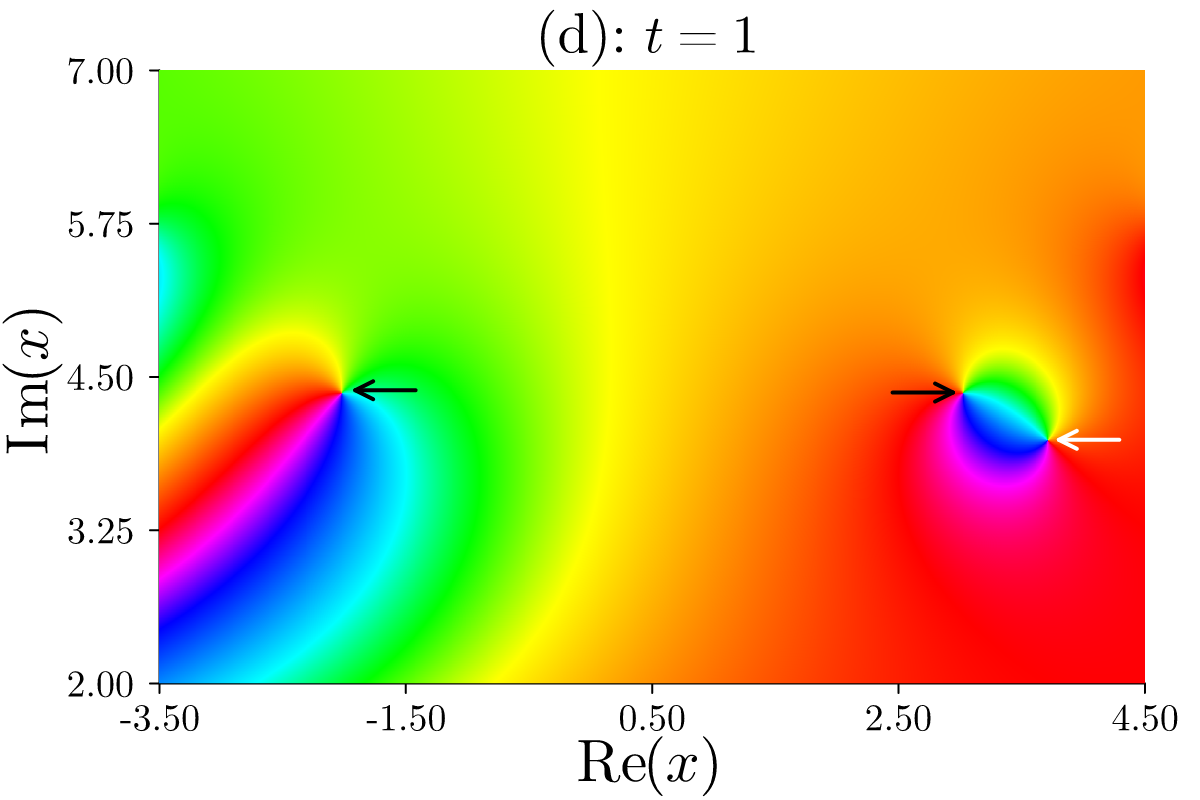}
        \end{subfigure}
        \caption{Colour scheme used for plotting the phase portraits \cite{nistDLMFAbout}. (a) Allowing $x$ to be complex, the function $f(x) = x$ acts as a colour wheel for our phase portraits, with red indicating positive real, yellow positive imaginary, aqua negative real, and dark blue negative imaginary.  Any simple zero will appear locally either like this function or a rotated version.  (b) The function $f(x) = -1/x$ shows the local behaviour of a simple pole with a negative real residue.  Note that in all of our phase portraits, a zoomed-in view near each singularity will appear locally like this, since all singularities of Burgers' equation are simple poles with residue $-2\mu$, with $\mu>0$ (see (\ref{eq:burgerpoleequation})). (c)-(d) Zoomed in versions of figure~\ref{fig:stepdowntotravwave}(f) and figure~\ref{fig:stepup}(e) to show coloured phases more clearly, with white arrows pointing to zeros and black arrows pointing to poles.}
        \label{fig:wheel}
    \end{figure}

In figure~\ref{fig:stepdowntotravwave}(a) we show solution profiles on the real line at a few representative times for the step-down initial condition (\ref{eq:stepdownIC}), computed using the formula (\ref{eq:stepdownsolution}).  In panel (b) we show an analytic landscape of (\ref{eq:stepdownsolution}), with the elevation of the surface representing the modulus of $u(x,t)$, drawn using a colour scheme to indicate the phase (the argument) of $u(x,t)$ according to the colour wheel in figure~\ref{fig:wheel}(a).  In panels (c)-(k) we show phase portraits at various times; these planar plots indicate the phase only via the colour scheme \cite{Wegert2011}.
In these and all of our phase portraits, we shall include the upper half planes only, as the structure of the singularities and zeros in the lower half plane is simply a reflection of the upper half plane.
We shall return to figure \ref{fig:stepdowntotravwave} below, where we describe the complex-plane behaviour in more detail, but for now it is worth noting that for this step-down initial condition, there is a clear structure of poles (and zeros) that has been established at the intermediate time $t = 1$. We wish to understand how these poles are born at $t = 0^+$, describe their small-time asymptotic trajectories and explain how they transition to their appropriate large-time behaviour, which in this case involves a travelling wave.

Throughout this paper we shall use domain colouring~\cite{ponce2021domain} to visualise singularities via the colour wheel in figure~\ref{fig:wheel}(a) with the NIST colour scheme~\cite{nistDLMFAbout}.  In our phase portraits, simple zeros of our solutions will appear locally as in figure~\ref{fig:wheel}(a), up to a rotation.  Since all singularities of Burgers' equation are simple poles with residue $-2\mu$ (see (\ref{eq:burgerpoleequation})), the singularities in our figures will appear locally like figure~\ref{fig:wheel}(b) (without any rotation).  To take an example, we show in figure~\ref{fig:wheel}(c) a zoomed in version of the phase portrait in figure~\ref{fig:stepdowntotravwave}(f).  On this finer scale it is easier to see the poles and zeros.  Another example is shown in figure~\ref{fig:wheel}(d), this time a zoomed in version of a phase portrait shown later in the paper, namely in figure~\ref{fig:stepup}(e).

As part of our study, we will use exact solutions like \eqref{eq:stepdownsolution} to plot solutions on the real line and phase portraits in the complex-$x$ plane, as we have done in figure~\ref{fig:stepdowntotravwave}.  However, we emphasise that, where possible, we shall develop some methodology in matched asymptotic expansions to describe the trajectories of singularities in the limit $t\rightarrow0^+$ that can potentially be applied for other piecewise-continuous initial conditions and to other nonlinear pdes without exact solutions.

\subsection{Step-down initial condition: small-time analysis}
\label{sec:smalltime}

In the spirit of matched asymptotic expansions, we shall treat (\ref{eq:burgers}) with (\ref{eq:stepdownIC}) by considering a number of different spatial regions in the complex plane in the small-time limit.

\subsubsection{Outer region, $x=\mathcal{O}(1)$}

For Burgers' equation \eqref{eq:burgers} with the step-down initial condition \eqref{eq:stepdownIC}, in lieu of an outer region that involves a power-series expansion of the form (\ref{eq:outergeneric}) (which is not appropriate as the initial condition $u_0(x)$ is not a analytic function of $x$), we first assume $x\in\mathbb{R}$.  Then, for $x=\mathcal{O}(1)$ we simply have $u\sim 1$ as $t\rightarrow 0^+$ for $x<0$ and $u\sim 0$ as $t\rightarrow 0^+$ for $x>0$.

\subsubsection{Inner region, $x=\mathcal{O}(t^{1/2})$}

Clearly there is an inner region near $x=0$ that describes the steep transition from $u\sim 1$ for $x<0$ to $u\sim 0$ for $x>0$.  This observation suggests we examine an inner region in the complex-$x$ plane with $u(x,t)=f(\xi,t)$, where $\xi=x/t^b$ is a similarity variable.  Rewriting Burgers' equation (\ref{eq:burgers0}) exactly using these variables gives
$t f_t-b\xi f_\xi + t^{1-b}ff_\xi=\mu t^{1-2b}f_{\xi\xi}$.  The only possible self-consistent leading-order balance comes from choosing $b=1/2$.  Thus, in the inner region, which is for $x=\mathcal{O}(t^{1/2})$, we write
\begin{equation}
u\left(x, t\right) = f\left(\xi, t\right), \quad \xi = x/t^{1/2}=\mathcal{O}(1),
\label{eq:similarity}
\end{equation}
where
\begin{equation}
    t f_t-\sfrac{1}{2}\xi f_\xi + t^{1/2}ff_\xi=\mu f_{\xi\xi}.
    \label{eq:asymptoticsol1}
\end{equation}
We expand $f$ as
\begin{equation}
    u\sim f_0\left(\xi\right) + t^{1/2}f_1\left(\xi\right) + \ldots \hspace{0.2cm} \mathrm{as} \hspace{0.2cm} t\rightarrow 0^+,
    \label{eq:asymptoticsol2}
\end{equation}
to give
\begin{eqnarray}
-\sfrac{1}{2}\xi f_0' & = & \mu f_0'',
\quad f_0\rightarrow 1 \quad\mbox{as}\quad \xi\rightarrow -\infty,
\quad f_0\rightarrow 0 \quad\mbox{as}\quad \xi\rightarrow +\infty,
\label{eq:asymptoticsol4}
\\
-\sfrac{1}{2}\xi f_1' + \frac{1}{2} f_1 + f_0f_0' & = &\mu f_1'',
\quad f_1\rightarrow 0 \quad\mbox{as}\quad \xi\rightarrow \pm\infty.
    \label{eq:asymptoticsol6}
\end{eqnarray}
These problems for $f_0$ and $f_1$ are both linear, with solutions
\begin{eqnarray}
f_0 & = & \frac{1}{2}-\frac{1}{2}\mathrm{erf}\! {\left(\frac{\xi}{2\sqrt{\mu}}\right)},
        \label{eq:asymptoticsolf0} \\
f_1 & = & \frac{1}{8\sqrt{\pi\mu}}\left( \mathrm{erf}\! {\left(\frac{\xi}{2\sqrt{\mu}}\right)}+1\right)
\left(\frac{\sqrt{\pi}\xi}{\sqrt{\mu}}
\left(-1+\mathrm{erf}\! {\left(\frac{\xi}{2\sqrt{\mu}}\right)}\right)
+2\exp{\left(-\frac{\xi^2}{4\mu}\right)}\right).
    \label{eq:asymptoticsolf1}
\end{eqnarray}
Further terms in (\ref{eq:asymptoticsol2}) could be considered, but these are not needed for what follows (cf.~Appendix~\ref{appendix:discderiv}, where $f_2$ and $f_3$ are needed to resolve the inner scalings).

We note that \eqref{eq:asymptoticsol4} comes from balancing the temporal derivative with the linear diffusion term in \eqref{eq:burgers}, as the nonlinear advection term does not contribute to leading order. Thus, our leading-order approximation $f\sim f_0(\xi)$ as $t\rightarrow0^+$ with $f_0$ given by \eqref{eq:asymptoticsolf0} is the same as the similarity solution for the linear heat equation. We have plotted this approximation in figure \ref{fig:stepdowntotravwave}(a) as a dot-dashed curve to show that it does a good job of approximating the solution on the real line at a moderately small time $t = 0.5$.  Indeed, the comparison is so good that on this scale is difficult to see the dot-dashed curve.  For smaller times, the approximation is better, as expected.

One of our goals is to provide an asymptotic description of the singularity structure in the complex-$x$ plane in the limit $t\rightarrow 0^{+}$; however, the terms $f_0$ and $f_1$ in the series \eqref{eq:asymptoticsol2} do not contain any singularities in the complex plane as they are entire functions of $x$ (as are the neglected terms $f_2$, $f_3$, \ldots).  The reason for the absence of singularities is that, as just mentioned, the dominant balance \eqref{eq:asymptoticsol4} is linear and homogeneous while the nonlinear effects in \eqref{eq:burgers} are relegated to the problem for $f_1$ in \eqref{eq:asymptoticsol6} via the term $f_0f_0'$.  All of these observations suggest that we need to rescale our problem for $f$ in regions in the upper half $\xi$ plane for which the nonlinearity appears in the dominant balance.

\begin{figure}[!h]
        \centering
        \begin{subfigure}[t]{0.4\textwidth}
            \includegraphics[width=\linewidth]{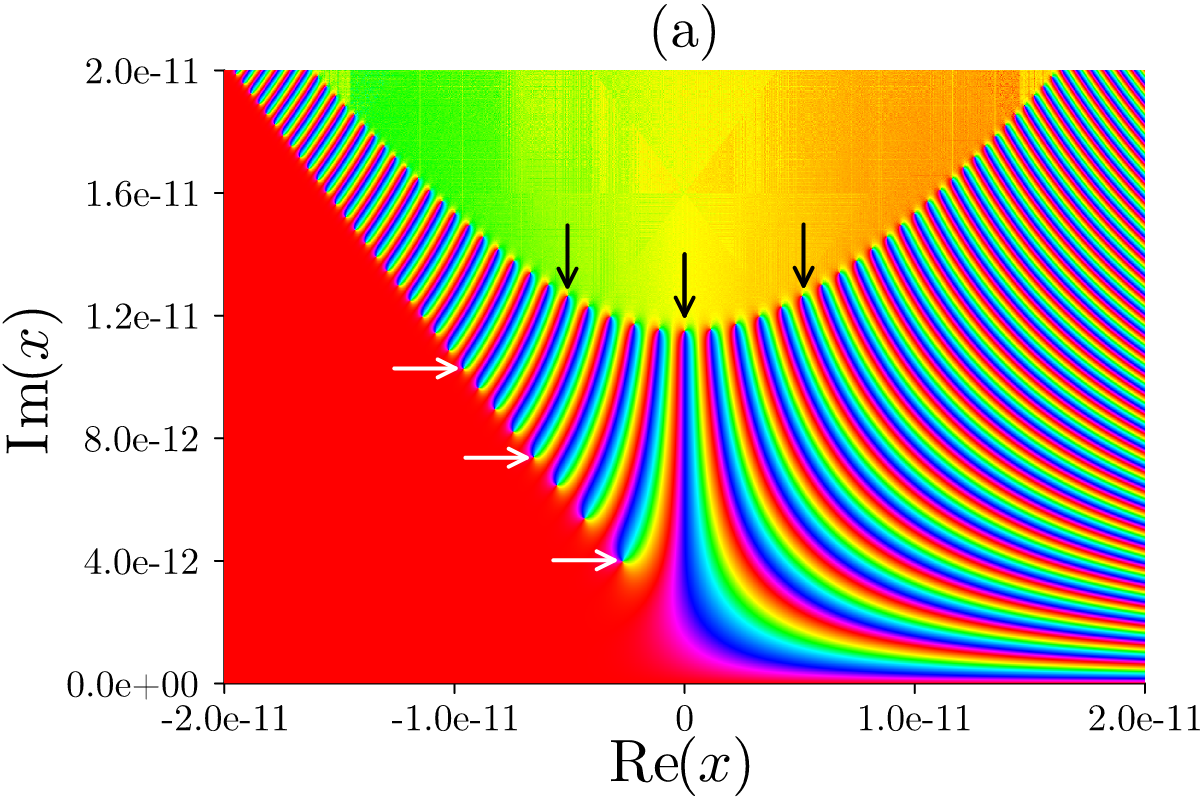}
        \end{subfigure}
        \begin{subfigure}[t]{0.4\textwidth}
            \includegraphics[width=\linewidth]{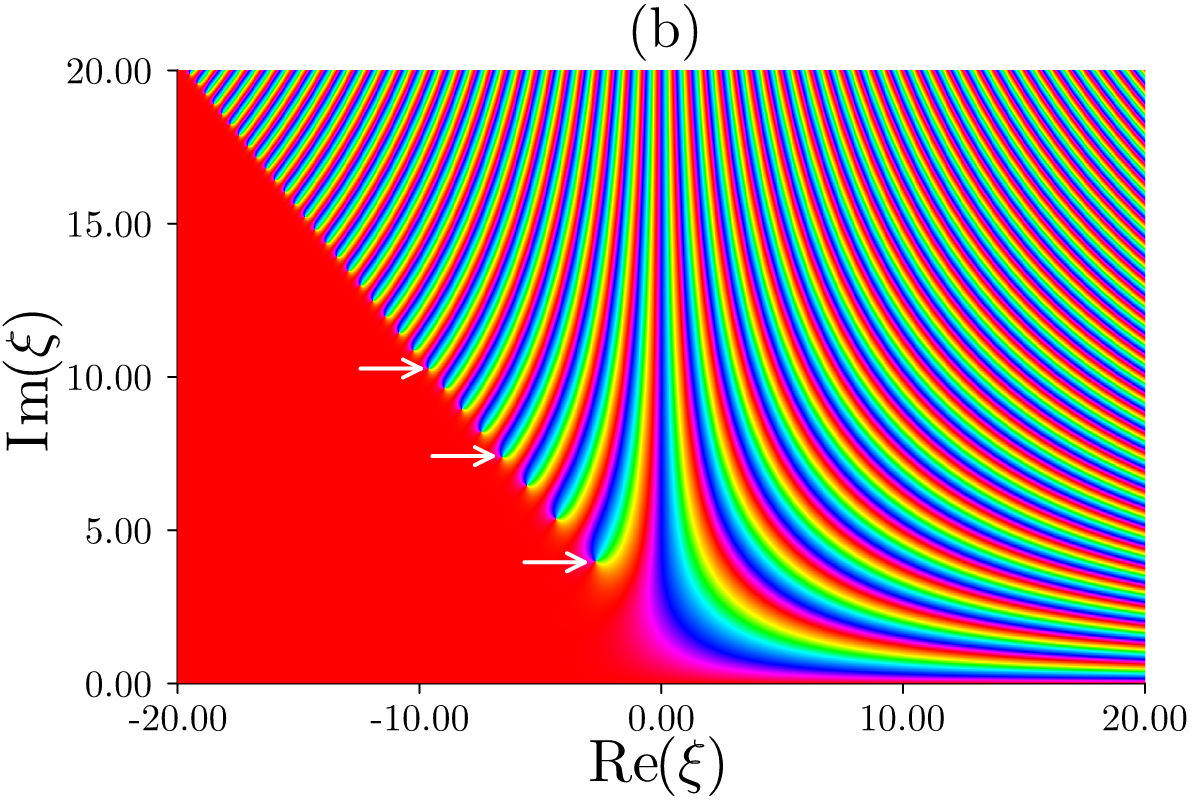}
        \end{subfigure}
        \caption{(a) Phase portrait of the exact solution (\ref{eq:stepdownsolution}) of
        Burgers' equation (\ref{eq:burgers}) with the step-down initial condition \eqref{eq:stepdownIC}, computed for $\mu = 1$ at the extremely small time $t=10^{-24}$.   (b) The leading-order approximation \eqref{eq:asymptoticsolf0} shown on a scale that aligns with part (a).  Black arrows point to poles, while white arrows point to zeros.}
        \label{fig:stepdownphaseandasymptotic}
    \end{figure}

Before doing that, it is worth visualising the solution $u(x,t)$ in the complex plane at an extremely small time.  In figure~\ref{fig:stepdownphaseandasymptotic}(a), we present a phase portrait of the exact solution (\ref{eq:stepdownsolution}) for $\mu=1$ at $t=10^{-24}$ on an extremely small spatial scale near $x=0$.  We have indicated the location of three of the poles with black arrows and three of the zeros with white arrows.  In figure~\ref{fig:stepdownphaseandasymptotic}(b) we plot only the leading-order term $f_0$ in (\ref{eq:asymptoticsol2}), given by (\ref{eq:asymptoticsolf0}).  The scales used in parts (a) and (b) are chosen so that, for this value of $t$, they represent the same spatial region via the relationship $\xi=x/t^{1/2}$.  Since $f_0$ is an entire function, there are no poles in figure~\ref{fig:stepdownphaseandasymptotic}(b), but again three of the zeros are indicated with white arrows.  It is clear from comparing parts (a) and (b) of this figure that the leading-order term $f_0$ does an excellent job of approximating the full solution between the real line and a convex-shaped curve on which the poles are located; on the other hand, above this curve the function $f_0$ is not at all close to the full solution $u(x,t)=f(\xi,t)$.

\subsubsection{Breakdown of inner expansion} \label{stepdownbreakdown}

Our expansion in \eqref{eq:asymptoticsol2} is well-ordered in regions in which $t^{1/2}f_1(\xi)$ is much smaller than $f_0(\xi)$, such as on the real line. As $\mathrm{Im}(\xi)$ increases, the leading-order term $f_0$ increases exponentially in size, but the first correction term $f_1$ increases even faster. This means that, for a fixed time $t\ll 1$,  \eqref{eq:asymptoticsol2} will remain valid only if we are not too far away from the real-$\xi$ axis.

In particular, \eqref{eq:asymptoticsol2} ceases to be well ordered where $f_0(\xi)$ is the same size as $t^{1/2}f_1(\xi)$, i.e. where
\begin{equation}
f_0 = \mathcal{O}(t^{1/2}f_1).
\label{eq:f0tf1}
\end{equation}
The smaller $t$ becomes, the further away this non-uniformity occurs in the $\xi$-plane, so we apply approximations of $f_0$ and $f_1$ as $\xi \rightarrow \mathrm{i}\infty$ and then solve for $\xi$ in (\ref{eq:f0tf1}).  Using the standard asymptotic behaviour of the error function for complex arguments \cite{nistDLMFAbout}, we find
\begin{equation}
    f_0 \sim \sqrt{\frac{\mu}{\pi}}\frac{1}{\xi}
    \,\mathrm{e}^{-\xi^2/4\mu}, \hspace{0.2cm} f_1 \sim -\frac{\mu}{\pi}\frac{1}{\xi^3}
    \,\mathrm{e}^{-\xi^2/2\mu}\hspace{0.2cm} \text{as} \hspace{0.2cm} \xi \rightarrow \mathrm{i}\infty,
    \label{eq:leadingorderexpansion}
\end{equation}
which suggests that (\ref{eq:f0tf1}) becomes
\begin{equation}
    -\sqrt{\frac{\pi}{\mu}}\xi^2
    \,\mathrm{e}^{\xi^2/4\mu}
        =\mathcal{O}(t^{1/2}).
    \label{eq:solveforxi}
\end{equation}
Solving (\ref{eq:solveforxi}) for $\xi$ will locate inner-inner regions in the $\xi$-plane in which we shall rescale our governing equations.

We may deal with (\ref{eq:solveforxi}) asymptotically by noting that, to leading order, $\mathrm{e}^{\xi^2/4\mu}$ balances $t^{1/2}$, which suggests $\xi^2\sim -2\mu\ln(1/t)$ as $t\rightarrow 0^+$.  Further terms must involve $\ln(\ln(1/t))$ in order to take account of the $\xi^2$ term on the left-hand side of (\ref{eq:solveforxi}).  By posing $\xi^2\sim -2\mu\ln(1/t)+a_0\ln\ln(1/t)+a_1$, substituting into (\ref{eq:solveforxi}), and solving for $a_0$ and $a_1$, we find
\begin{equation}
    \frac{\xi^2}{4\mu}\sim -\sfrac{1}{2}\ln(1/t) -\ln\left(\ln(1/t)\right)-\ln 2\sqrt{\mu\pi}+2m\pi\mathrm{i}, \quad m\in\mathbb{Z}.
    \label{eq:xi_1_form}
\end{equation}
Here the counter $m$ indexes the location of inner-inner regions, which we describe shortly.  In the expansion (\ref{eq:xi_1_form}), for a fixed $m$, the size of the terms are ordered by $\sfrac{1}{2}\ln(1/t)\gg \ln\left(\ln(1/t)\right)\gg 2m\pi$, provided $t$ is sufficiently small.  This is quite a restrictive condition; for example, if $t=10^{-3}$, then $\sfrac{1}{2}\ln(1/t)\approx 3.45$ and $\ln\left(\ln(1/t)\right)\approx 1.93$, neither of which would normally be considered large, and certainly not even larger than $2m\pi$ for $m=1$.  For this reason, we shall focus on the cases $m=\mathcal{O}(1)$ (for extremely small times) and $m=\mathcal{O}(\ln(1/t))$ (for moderately small times) separately.

\subsubsection{Inner-inner regions, $\xi=\mathcal{O}\left(\mathrm{i}\sqrt{2\mu}\ln^{1/2}(1/t)\right)$, $m=\mathcal{O}(1)$}\label{sec:innerinnerm1}

For each $m=\mathcal{O}(1)$, there will be an inner-inner region with a new spatial variable $X$ defined by
\begin{equation}
    \frac{\xi}{\sqrt{2\mu}} = \mathrm{i}\ln^{1/2}\left(1/t\right)
    +\frac{\mathrm{i}\ln\left(\ln\left(1/t\right)\right) + 2m\pi +\mathrm{i}\ln\left(2\sqrt{\mu\pi}\right)+X}{\ln^{1/2}\left(1/t\right)},
    \quad X=\mathcal{O}(1).
    \label{eq:inner_regions3}
\end{equation}
By writing the limiting behaviour of the inner solution, (\ref{eq:leadingorderexpansion}), in terms of this new inner-inner variable $X$, we find
\begin{equation}
    f_0 + t^{1/2}f_1\sim \mu^{1/2}\,\frac{\ln^{1/2}\left(1/t\right)}{(2t)^{1/2}}
    \left(-2\mathrm{i}\mathrm{e}^{-\mathrm{i}X}-2\mathrm{i}\mathrm{e}^{-2\mathrm{i}X}\right),
    \label{eq:inner_regions5}
\end{equation}
which has two implications.  First, the temporal scaling at the front of the right-hand side of (\ref{eq:inner_regions5}) suggests we pose the ansatz
\begin{equation}
    f = \mu^{1/2} \,\frac{\ln^{1/2}\left(1/t\right)}{(2t)^{1/2}}F\left(X, t\right)
    \label{eq:inner_regions6}
\end{equation}
for each inner-inner region (i.e., for each $m$). Thus, our inner-inner regions
are defined by (\ref{eq:inner_regions6}) with (\ref{eq:inner_regions3}).
Second, the term $-2\mathrm{i}\mathrm{e}^{-\mathrm{i}X}-2\mathrm{i}\mathrm{e}^{-2\mathrm{i}X}$
on the right-hand side of (\ref{eq:inner_regions5}) gives far-field conditions for $F$ below in the limit $X\rightarrow -\mathrm{i}\infty$ (the reason for matching in this direction is that we need to link back to the outer region on the real line, given we are in the upper half plane).  With all of this in mind, a singularity $X=X_s$ of $F(X,t)$ corresponds to a singularity $x=s_m(t)$ of the original dependent variable $u(x,t)$, where
\begin{equation}
s_m(t) \sim \mathrm{i}\sqrt{2\mu}\,t^{1/2}\ln^{1/2}\left(1/t\right)
+
\frac{\sqrt{2\mu}\,t^{1/2}\left(
\mathrm{i}\ln\left(\ln\left(1/t\right)\right) + 2m\pi +\mathrm{i}\ln\left(2\sqrt{\mu\pi}\right)\right)}{\ln^{1/2}\left(1/t\right)}
+\mu^{1/2}\,\frac{(2t)^{1/2}}{\ln^{1/2}\left(1/t\right)}\,X_s
\label{eq:inner_regions11}
\end{equation}
as $t\rightarrow 0^+$.

Given (\ref{eq:inner_regions6}), the first term on the left-hand side of (\ref{eq:asymptoticsol1}) is of $\mathcal{O}(t^{-1/2}\ln^{1/2}(1/t))$, while the remaining terms are of
$\mathcal{O}(t^{-1/2}\ln^{3/2}(1/t))$.
By writing $F\sim F_0$ as $t\rightarrow0^+$ and collecting leading-order terms, we find
\begin{align}
    &\mathrm{i}F_0' + F_0 F_0' = F_0'', \label{eq:inner_regions7A} \\
    F_0 &\sim-2\mathrm{i}\,\mathrm{e}^{-\mathrm{i}X}
    - 2\mathrm{i}\,\mathrm{e}^{-2\mathrm{i}X} \hspace{0.2cm} \text{as} \hspace{0.2cm} X \to -\mathrm{i}\infty. \label{eq:inner_regions7B}
\end{align}
We can integrate \eqref{eq:inner_regions7A} once to arrive at the first-order separable equation
\begin{equation}
    -\mathrm{i}F_0+\sfrac{1}{2}F_0^2=F_0',
    \label{eq:inner_regions8}
\end{equation}
whose solution subject to \eqref{eq:inner_regions7B} is simply
\begin{equation}
    F_0=\frac{2\,\mathrm{i}}{1-\mathrm{e}^{\mathrm{i}X}}.
    \label{eq:inner_regions9}
\end{equation}
Therefore, the inner-inner region for each $m$ has
\begin{equation}
    f\sim \mu^{1/2}\, \frac{\ln^{1/2}\left(1/t\right)}{(2t)^{1/2}}\frac{2\,\mathrm{i}}
    {1-\mathrm{e}^{\mathrm{i}X}}
    \hspace{0.2cm} \text{as} \hspace{0.2cm} t\rightarrow0^+,
    \label{eq:inner_regions10}
\end{equation}
provided $m = \mathcal{O}(1)$.

Clearly, \eqref{eq:inner_regions9} has a simple pole at $X = X_s$ for each $X_s=2m\pi$ and, in fact, from \eqref{eq:inner_regions10} we have
\begin{equation*}
        f \sim \mu^{1/2} \,\frac{\ln^{1/2}\left(1/t\right)}{(2t)^{1/2}}
        \left(-\frac{1}{X-X_s}\right) \hspace{0.2cm} \text{as} \hspace{0.2cm} X\rightarrow X_s.
\end{equation*}
In terms of the original variables, after noting that
\begin{equation*}
x-s_m(t)\sim \mu^{1/2} \,\frac{\ln^{1/2}\left(1/t\right)}{(2t)^{1/2}}(X-X_s),
\end{equation*}
we have $u\sim -2\mu/(x - s_m(t))$ as $x\rightarrow s_m(t)$.
Thus, provided $m=\mathcal{O}(1)$, we have simple poles with the required residue (see \eqref{eq:burgerpoleequation}) at (\ref{eq:inner_regions11}) with $X_s=2m\pi$.

\begin{figure}[!h]
        \centering
        \begin{subfigure}[t]{0.4\textwidth}
            \includegraphics[width=\linewidth]{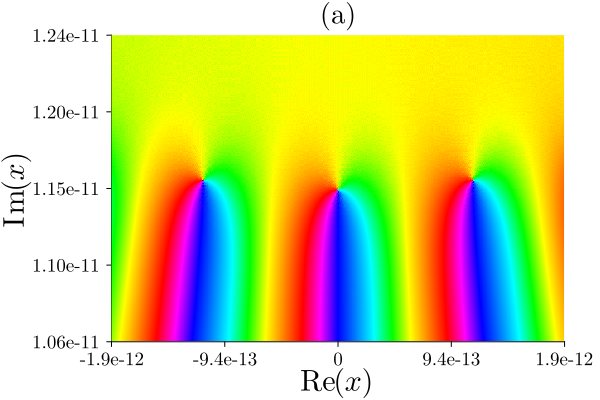}
        \end{subfigure}
        \begin{subfigure}[t]{0.4\textwidth}
            \includegraphics[width=\linewidth]{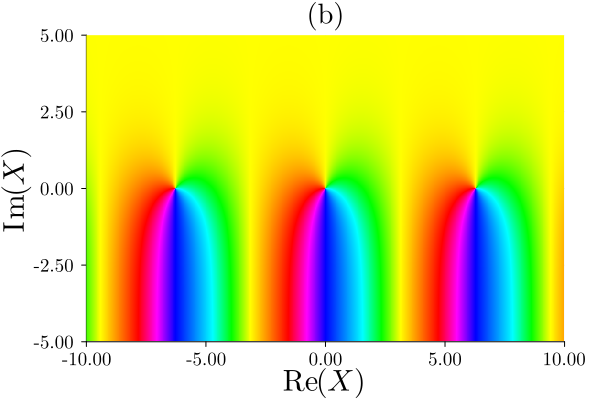}
        \end{subfigure}

        \begin{subfigure}[t]{0.4\textwidth}
            \includegraphics[width=\linewidth]{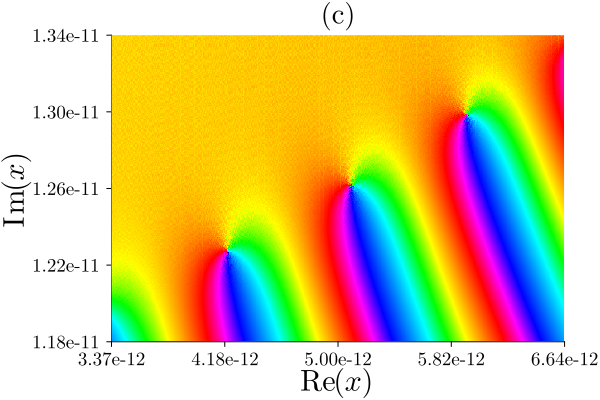}
        \end{subfigure}
        \begin{subfigure}[t]{0.4\textwidth}
            \includegraphics[width=\linewidth]{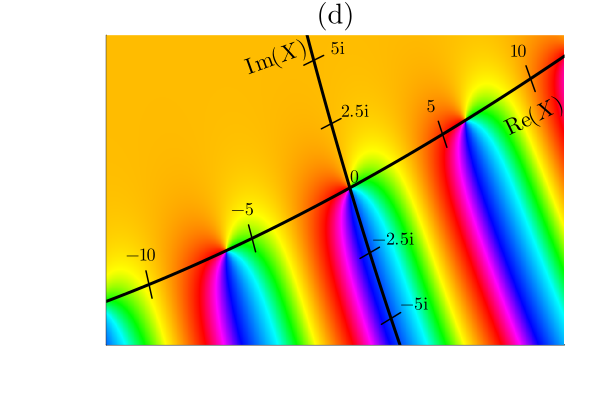}
        \end{subfigure}
        \caption{(a) Zoomed-in portion of the phase portrait shown in figure~\ref{fig:stepdownphaseandasymptotic}(a).  This is the exact solution (\ref{eq:stepdownsolution}) computed for $\mu = 1$ at the extremely small time $t=10^{-24}$.  (b) The inner-inner solution \eqref{eq:inner_regions10}, which valid in this neighbourhood of the closest poles to the real axis.  The variable $X$ in \eqref{eq:inner_regions10} is related to $x=t^{1/2}\xi$, where $\xi$ is given by (\ref{eq:inner_regions3}) for $m=0$.  (c) Zoomed-in portion of figure~\ref{fig:stepdownphaseandasymptotic}(a) near the fifth pole from the centre, $m=5$.  (d) The inner-inner solution (\ref{eq:inner_regions10analogue}) centred at the pole for $m=5$, where $X$ is related to $x=t^{1/2}\xi$ via (\ref{eq:innerregionLambert}).}
        \label{fig:innerregioncompare}
    \end{figure}

In figure \ref{fig:innerregioncompare}(a)-(b) we illustrate the inner solution \eqref{eq:inner_regions9} by comparing it with the exact solution (\ref{eq:stepdownsolution}).  In part (a) of the figure is the exact solution plotted at the extremely small time $t=10^{-24}$ in an extremely small spatial region that includes the closest three singularities to the real-$x$ axis. Note this plot in figure \ref{fig:innerregioncompare}(a) is a zoomed-in version of the phase portrait in figure \ref{fig:stepdownphaseandasymptotic}(a). In part (b) of figure~\ref{fig:innerregioncompare} is the inner solution (\ref{eq:inner_regions9}).  Here, the independent variable $X$ is related to $x$ via the complicated scaling \eqref{eq:inner_regions3}, where $\xi=x/t^{1/2}$. One observation is that the plots in parts (a) and (b) of figure~\ref{fig:innerregioncompare} are in very close agreement, providing strong confidence in the asymptotic analysis. Another is that, for extremely early times, the local solution \eqref{eq:inner_regions9} near each pole resembles a finger-like structure with a simple-pole singularity at the tip of each finger. This finger-like solution comes from (\ref{eq:inner_regions8}), a rescaled version of the classical travelling wave ordinary differential equation (ode), subject to the far-field condition $F_0\rightarrow 0$ as $X\rightarrow -\mathrm{i}\infty$. We discuss this observation in more detail in Appendix~\ref{appendix:inner} and return to the panels in figure~\ref{fig:innerregioncompare}(c),(d) below.

\subsubsection{Inner-inner regions, $\xi=\mathcal{O}\left(\mathrm{i}\sqrt{2\mu}
(\ln(1/t)+2\ln(2\sqrt{\mu\pi})-4m\pi\mathrm{i}))^{1/2}\right)$, $m=\mathcal{O}(\ln(1/t))$}\label{sec:stepdownagain}

As mentioned above, our analysis in section~\ref{sec:innerinnerm1} holds in the limit $t\rightarrow 0^+$ for $m=\mathcal{O}(1)$ only.  With these assumptions, we have successfully predicted the location of the closest singularity to the real axis (which is also closest to the imaginary axis) and perhaps a couple of the neighbouring poles, depending on the size of $t$.  However, from figure~\ref{fig:stepdownphaseandasymptotic}(a) it is clear that, even for an extremely small value of $t$, the string of infinitely many poles follows a convex-shaped curve which asymptotes towards rays at angles $\pi/4$ and $3\pi/4$.  To capture the poles that are much further away from the imaginary axis, we arrange our terms to allow for the integer $m$ to be of $\mathcal{O}(\ln(1/t))$ or smaller.

One way of proceeding is to revisit (\ref{eq:solveforxi}) and note that we can solve $-\sqrt{\pi/\mu}\,\xi^2\,\mathrm{exp}(\xi^2/4\mu)=t^{1/2}$ exactly to give
\begin{equation}
    \xi = 2\sqrt{\mu}\left(W_{n}\left(\frac{-t^{1/2}}{4\sqrt{\mu \pi}}\right)\right)^{1/2}, \hspace{0.5cm} n = \pm1, \pm2, ...
    \label{eq:lambertscaled}
\end{equation}
where each $W_n(z)$ denotes one of the infinite number of branches the Lambert-$W$ function~\cite{corless1996lambert,nistDLMFAbout}. The principle branch $W_0(z)$ is not relevant here (as solutions of
(\ref{eq:solveforxi}) on the principle branch behave as $|\xi|\rightarrow 0$ instead of $|\xi|\rightarrow\infty$ as $t\rightarrow 0^+$).  The branch of the one-half power in \eqref{eq:lambertscaled} is chosen such that the $\xi$ values lie in the upper half $\xi$ plane.  With these conventions, equation (\ref{eq:lambertscaled}) locates all of the inner-inner regions near the singularities (see the phase portrait in figure~\ref{fig:stepdownphaseandasymptotic}(a), for example, with the poles arranged along a convex-shaped curve), with $n=-1$ for the closest inner-inner region to the origin in the $\xi$-plane, while the values $n=-2, -3, ...$, locate the inner-inner regions in the second quadrant and the values $n = 1, 2, ...$, are for inner-inner regions in the first quadrant.

We make use of the asymptotic expansion
$$
W_n(-\epsilon)\sim -\ln(1/\epsilon)+(2n+1)\pi\mathrm{i}-\mathrm{Log}\left(
-\ln(1/\epsilon)+(2n+1)\pi\mathrm{i}\right)
\quad\mbox{as}\quad \epsilon\rightarrow 0^+
$$
\cite{corless1996lambert,nistDLMFAbout}, where here we are using $\mathrm{Log}(z)$ to denote the principal branch of the complex logarithm.  Thus, from (\ref{eq:lambertscaled}) we define our inner-inner regions by
\begin{equation}
\frac{\xi^2}{4\mu} =-\sfrac{1}{2}\ln(1/t)-\ln (2\sqrt{\mu\pi})+2m\pi\mathrm{i}
-\mathrm{Log}\left(\ln(1/t)+2\ln (2\sqrt{\mu\pi})-4m\pi\mathrm{i}
\right)+\mathrm{i}X, \quad X=\mathcal{O}(1).
\label{eq:innerregionLambert}
\end{equation}
This relationship between $\xi$ and the new inner variable $X$, valid for $m=\mathcal{O}(\ln(1/t))$ as $t\rightarrow 0^+$, is analogous to (\ref{eq:inner_regions3}), which was for $m=\mathcal{O}(1)$ only.
Thus, from (\ref{eq:leadingorderexpansion}), we have
$$
f_0+t^{1/2}f_1\sim
\frac{\mu^{1/2}}{(2t)^{1/2}}
\left(\ln(1/t)+2\ln (2\sqrt{\mu\pi})-4m\pi\mathrm{i}
\right)^{1/2}
\left(-2\mathrm{i}\,\mathrm{e}^{-\mathrm{i}X}
-2\mathrm{i}\,\mathrm{e}^{-2\mathrm{i}X}\right),
$$
where we are deliberately grouping the terms $\ln(1/t)$, $2\ln (2\sqrt{\mu\pi})$ and $4m\pi\mathrm{i}$ together. These scalings suggest that for this inner-inner region we define
\begin{equation}
f=\mu^{1/2}\,\frac{\left(\ln(1/t)+2\ln (2\sqrt{\mu\pi})-4m\pi\mathrm{i}
\right)^{1/2}}{(2t)^{1/2}}
F(X,t)
\label{eq:innerinnerdefnf}
\end{equation}
(cf.~(\ref{eq:inner_regions6})), so if we write $F\sim F_0(X)$ as $t\rightarrow 0^+$ for $m=\mathcal{O}(\ln(1/t))$, then we find $F_0$ satisfies (\ref{eq:inner_regions7A})-(\ref{eq:inner_regions7B}) with solution (\ref{eq:inner_regions9}), so that
\begin{equation}
f\sim \mu^{1/2}\,\frac{\left(\ln(1/t)+2\ln (2\sqrt{\mu\pi})-4m\pi\mathrm{i}
\right)^{1/2}}{(2t)^{1/2}}
\frac{2\,\mathrm{i}}{1-\mathrm{e}^{\mathrm{i}X}}
\quad\mbox{as}\quad t\rightarrow 0^+.
\label{eq:inner_regions10analogue}
\end{equation}
Again, $F_0$ has a simple pole at $X=0$ and, in fact, has poles at $X=2k\pi$ for $k\in\mathbb{Z}$, all of which have the required residue (given by \eqref{eq:burgerpoleequation}).  Note the integer $k$ can be absorbed into $m$ in (\ref{eq:innerregionLambert}), so there is no need to keep track of both $m$ and $k$.

In this way, we are able to describe inner-inner regions around each of the singularities, located asymptotically, in original variables, at
\begin{align}
s_m(t) \sim & \,\,\mathrm{i}\sqrt{2\mu}\,t^{1/2}
\left(\ln(1/t)+2\ln (2\sqrt{\mu\pi})-4m\pi\mathrm{i}\right)^{1/2}
+ \mathrm{i}\sqrt{2\mu}\,t^{1/2}
\frac{\mathrm{Log}\left(\ln(1/t)+2\ln (2\sqrt{\mu\pi})-4m\pi\mathrm{i}\right)}
{\left(\ln(1/t)+2\ln (2\sqrt{\mu\pi})-4m\pi\mathrm{i}\right)^{1/2}}.
\label{eq:innerextended}
\end{align}
This formula provides an excellent approximation for all of the poles for small time, as we show in figure~\ref{fig:stepdowntotravwave}(c), which is drawn for an extremely small time.  The black dots in this image show the predicted location of the poles according to (\ref{eq:innerextended}).  The approximation clearly picks up all of the poles along a convex-shaped curve, providing confidence that our asymptotics are correct.  Importantly, if we instead plot the black dots using the Lambert-$W$ function from
(\ref{eq:lambertscaled}), then the results would appear the same on this scale.  In part (e) of  figure~\ref{fig:stepdowntotravwave}, we also plot black dots to indicate the predicted pole locations.  Even for this much larger time, $t=0.1$, the approximation is still excellent.

Returning to figure~\ref{fig:innerregioncompare}, in panel (c) we show a phase portrait of the exact solution (\ref{eq:stepdownsolution}) drawn for the extremely small time $t=10^{-24}$ close to the fifth singularity to the right of the imaginary axis.   Then, in panel (d) we plot the inner-inner approximation (\ref{eq:inner_regions10analogue}) versus the variable $X$.  To make a clear comparison with the exact solution in (c), we have presented the image so that the horizontal and vertical directions are the real and imaginary part of the original variable $x$, which means we have had to relate $X$ to $x=t^{1/2}\xi$ by the change of variables (\ref{eq:innerregionLambert}).
We see from the colour pattern in (c) and (d) that the inner-inner solution is doing an excellent job of approximating the exact solution.

Note that the predicted pole location (\ref{eq:innerextended}) appears to work well even for $|m|\gg \ln(1/t)$.  Indeed, if we take $m=\mathcal{O}(1/t)$ in (\ref{eq:innerextended}) then we find $s_m\sim 2\sqrt{2\mu\pi}(mt)^{1/2} \mathrm{e}^{\pi\mathrm{i}/4}$ for $m>0$ and $s_m\sim 2\sqrt{2\mu\pi}(-mt)^{1/2} \mathrm{e}^{3\pi\mathrm{i}/4}$ for $m<0$.  In both cases, the distance between the poles decays like $\sqrt{2\mu\pi}(t/m)^{1/2}$.  Thus, we see that our asymptotics demonstrates that the closest poles to the real axis (small $|m|$) move with a speed of $\mathcal{O} (t^{-1/2}\ln^{1/2}(1/t))$ as $t\rightarrow 0^+$, while, as we move further way (large $|m|$), the speed transitions to be of $\mathcal{O}(t^{-1/2})$.

\subsection{Step-down initial condition: evolution to a travelling wave}
\label{stepdowntotravwave}

It is well known that, for certain initial conditions with $u_L>u_R$ that satisfy $u_0\rightarrow u_L$ as $x\rightarrow -\infty$ and $u_0\rightarrow u_R$ as $x\rightarrow\infty$, solutions of (\ref{eq:burgers}) evolve to the travelling wave profile
\begin{equation}
     U_{\mathrm{tw}}(z) = u_R+\frac{u_L-u_R}{1+
     \mathrm{e}^{(u_L-u_R)z/2\mu}}
\label{eq:travwavesolution}
\end{equation}
in the long-time limit, where $z=x-ct-x_0$ is the travelling wave variable and the travelling wave speed is given by $c=(u_L+u_R)/2$. Here, the parameter $x_0$ is a real-valued shift that, in general, depends on the initial condition in a complicated way.  For the step-down initial condition \eqref{eq:stepdownIC} with $u_L=1$ and $u_R=0$, the travelling wave solution (\ref{eq:travwavesolution}) reduces to
\begin{equation}
U_{\mathrm{tw}}(z)=\frac{1}{1+\mathrm{e}^{(x-t/2-x_0)/2\mu}}.
\label{eq:travwavesolution2}
\end{equation}
We have plotted (\ref{eq:travwavesolution2}) as a dashed purple curve in figure~\ref{fig:stepdowntotravwave}(a) using $x_0=0$, $\mu=1$ and $t=15$.  We see that, at this time, the time-dependent solution (\ref{eq:stepdownsolution}), plotted as the solid purple curve, is quite close to the travelling wave profile.  For later times, the comparison is even better, as expected.

In the complex plane, the travelling wave solution (\ref{eq:travwavesolution2}) has a vertical array of simple poles located at $x=x_0+\sfrac{1}{2}t+2\mu(2j-1)\pi\mathrm{i}$.  In the upper half plane, $j=1,2,\ldots$.  The phase portrait of (\ref{eq:travwavesolution2}) involves an array of horizontal fingers, each with a tip at one of the singularities.  A very good representation of this phase portrait is illustrated in figure~\ref{fig:stepdowntotravwave}(k), where a solution is presented for a large time, $t=300$ (with this large time, the time-dependent solution has effectively evolved to the travelling wave solution on this scale).

Reflecting on the phase portraits in figure~\ref{fig:stepdowntotravwave}(d)-(f), computed for small to intermediate times, it is not immediately clear how this $V$-shaped arrangement of poles ends up approaching a vertical array that comes from the travelling wave limit.  To explain this process, we show in figure~\ref{fig:stepdowntotravwave}(g)-(k) five further phase portraits plotted for larger times.  The first observation is that in parts (d)-(f), there is a $V$-shaped configuration of poles centred at $\mathrm{Re}(x)=0$, a few of which are indicated by black arrows in (d).  Each pole of $u(x,t)$ on the left arm of the $V$ is close to a zero of $u(x,t)$, forming a dipole-like arrangement, while each pole on the right arm of the $V$ is the tip of a finger-like structure.

The second observation is that the window in which we present the phase portraits moves to the right with speed approximately $1/2$, which is the travelling wave speed for this solution.  For example, the window for $t=25$ in figure~\ref{fig:stepdowntotravwave}(g) is centred at $\mathrm{Re}(x)=12.5$.  Thus, the $V$-shaped configuration is clearly moving to the right.  However, the pairs of poles and zeros on the left arm of the $V$ are travelling faster to the right than the poles on the right arm.  This difference in speed causes pole-zero pairs on the left arm to ultimately collide with poles on the right arm.

More specifically, the pole that is closest to the imaginary axis in figure~\ref{fig:stepdowntotravwave}(d)-(f) does not collide with any other pole or zero; it simply evolves to be the pole closest to the real line in parts (g)-(k).  It is attached to a curved finger-like structure that eventually flattens out and becomes horizontal.  This is the lowest horizontal finger in part (k).  The remaining poles behave in a more complicated way.  The lowest pole-zero pair on the left arm in figure~\ref{fig:stepdowntotravwave}(d)-(f) collides with the lowest pole on the right arm, connected to the second-lowest finger on the right, at roughly $t=30$.  Thus, part (g) shows the phase portrait just before this collision.  After the collision, the pole-zero pair deflects off the finger-tip pole and starts a new finger of its own.  By the time we reach $t=70$ in part (i), we can see this new finger developing, which will end up being the third lowest horizontal finger in part (k).

This pattern continues, so that the second-lowest pole and its associated finger on the right arm in figure~\ref{fig:stepdowntotravwave}(d)-(f) becomes the fourth lowest pole and finger in part (k), while the second-lowest pole-zero pair on the left arm in figure~\ref{fig:stepdowntotravwave}(d)-(f) becomes the fifth lowest horizontal finger in part (k), after the collision at roughly $t=75$, and so on.  This series of collisions between pole-zero pairs and poles that lie at the tips of fingers is best observed by slowing down the accompanying video in the supporting materials for this manuscript.

\subsection{Step-down initial condition: special case $u_R=-u_L$}
\label{sec:stepdown}

A special case of the initial condition (\ref{eq:heaviside}) arises when $u_R=-u_L$ for $u_L>u_R$. Here, the real-valued solution remains odd for all time and ultimately evolves to a ``travelling wave’’ solution with speed $(u_L+u_R)/2=0$.  That is, the time-dependent solution approaches a steady-state solution profile, which is simply
\begin{equation}
   U_{\mathrm{tw}}(x) = u_L\tanh\left(\dfrac{u_L x}{2\mu}\right).
    \label{eq:oddtravwave}
\end{equation}
This steady-state solution has poles at $x=(2j-1)\mu\pi\mathrm{i}/u_L$, $j\geq 1$, and zeros at $x=2j\mu\pi\mathrm{i}/u_L$ for $j\geq 0$.  We have two motivations for highlighting this special case.  First, the odd-ness of the time-dependent solution forces poles and zeros to collide on the imaginary axis and to subsequently remain on this axis.  We can therefore easily plot their speed and location versus time on a single figure.  Second, the steady-state solution (\ref{eq:oddtravwave}) is qualitatively different from the travelling wave solution considered in subsection \ref{stepdowntotravwave} in the sense that it has a string of zeros that lie in between the poles, whereas the travelling wave solution (\ref{eq:travwavesolution2}) does not have any zeros in the plane (its zeros appear only as $\mathrm{Re}(x)\rightarrow\infty$).

\begin{figure}[!h]
        \centering
        \begin{subfigure}[t]{0.32\textwidth}
            \includegraphics[width=\linewidth]{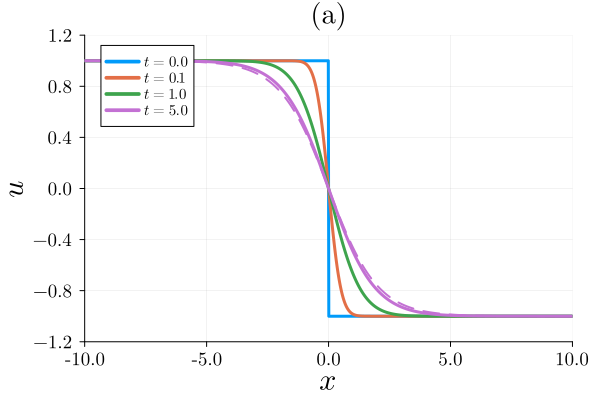}
        \end{subfigure}
        \begin{subfigure}[t]{0.32\textwidth}
            \includegraphics[width=\linewidth]{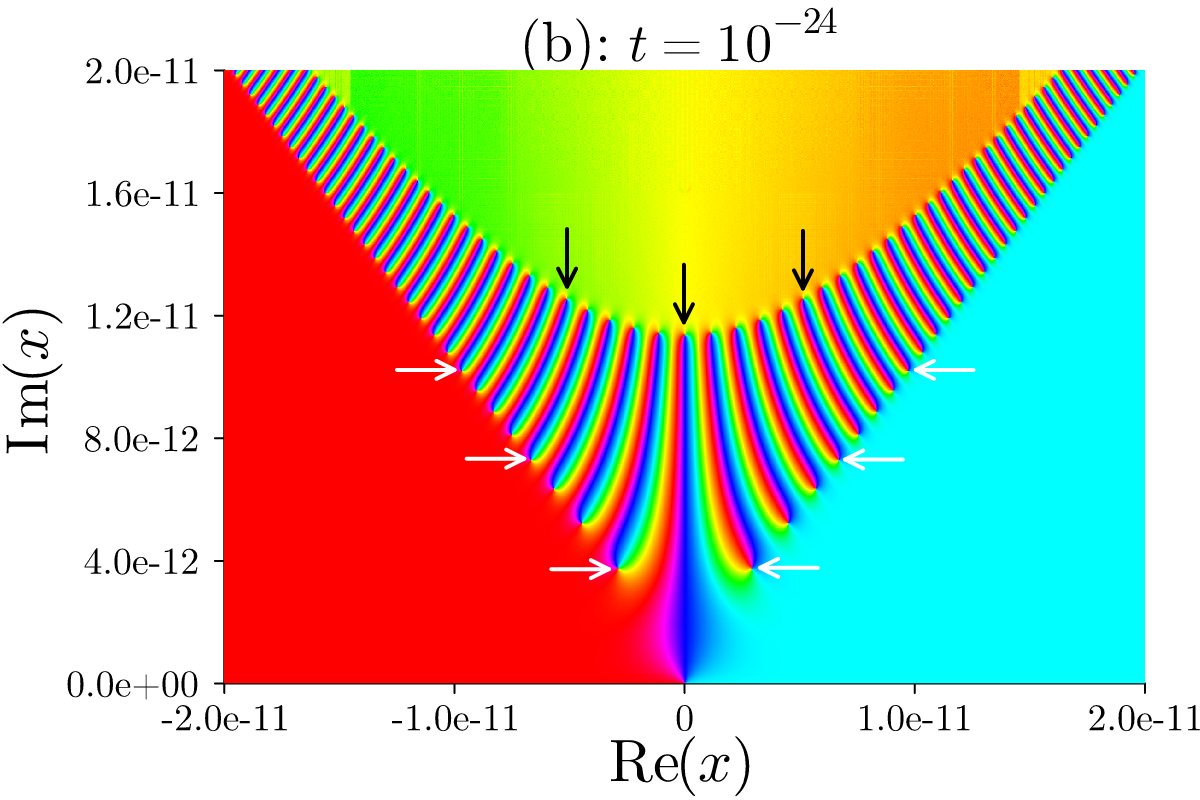}
        \end{subfigure}
        \begin{subfigure}[t]{0.32\textwidth}
            \includegraphics[width=\linewidth]{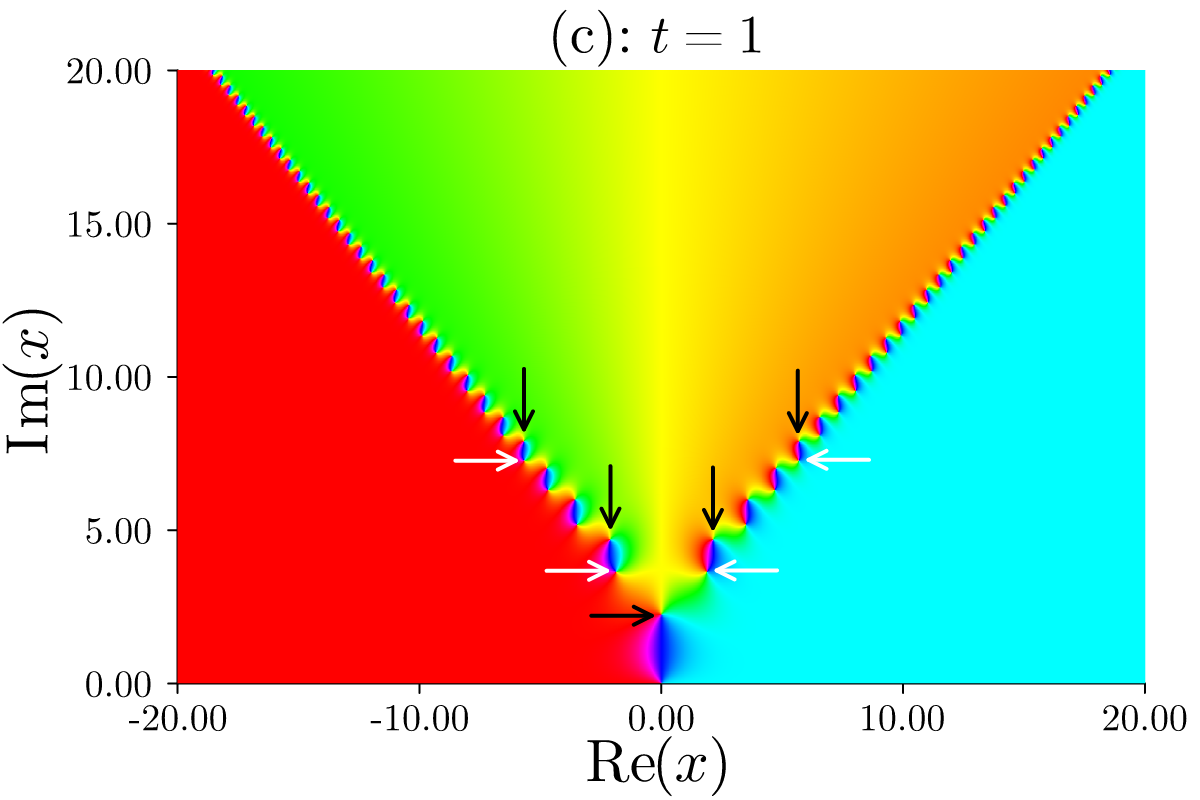}
        \end{subfigure}
        \begin{subfigure}[t]{0.32\textwidth}
            \includegraphics[width=\linewidth]{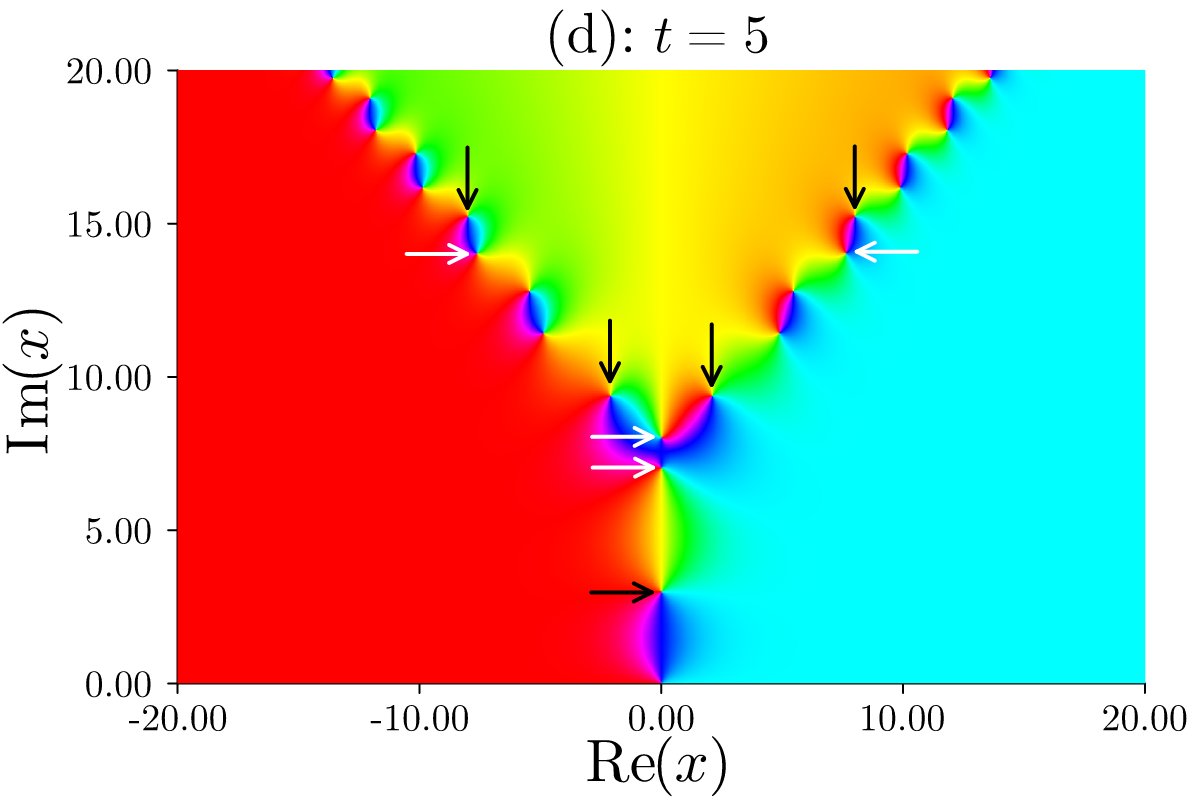}
        \end{subfigure}
        \begin{subfigure}[t]{0.32\textwidth}
            \includegraphics[width=\linewidth]{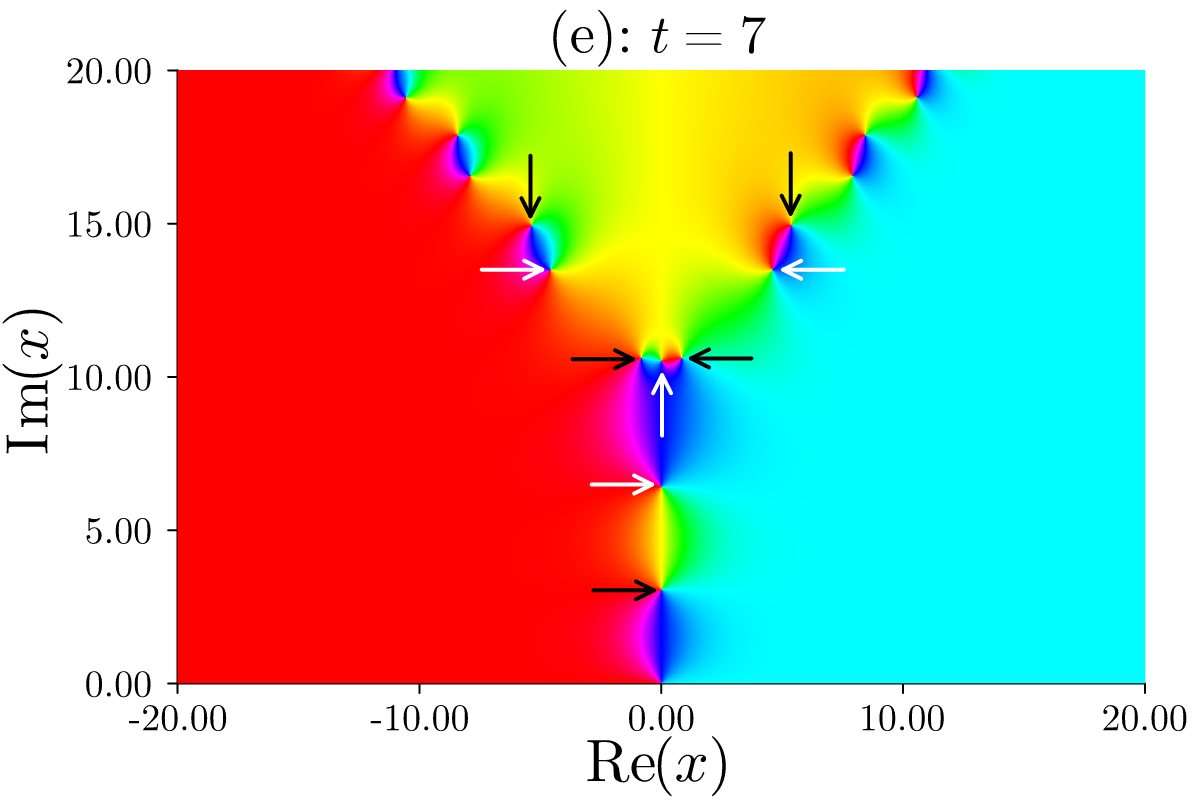}
        \end{subfigure}
        \begin{subfigure}[t]{0.32\textwidth}
            \includegraphics[width=\linewidth]{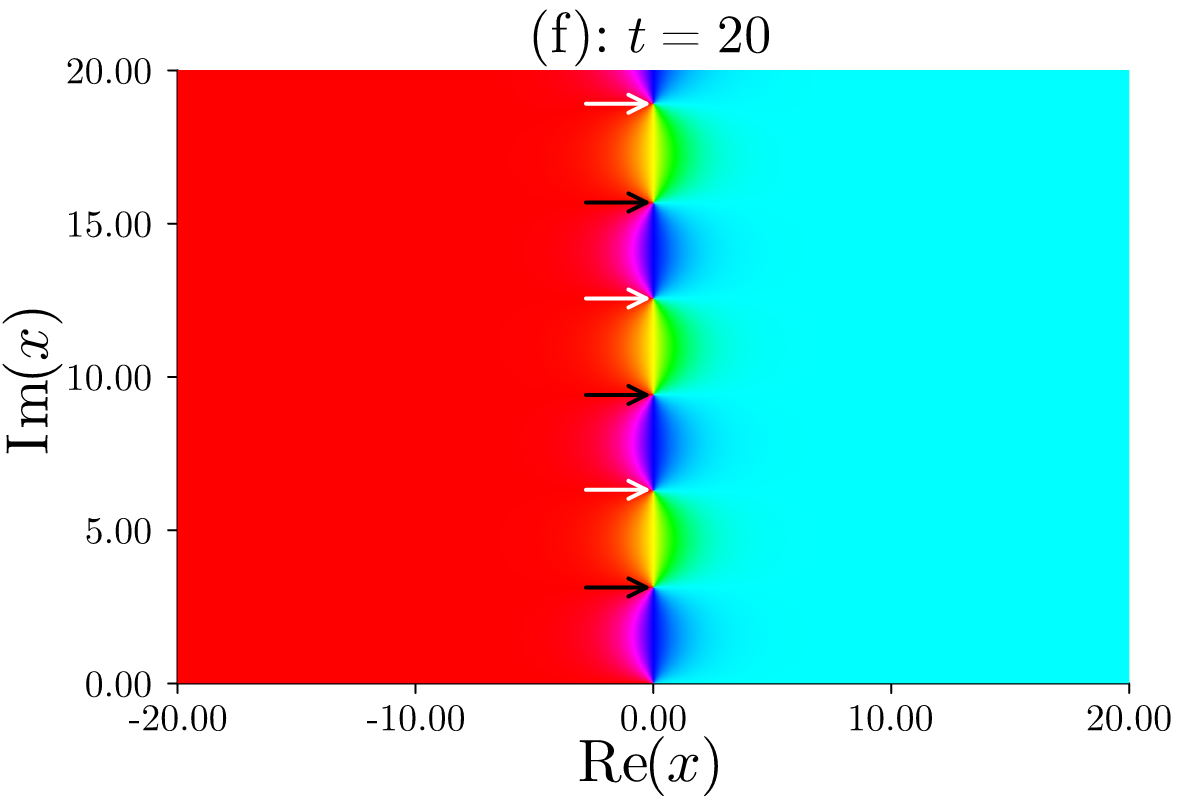}
        \end{subfigure}
        \caption{(a) Real-valued solution profiles of Burgers' equation \eqref{eq:burgers} with $\mu = 1$ for various times using initial condition \eqref{eq:stepdownoddIC} (solid lines), together with the odd travelling wave solution \eqref{eq:oddtravwave} (dashed line). (b)-(f) Phase portraits of the exact solution (\ref{eq:stepdownoddsolution}) of Burgers' equation with initial condition \eqref{eq:stepdownoddIC} at various times, where black arrows point to poles and white arrows point to zeros.
        Note that, while this last image in panel (f) is drawn for a finite time $t=20$, it is visually undistinguishable from the travelling wave solution \eqref{eq:oddtravwave} on this scale.}
        \label{fig:oddstepdowntotravwave}
    \end{figure}

We illustrate the general solution behaviour in figure~\ref{fig:oddstepdowntotravwave},
where we show real-valued solutions of (\ref{eq:burgers}) in (a) and phase portraits in (b)-(f) for the specific initial condition
\begin{equation}
u\left(x, 0\right) = \begin{cases}
      1 & x < 0 \\
      -1 & x > 0.
   \end{cases}
   \label{eq:stepdownoddIC}
\end{equation}
In this case, the exact solution (\ref{eq:colehopf1})--(\ref{eq:colehopf3}) reduces to
\begin{equation}
u\left(x, t\right) =
\frac{\exp{\left(\displaystyle\frac{-x}{2\mu}\right)}
\erfc{\left(\displaystyle\frac{x-t}{2\sqrt{\mu t}}\right)}
-
\exp{\left(\displaystyle\frac{x}{2\mu}\right)}
\erfc{\left(\displaystyle-\frac{x+t}{2\sqrt{\mu t}}\right)}}
{\exp{\left(\displaystyle\frac{-x}{2\mu}\right)}
\erfc{\left(\displaystyle\frac{x-t}{2\sqrt{\mu t}}\right)}
+
\exp{\left(\displaystyle\frac{x}{2\mu}\right)}
\erfc{\left(\displaystyle-\frac{x+t}{2\sqrt{\mu t}}\right)}}.
   \label{eq:stepdownoddsolution}
\end{equation}
We include in part (a) of this figure the steady-state solution (\ref{eq:oddtravwave}) as a purple dashed line.  We see that on this scale, the real-valued solution at $t=5$ is already close to the steady state.

Turning to the phase portraits in figure~\ref{fig:oddstepdowntotravwave}, we see the solution in part (b) plotted for the extremely small time $t=10^{-24}$ exhibits many of the same characteristics at the phase portrait in figure~\ref{fig:stepdownphaseandasymptotic}(a).  Importantly, the singularities are arranged along the same convex-shaped configuration that is approximated by (\ref{eq:innerextended}).  The reason is that the small-time asymptotics for these two examples are almost exactly the same.  One difference between figure~\ref{fig:oddstepdowntotravwave}(b) and figure~\ref{fig:stepdownphaseandasymptotic}(a) is that the aqua region (indicating solution values close to being real and negative) in the right-hand part of \ref{fig:oddstepdowntotravwave}(b) does not appear in \ref{fig:stepdownphaseandasymptotic}(a), which instead has a fingering structure.  This is due to the leading-order term $f_0$, given by
$f_0=-\mathrm{erf}(\xi/2\sqrt{\mu})$ for the initial condition (\ref{eq:stepdownoddIC}) (cf.~(\ref{eq:asymptoticsolf0})).

The complex-plane behaviour of the solution (\ref{eq:stepdownoddsolution}) for moderate times is very interesting, as we see from the phase portraits in   figure~\ref{fig:oddstepdowntotravwave}(c)-(e).  Looking at the phase portrait for $t=1$ in (c), we note that, apart from a single zero at $x=0$ and a single pole on the imaginary axis, each of the poles and zeros is arranged to lie close to rays at angles $\pi/4$ and $3\pi/4$.  The pole and zero configuration is symmetric about the imaginary axis.  As time increases from $t=1$, the poles and zeros along the ray at angle $\pi/4$ move closer to those on the ray at $3\pi/4$ and ultimately collide on the imaginary axis.  After each collision, the pair of poles (zeros) breaks up into two poles (zeros), both of which remain on the imaginary axis, with one moving upwards and the other moving down.

The subsequent motion is very subtle, as the poles then slowly approach their appropriate long-time limit along the imaginary axis. For example, there is a pole on the imaginary axis from $t=0^+$ that evolves to $\pi\mathrm{i}$ as $t\rightarrow\infty$.  This is the closest pole to the real axis. The next pair of poles collide on the imaginary axis at approximately $t=7.4$, and subsequently break up into two poles on the imaginary axis that evolve to $3\pi\mathrm{i}$ and $5\pi\mathrm{i}$ in the long-time limit. The next pair collide at approximately $t=18.9$ and break up into two that approach $7\pi\mathrm{i}$ and $9\pi\mathrm{i}$, and so on. Again, this series of collisions is easily observed by viewing the accompanying video in the supporting materials for this manuscript.

By observing each pole-pole collision very carefully (e.g., by zooming in on the videos and slowing them down), we see the collision also involves a single zero that is propagating up the imaginary axis.  For example, in figure~\ref{fig:oddstepdowntotravwave}(e), there are two poles close to the imaginary axis that are about to collide and a zero on the imaginary axis that is also about to collide with both of the poles.  After the pole-pole-zero collision, the zero continues to propagate up the imaginary axis seemingly unaffected in between two poles, one of which propagates up and the other propagates down.

\begin{figure}[!h]
        \centering
        \begin{subfigure}[t]{0.49\textwidth}
            \includegraphics[width=\linewidth]{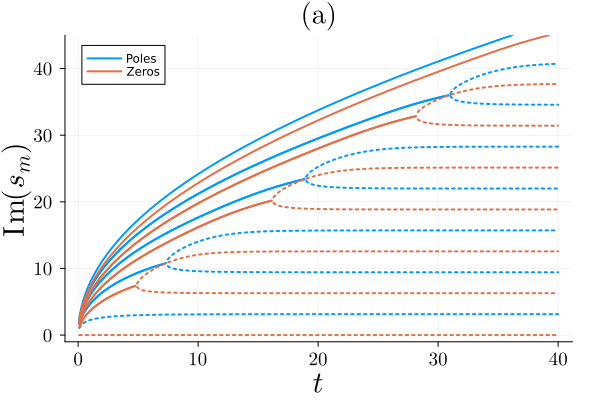}
        \end{subfigure}
        \begin{subfigure}[t]{0.49\textwidth}
            \includegraphics[width=\linewidth]{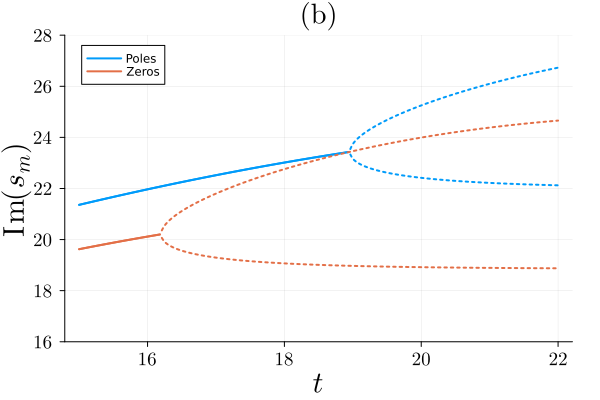}
        \end{subfigure}
        \caption{(a) Tracking the imaginary part of the closest poles and zeros for Burgers' equation \eqref{eq:stepdownoddsolution} with the special step-down initial condition \eqref{eq:stepdownoddIC}. Here, $\mathrm{Im}(s_m)$ represents the distance to the real line, solid curves represent poles and zeros that are off the imaginary axis, while dash curves represent poles and zeros on the imaginary axis. (b) Zoomed-in plot of the second zero-zero collision and the second pole-pole-zero collision.}
        \label{fig:oddpoletrack}
    \end{figure}

In figure~\ref{fig:oddpoletrack} we plot the imaginary parts of the location of the poles (in blue) and zeros (in orange) versus time.  For poles (zeros) that are off the imaginary axis, there is another pole (zero) with the same imaginary part whose location is a reflection about the imaginary axis.  The curves in figure~\ref{fig:oddpoletrack} for these poles (zeros) are drawn as solid lines.  At the time at which these pairs of poles (zeros) collide, the relevant curve bifurcates into two branches, drawn as a dashed line.  Thus, we can see that, as time increases, there are alternating events of pole-pole collisions and zero-zero collisions.  As noted above, each of the zeros on the upper dashed branches meets a pole-pole collision on its upward journey but appears to continue as if unaffected.  One of these events is shown more clearly in figure~\ref{fig:oddpoletrack}(b), with the relevant pole-pole-zero collision occurring at roughly $t=18.9$.

We close this subsection by very briefly describing the asymptotic structure of each pole-pole collision.  Suppose a pole-pole collision occurs at $t=t_c$ and $x=x_c$, then the solution of Burgers' equation at the collision time has the local behaviour
\begin{equation}
u(x,t_c)\sim -\frac{4\mu}{x-x_c}+A_1
\quad\mbox{as}\quad x\rightarrow x_c,
\label{eq:4mu}
\end{equation}
where $A_1$ is a complex constant.  Note the residue in (\ref{eq:4mu}) is $-4\mu$ at the moment of a two-pole collision (see the example in section 8.3 of \cite{crowdy2021}, for example; we expect it would be $-2N\mu$ for an $N$-pole collision).  If we pose the similarity ansatz $u\sim F(\eta)/(t-t_c)^{1/2}$, $\eta=(x-x_c)/(t-t_c)^{1/2}$, subject to the far-field condition $F\sim -4\mu/\eta$ as $\eta\rightarrow -\mathrm{i}\infty$, then the solution is the rational function $F=-4\mu\eta/(\eta^2+2\mu)$.  By applying a Galilean boost $x\rightarrow x-A_1t$, $u\rightarrow u+A_1$, we arrive at the inner description
\begin{equation}
u\sim \frac{1}{(t-t_c)^{1/2}}F(\eta-A_1(t-t_c)^{1/2})+A_1
=-\frac{-4\mu(x-x_c-A_1(t-t_c))}{(x-x_c-A_1(t-t_c))^2+2\mu (t-t_c)}+A_1
\quad\mbox{as}\quad t\rightarrow t_c.
\label{eq:collision}
\end{equation}
For left-right symmetric collisions that occur on the imaginary axis in the upper half plane (such as in the example of this subsection), we write $x_c=\mathrm{i}y_c$ and $A_1=\mathrm{i}\delta$, where both $y_c>0$ and $\delta>0$ (the value of $\delta$ depends on conditions away from the collision and is not determined by a local analysis).

The asymptotic solution (\ref{eq:collision}) holds for both $t<t_c$ and $t>t_c$, namely pre- and post-collision. Labelling the poles as $x=s(t)$, then pre-collision we have
$$
s(t)\sim \mathrm{i}y_c \pm \sqrt{2\mu}(t_c-t)^{1/2} + \mathrm{i}\delta(t-t_c)
\quad\mbox{as}\quad t\rightarrow t_c^-,
$$
while post-collision we have
$$
s(t)\sim \mathrm{i}\left(y_c\pm \sqrt{2\mu}(t-t_c)^{1/2}+ \mathrm{i}\delta(t-t_c)\right)
\quad\mbox{as}\quad t\rightarrow t_c^+.
$$
That is, two poles move towards each other with increasing speed so that their direction is asymptotically horizontal and their speed is of $\mathcal{O}((t_c-t)^{-1/2})$ in the collision limit.  Then, both post-collision poles remain on the imaginary axis, one moving up and the other down, both with a speed that is of $\mathcal{O}((t-t_c)^{-1/2})$.  Interestingly, this type of generic collision between two poles always involves a simple zero, $s=z(t)$, that moves vertically as
$$
z(t)\sim \mathrm{i} y_c+\sfrac{3}{2}\mathrm{i}\delta (t-t_c)
\quad\mbox{as}\quad t\rightarrow t_c
\quad(\mbox{for}\,\,\, t<t_c\,\,\,\mbox{and}\,\,\, t>t_c).
$$
Thus, while the two pre-collision poles are heading towards each other horizontally with increasing speed, a simple zero moves in from below at an $\mathcal{O}(1)$ speed.  The two poles and the zero all meet at the collision time $t=t_c$ at $x=\mathrm{i}y_c$.  After the collision, the zero continues is trajectory unaffected.  We see this behaviour illustrated in figure~\ref{fig:oddpoletrack}(b).

\subsection{Step-up initial condition \eqref{eq:stepupIC}}

\begin{figure}[!h]
\centering
        \begin{subfigure}[t]{0.4\textwidth}
            \includegraphics[width=\linewidth]{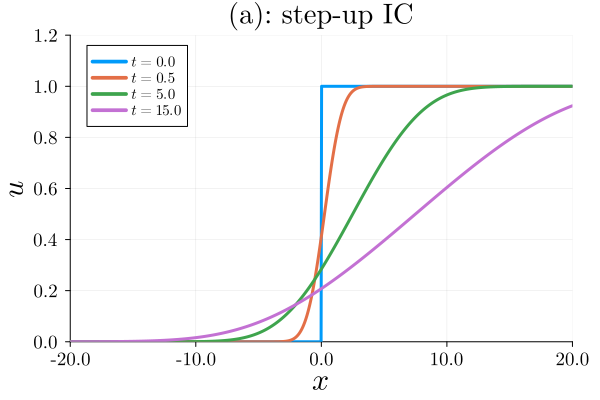}
        \end{subfigure}
        \begin{subfigure}[t]{0.35\textwidth}
            \includegraphics[width=\linewidth]{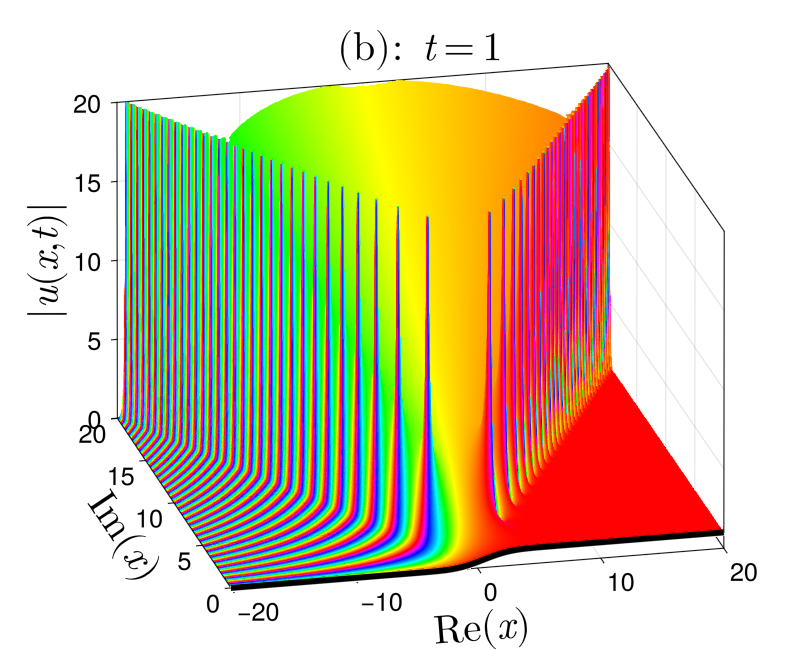}
        \end{subfigure}

        \begin{subfigure}[t]{0.32\textwidth}
            \includegraphics[width=\linewidth]{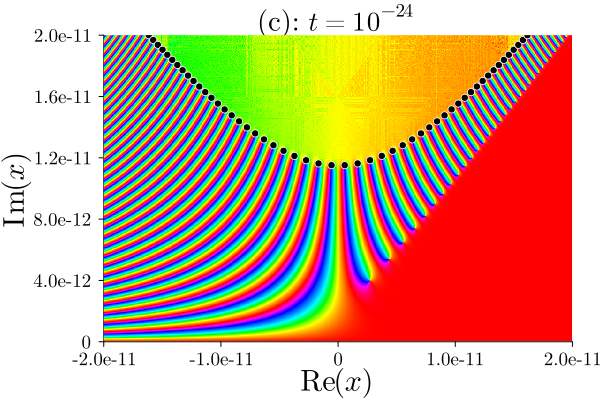}
        \end{subfigure}
        \begin{subfigure}[t]{0.32\textwidth}
            \includegraphics[width=\linewidth]{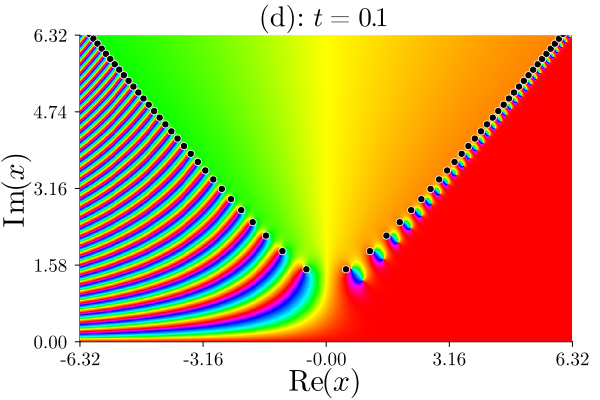}
        \end{subfigure}
        \begin{subfigure}[t]{0.32\textwidth}
            \includegraphics[width=\linewidth]{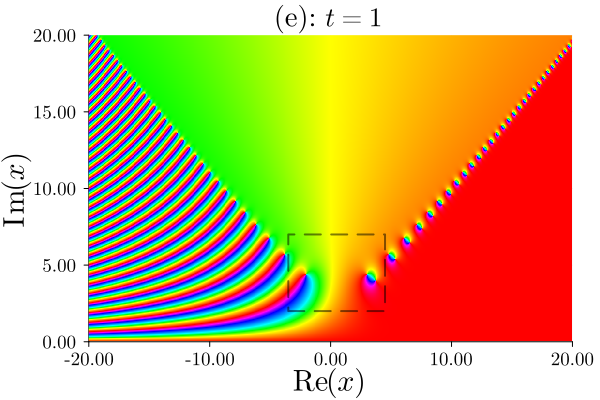}
        \end{subfigure}
        \caption{(a) Real-valued solution profiles of Burgers' equation (\ref{eq:burgers}) with $\mu=1$ at $t=0$, $0.5$, $5$ and $15$ using the step-up initial condition \eqref{eq:stepupIC}.  (b) Analytic landscape plot of the solution at $t=1$ with the black curve along $\mathrm{Im}(x)=0$ indicating the real-valued solution on the real line.  (c)-(e) Phase portraits plotted for times $t=10^{-24}$, $0.1$ and $1$.  In (c),(d), black dots that indicate the approximate locations of the simple poles according to (\ref{eq:innerextended_v2}).  Equivalent approximations on this scale could be drawn using the Lambert-$W$ function (\ref{eq:lambertscaled_v2}).}
        \label{fig:stepup}
    \end{figure}

We summarise here the small-time asymptotics for Burgers’ equation (\ref{eq:burgers}) with the step-up initial condition
\begin{equation}
u\left(x, 0\right) = \begin{cases}
      0 & x < 0 \\
      1 & x > 0,
   \end{cases}
   \label{eq:stepupIC}
\end{equation}
whose exact solution is given by
\begin{equation}
\begin{aligned}
    u\left(x, t\right) &= \frac{\mathrm{erf}\! {\left(\dfrac{t-x}{2\sqrt{\mu t}}\right)} - 1}{\mathrm{erf}\! {\left(\dfrac{t-x}{2\sqrt{\mu t}}\right)} - 1 + \exp{\left(-\dfrac{t-2x}{4\mu}\right)}\left(\mathrm{erf}\! {\left(\dfrac{x}{2\sqrt{\mu t}}\right)} -1\right)}.
\end{aligned}
\label{eq:stepupsolution}
\end{equation}
We start by showing real-valued solution profiles, an analytic landscape plot and three phase portraits in figure~\ref{fig:stepup}, all drawn using the exact solution (\ref{eq:stepupsolution}).  At first glance, the behaviour of this solution in the complex plane for sufficiently small time appears to be roughly a reflection about the imaginary axis of the solution with the step-down initial condition shown in figure~\ref{fig:stepdowntotravwave}.  However, there is a minor difference, as we now explain.

By writing $u(x,t)=f(\xi,t)$ as in (\ref{eq:similarity}), a small-time expansion of the form (\ref{eq:asymptoticsol2}) gives
\begin{eqnarray}
f_0 & = & \frac{1}{2}+\frac{1}{2}\mathrm{erf}\! {\left(\frac{\xi}{2\sqrt{\mu}}\right)},
        \label{eq:asymptoticsolf0_v2} \\
f_1 & = & \frac{1}{8\sqrt{\pi\mu}}\left( \mathrm{erf}\! {\left(\frac{\xi}{2\sqrt{\mu}}\right)}-1\right)
\left(\frac{\sqrt{\pi}\xi}{\sqrt{\mu}}
\left(1+\mathrm{erf}\! {\left(\frac{\xi}{2\sqrt{\mu}}\right)}\right)
+2\exp{\left(-\frac{\xi^2}{4\mu}\right)}\right).
    \label{eq:asymptoticsolf1_v2}
\end{eqnarray}
(cf.~(\ref{eq:asymptoticsolf0})-(\ref{eq:asymptoticsolf1})), with the far field behaviours
\begin{equation}
    f_0 -\sim \sqrt{\frac{\mu}{\pi}}\frac{1}{\xi}
    \,\mathrm{e}^{-\xi^2/4\mu}, \hspace{0.2cm} f_1 \sim -\frac{\mu}{\pi}\frac{1}{\xi^3}
    \,\mathrm{e}^{-\xi^2/2\mu}\hspace{0.2cm} \text{as} \hspace{0.2cm} \xi \rightarrow \mathrm{i}\infty,
    \label{eq:leadingorderexpansion_v2}
\end{equation}
(cf.~(\ref{eq:leadingorderexpansion})).  Thus, the series (\ref{eq:asymptoticsol2}) breaks down when
\begin{equation}
    \sqrt{\frac{\pi}{\mu}}\xi^2
    \,\mathrm{e}^{\xi^2/4\mu}
        =\mathcal{O}(t^{1/2}).
    \label{eq:solveforxi_v2}
\end{equation}
(cf.~(\ref{eq:solveforxi})).  We could solve  $\sqrt{\pi/\mu}\,\xi^2\,\mathrm{exp}(\xi^2/4\mu)=t^{1/2}$ exactly to give
\begin{equation}
    \xi = 2\sqrt{\mu}\left(W_{n}\left(\frac{t^{1/2}}{4\sqrt{\mu \pi}}\right)\right)^{1/2},
    \hspace{0.5cm} n = \pm1, \pm2, ...
    \label{eq:lambertscaled_v2}
\end{equation}
where, just like (\ref{eq:lambertscaled}), $W_n(z)$ represents a branch of the Lambert-$W$ function, and the branch of the $1/2$ power is taken so that $\xi$ lies in the upper half plane.  The only difference between equation (\ref{eq:lambertscaled_v2}), which locates the inner-inner regions for the step-up initial condition (\ref{eq:stepupIC}), and (\ref{eq:lambertscaled}) for the step-down initial condition, is the sign of the argument of the Lambert-$W$ function.  The consequence is that the poles for (\ref{eq:stepupIC}) lie in between those for (\ref{eq:stepdownIC}).  For example, one of the values of (\ref{eq:lambertscaled}) lies on the imaginary $\xi$ axis, while none of the values (\ref{eq:lambertscaled_v2}) have this property.

We can conclude that the way to adapt the working in subsubsection~\ref{sec:stepdownagain} to hold for the step-up initial condition (\ref{eq:stepupIC}) is to replace $m$ in equations (\ref{eq:innerregionLambert}), (\ref{eq:innerinnerdefnf}), (\ref{eq:inner_regions10analogue}) and (\ref{eq:innerextended}) with $m-1/2$.  For example, the key result for the location of the poles is that
\begin{align}
s_m(t) \sim & \,\,\mathrm{i}\sqrt{2\mu}\,t^{1/2}
\left(\ln(1/t)+2\ln (2\sqrt{\mu\pi})-2(2m-1)\pi\mathrm{i}\right)^{1/2}
\\
& +
\mathrm{i}\sqrt{2\mu}\,t^{1/2}
\frac{\mathrm{Log}\left(\ln(1/t)+2\ln (2\sqrt{\mu\pi})-2(2m-1)\pi\mathrm{i}\right)}
{\left(\ln(1/t)+2\ln (2\sqrt{\mu\pi})-2(2m-1)\pi\mathrm{i}\right)^{1/2}},
\label{eq:innerextended_v2}
\end{align}
where here $m=1,2,3,\ldots$ are for poles in the first quadrant and $m=0,-1,-2,\ldots$ are for poles in the second quadrant.  This formula is used to plot the black dots in figure~\ref{fig:stepup}(c)-(d), which appears to be providing an excellent approximation for the actual pole locations for both $t=10^{-24}$ and $0.1$.

Thus, we summarise that the small-time asymptotics for the step-up initial condition (\ref{eq:stepupIC}) is almost a reflection of that for the step-down initial condition (\ref{eq:stepdownIC}) about the imaginary axis.   The difference is that each of the poles for (\ref{eq:stepupIC}) is shifted slightly along the convex-shaped curve defined by the Lambert-$W$ function when compared to the poles for (\ref{eq:stepdownIC}), so that the poles for (\ref{eq:stepupIC}) sit in between those for (\ref{eq:stepdownIC}).  Note that, for (\ref{eq:stepupIC}) there is a pole that is initially travelling perfectly upwards in the imaginary direction, whereas for (\ref{eq:stepdownIC}) there is no such pole.

The large-time asymptotics for (\ref{eq:stepupIC}) are less interesting than that for (\ref{eq:stepdownIC}), since there are no pole collisions.  Instead, by plotting the exact solution (\ref{eq:stepupsolution}), we find (not shown here) that the structure on the left remains intact (see the left-half of the image in figure~\ref{fig:stepup}(e), for example), albeit expanding at a rate $t^{1/2}$ for large time.  On the other hand, the poles on the right-hand side of the phase portrait travel more quickly to the right and move out of frame.  Provided $\mathrm{Re}(x)<t^{1/2}$, the relevant approximation for this smoothed rarefaction fan limit (that retains the correct singularity structure in the complex plane) is
$$
u(x,t)\sim 2\sqrt{\frac{\mu}{\pi t}}\,
\frac{\mathrm{e}^{-x^2/4\mu t}}{\mathrm{erfc}(x/2\sqrt{\mu t})}
\quad\mbox{as}\quad t\rightarrow\infty,
$$
which, incidently, comes from (\ref{eq:colehopf1}) with $\phi\sim \mathrm{erfc}(x/2\sqrt{\mu t})/2$.

\section{Other initial conditions}\label{sec:other}

We believe that the examples covered in section~\ref{Heaviside_initial_conditions} illustrate the key asymptotic behaviours in the limit $t\rightarrow 0^+$ for initial conditions with a discontinuity on the real line.  For other examples with more than one discontinuity, our conjecture is that the solutions in section~\ref{Heaviside_initial_conditions} will be relevant close to each individual discontinuity, depending on whether the real-valued initial condition steps down or up at the discontinuity.

In this section, we explore this conjecture by considering two further examples, each with two discontinuities in their initial conditions. We encounter a new phenomenon whereby the initial trajectories of poles emerging from different discontinuities cross paths.  Further, these two examples are interesting for moderate to large times as the solutions evolve to long-time limits that are different to the examples in section~\ref{Heaviside_initial_conditions}.

\subsection{Top-hat initial condition}\label{sec:rectangular}


We consider here a rectangular-type ``top-hat'' initial condition
\begin{equation}
u\left(x, 0\right) = \begin{cases}
      0 & x < -\frac{1}{2} \\
      a & -\frac{1}{2}\leq x\leq \frac{1}{2} \\
      0 & x\ > \frac{1}{2}
   \end{cases},
   \label{eq:tophatIC}
\end{equation}
which has two points of discontinuity, namely at $x=-1/2$ and $x=1/2$.  For this example, there will be an infinite number of singularities born at both $x=-1/2$ and $x=1/2$, whose small-time trajectories are mostly described in the previous section.  There is, however, one key difference, which occurs when poles from $x=-1/2$ propagating out at an angle $\pi/4$ meet with those from $x=1/2$ propagating out at $3\pi/4$.  This phenomenon can only occur for initial conditions with more than one point of discontinuity, which is one of the motivations for considering (\ref{eq:tophatIC}). The other motivation is to demonstrate how the poles and zeros rearrange themselves to ultimately approach the appropriate long-time limit for this example, which comes from the well-known similarity solution of Burgers' equation.

\begin{figure}[!h]
        \centering
        \begin{subfigure}[t]{0.32\textwidth}
            \includegraphics[width=\linewidth]{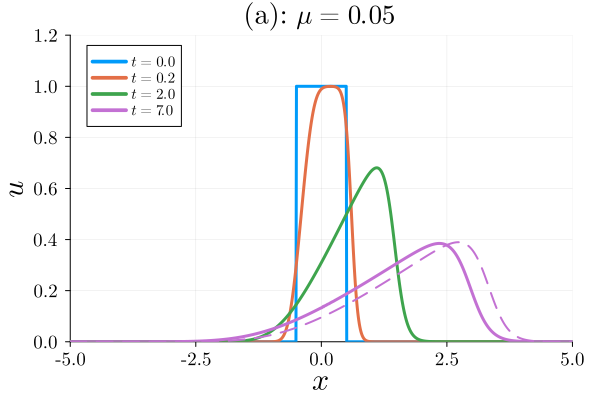}
        \end{subfigure}
        \begin{subfigure}[t]{0.32\textwidth}
            \includegraphics[width=\linewidth]{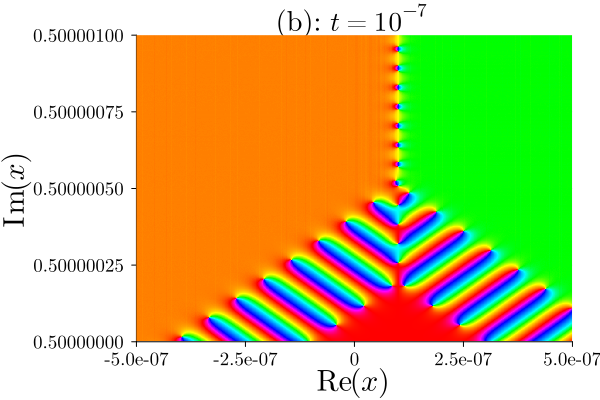}
        \end{subfigure}
        \begin{subfigure}[t]{0.32\textwidth}
            \includegraphics[width=\linewidth]{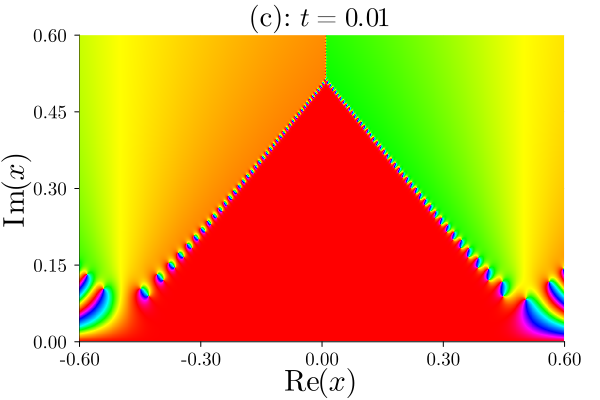}
        \end{subfigure}
        \begin{subfigure}[t]{0.32\textwidth}
            \includegraphics[width=\linewidth]{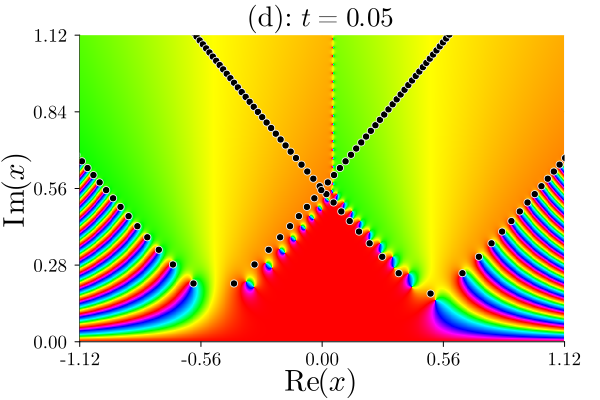}
        \end{subfigure}
        \begin{subfigure}[t]{0.32\textwidth}
            \includegraphics[width=\linewidth]{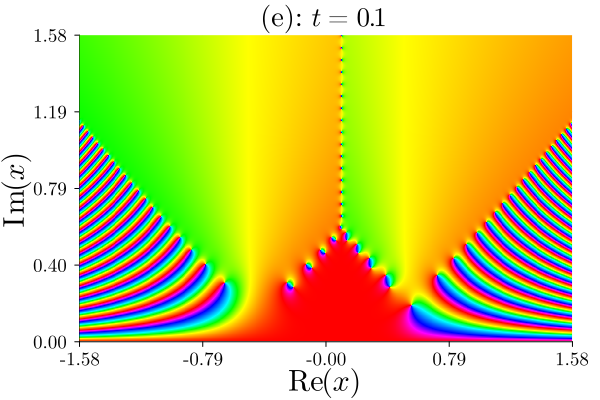}
        \end{subfigure}
        \begin{subfigure}[t]{0.32\textwidth}
            \includegraphics[width=\linewidth]{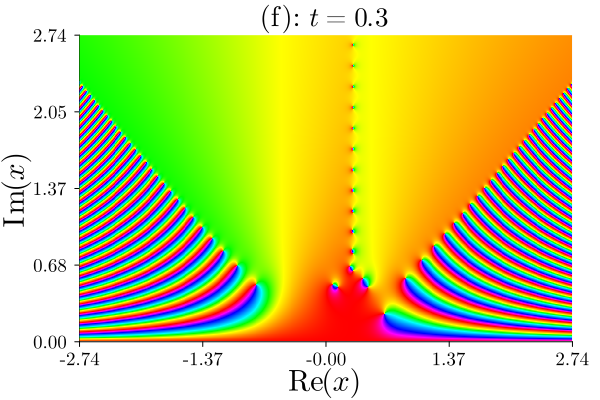}
        \end{subfigure}
        \begin{subfigure}[t]{0.32\textwidth}
            \includegraphics[width=\linewidth]{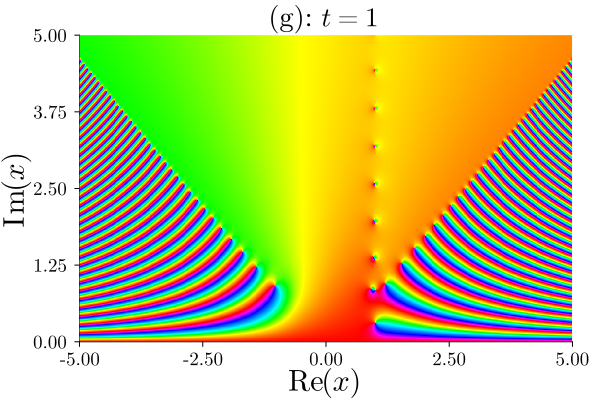}
        \end{subfigure}
        \begin{subfigure}[t]{0.32\textwidth}
            \includegraphics[width=\linewidth]{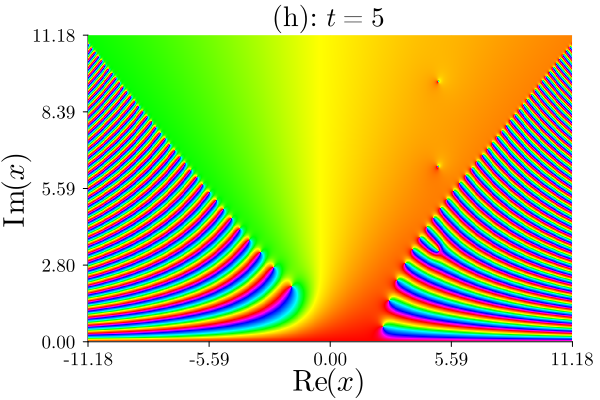}
        \end{subfigure}
        \begin{subfigure}[t]{0.32\textwidth}
            \includegraphics[width=\linewidth]{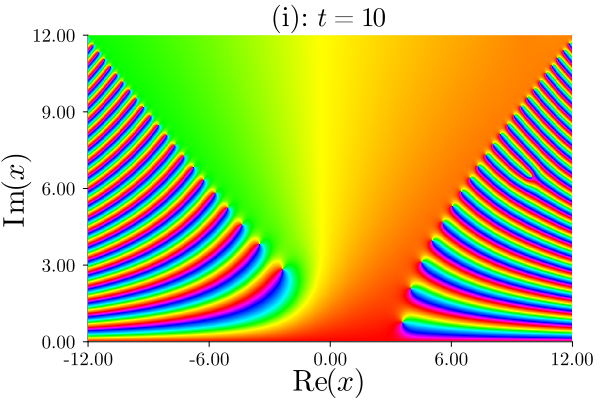}
        \end{subfigure}
        \caption{(a) Real-valued solution profiles of Burgers' equation (\ref{eq:burgers}), (\ref{eq:tophatIC}) with $\mu = 0.05$ for various times using the top-hat initial condition \eqref{eq:tophatsolution} (solid lines), together with the similarity solution (\ref{eq:simlarget}) (dashed line) computed at $t=7$.  (b)-(i) Phase portraits of the exact solution (\ref{eq:tophatsolution}) with $\mu=0.05$ at various times.  The black dots in (d) indicate approximate location of the simple poles according to (\ref{eq:innerextended}) (shifted to the right by $1/2$) and  (\ref{eq:innerextended_v2}) (shifted to the left by $1/2$).
        Note that the last image in (i) is very similar to the similarity solution (\ref{eq:simlarget}) when viewed on this scale.}
        \label{fig:tophatphasenew}
    \end{figure}

Burgers' equation (\ref{eq:burgers}) with (\ref{eq:tophatIC}) has the exact solution
\begin{equation}
    u(x,t)=\frac{a\exp(A)\left[\mathrm{erf}(B_{-})-\mathrm{erf}(B_{+})\right]
    }{
    -\exp\left(-\frac{a}{2\mu}\right)\left[1+\mathrm{erf}(C_{-})\right]+\exp(A)\left[\mathrm{erf}(B_{-})-\mathrm{erf}(B_{+})\right]+\mathrm{erf}(C_{+})-1},
    \label{eq:tophatsolution}
\end{equation}
where
$$
A=\frac{a(at-2x-1)}{4\mu},
\quad
B_{\pm}=\frac{2at-2x\pm1}{4\sqrt{\mu t}},
\quad
C_{\pm}=\frac{2x\pm1}{4\sqrt{\mu t}}.
$$
We plot real-valued solution profiles of this solution in figure~\ref{fig:tophatphasenew}(a) for $\mu=0.1$.  For the small-time profile at $t=0.2$, we see the two corners of the initial condition have smoothed off, as expected.  As time increases through $t=2$ and $7$, we observe the profile both spread out and bend to the right, as it ultimately tends towards a distorted Gaussian, as is well known to occur for this initial condition.

The phase portraits in figure~\ref{fig:tophatphasenew}(b) and (c) illustrate the small-time behaviour of the exact solution (\ref{eq:tophatsolution}).  For small times, we have observed (not shown here) that the local phase portraits near $x=-1/2$ are essentially the same as those found for the step-up initial condition (for example, shown in figure~\ref{fig:stepup}(c)), except shifted by 1/2 to the left, while the local phase portraits near $x=1/2$ are essentially the same as those found for the step-down initial condition (see figure~\ref{fig:stepdowntotravwave}(c)), this time shifted by 1/2 to the right.  The new behaviour exhibited in this example results from poles propagating out of $x=-1/2$ at the angle $\pi/4$ crossing the path of poles propagating out of $x=1/2$ at the angle $3\pi/4$.  For extremely small times, like in figure~\ref{fig:tophatphasenew}(b), these two paths meet each other at approximately $x=\mathrm{i}/2$.  Crucially, the poles emerging from $x=-1/2$ are out of phase compared to those emerging from $x=1/2$; therefore, even though the two paths meet each other, none of the poles undergo a collision.  Instead, each pole rapidly changes direction and begins to propagate upwards, forming part of a vertical array moving in the positive imaginary direction.

Studying this vertical array of poles more carefully
(e.g., via the videos in the supplementary material), we see that every second pole in this array was born at $x=-1/2$, while every other pole was born at $x=1/2$.  Thus, when they enter this vertical array, the poles weave themselves, whereby one enters from the left, then one from the right, then the left, and so on.
The poles in the vertical array appear to be equally spaced, with a separation distance that is approximately $4\pi\mu t$.  Note that the poles along the $45^\circ$ rays from $x=-1/2$ and $x=1/2$ have an approximate separation distance of $4\sqrt{2}\mu\pi t$ by the time they reach $x=\mathrm{i}/2$.

Returning to figure~\ref{fig:tophatphasenew}, even for $t=0.05$, which is not a particularly small time for this initial condition, we see in panel (d) the local behaviours near $x=-1/2$ and $x=1/2$ described above, as well as the new behaviour near the crossover point close to $x=\mathrm{i}/2$.  We have included in this image the approximate locations of the poles that we calculated in section~\ref{Heaviside_initial_conditions} for the step-up (near $x=-1/2$) and step-down (near $x=1/2$) initial conditions to show that the agreement is still very good for this moderately small time.  For later times, in figure~\ref{fig:tophatphasenew}(e)-(h), this vertical array of poles appears to drift to the right; otherwise, the phase portraits evolve towards the well-known late-time similarity solution (described shortly), which is closely approximated by the panel (i).



We now describe the late-time behaviour for this example. As is well known \cite{whitham1974linear}, for an initial condition in (\ref{eq:burgers}) with a finite mass $M$ on the real line, the solution evolves to the similarity solution
\begin{equation}
    u\left(x, t\right) \sim \frac{1}{t^{1/2}}\Phi\left(\xi\right)
    \quad\mbox{as}\quad t\rightarrow\infty,
    \hspace{0.5cm} \xi = \frac{x}{t^{1/2}},
    \label{eq:sim1}
\end{equation}
where $\Phi$ satisfies
\begin{equation}
    -\sfrac{1}{2}(\Phi+\xi\Phi') + \Phi\Phi' = \mu \Phi'',
    \quad \Phi\rightarrow 0 \quad \mbox{as}\quad \xi\rightarrow\pm\infty,
\label{eq:phi}
\end{equation}
and the primes denote differentiation with respect to $\xi$.  The relevant exact solution of (\ref{eq:phi}) is
$$
\Phi=\frac{2\sqrt{\mu}\mathrm{e}^{-\xi^2/4\mu}}{\sqrt{\pi}\left(\gamma-
\erf\left(\dfrac{\xi}{2\sqrt{\mu}}\right)\right)},
$$
where $\gamma$ is related to the mass $M$ via $M=2\mu\ln((\gamma+1)/(\gamma-1))$.  Or, in original variables, $u\sim U_\mathrm{sim}(x,t)$ as $t\rightarrow\infty$, where
\begin{equation}
U_\mathrm{sim}=
2\sqrt{\frac{\mu}{\pi t}}\,\frac{\mathrm{e}^{-x^2/4\mu t}}
{1+2(\mathrm{e}^{M/2\mu}-1)^{-1}-\erf\left(x/2\sqrt{\mu t}
\right)}.
\label{eq:simlarget}
\end{equation}
The initial condition (\ref{eq:tophatsolution}) has a mass $M=a$.  As noted above, we can see in figure~\ref{fig:tophatphasenew}(f)-(i) how the solution evolves to the similarity solution (\ref{eq:simlarget}), remembering that the image in panel (i) appears visually indistinguishable from the actual similarity solution (\ref{eq:simlarget}) on this scale.

\subsection{Saw-tooth initial condition}

The last example we consider in the main body of this paper uses the saw-tooth initial condition
\begin{equation}
u\left(x, 0\right) = \begin{cases}
      0 & x < -1 \\
      x & -1<x<1 \\
      0 & x > 1.
   \end{cases}
   \label{eq:nwaveIC}
\end{equation}
Our motivation for this initial condition is two-fold.  First, like the previous example in subsection~\ref{sec:rectangular} with a top-hat initial condition, this example involves an initial condition with two points of discontinuity.  However, unlike the top-hat initial condition, the initial condition (\ref{eq:nwaveIC}) is an odd function, which means the solution of Burgers' equation (\ref{eq:burgers}) remains odd for all time and so the singularity structure is symmetric about the imaginary axis.  Thus, in this case, poles that emerge from the points of discontinuity along paths that meet at $\mathrm{Re}(x)=0$ will undergo collision events before shooting upwards in a vertical array.  Second, the long-time behaviour of the solution with (\ref{eq:nwaveIC}) is the well-known $N$-wave solution \cite{whitham1974linear}
\begin{equation}
    U_\mathrm{nw}(x,t) = \frac{x}{t}\,
    \frac{\sqrt{a/t}\,\mathrm{e}^{-x^2/4\mu t}}{1+\sqrt{a/t}\,\mathrm{e}^{-x^2/4\mu t}},
    \label{eq:nwave}
\end{equation}
which is different from the similarity solution (\ref{eq:simlarget}).  Thus, we use this example to cover a long-time limiting behaviour that has not been considered so far.

\begin{figure}[!h]
        \centering
        \begin{subfigure}[t]{0.32\textwidth}
            \includegraphics[width=\linewidth]{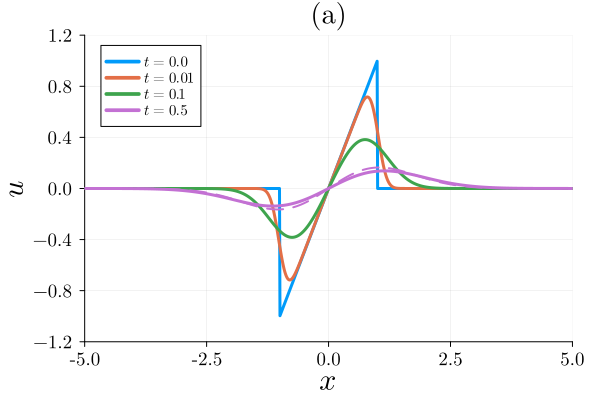}
        \end{subfigure}
        \begin{subfigure}[t]{0.32\textwidth}
           \includegraphics[width=\linewidth]{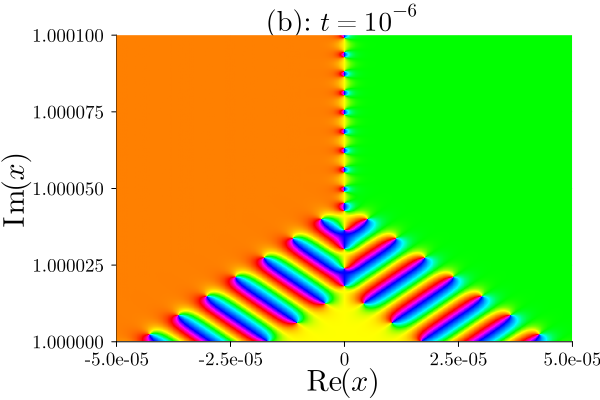}
        \end{subfigure}
        \begin{subfigure}[t]{0.32\textwidth}
           \includegraphics[width=\linewidth]{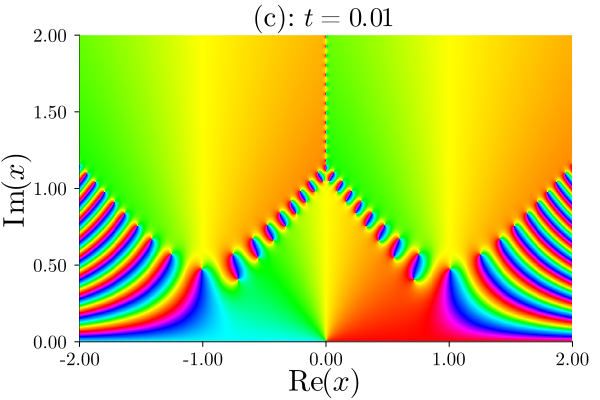}
        \end{subfigure}
        \begin{subfigure}[t]{0.32\textwidth}
           \includegraphics[width=\linewidth]{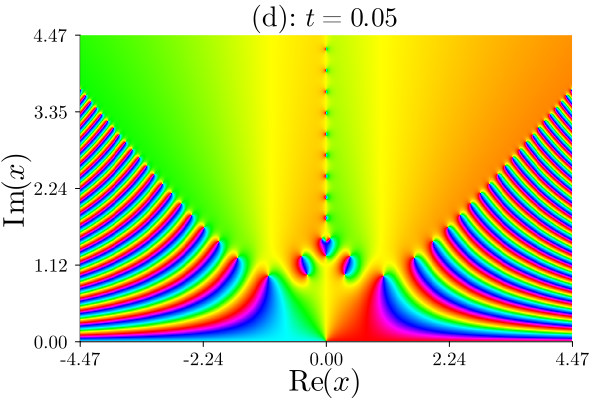}
        \end{subfigure}
        \begin{subfigure}[t]{0.32\textwidth}
          \includegraphics[width=\linewidth]{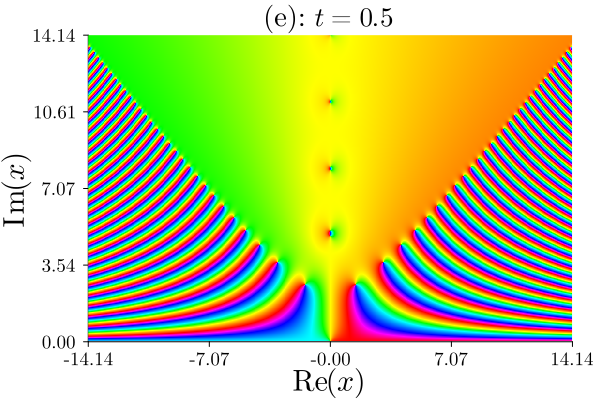}
        \end{subfigure}
        \begin{subfigure}[t]{0.32\textwidth}
         \includegraphics[width=\linewidth]{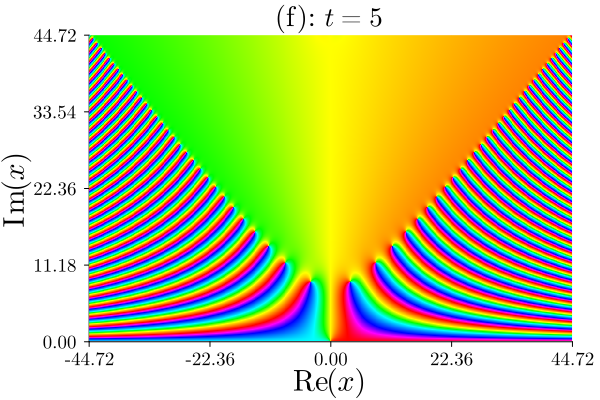}
        \end{subfigure}
        \caption{Real-valued solution profiles of Burgers' equation (\ref{eq:burgers})  with $\mu=1$ for various times using the initial condition (\ref{eq:nwaveIC}) (solid lines), together with the $N$-wave solution \eqref{eq:nwave} (dashed line).  (b)-(f) Phase portraits of the exact solution at various times.  Note that, while this last image in panel (f) is drawn for a finite time $t=5$, it is visually undistinguishable from the $N$-wave solution \eqref{eq:nwave} on this scale.}
        \label{fig:Nwave}
    \end{figure}

To calculate the exact solution of Burgers' equation (\ref{eq:burgers}) with the saw-tooth initial condition (\ref{eq:nwaveIC}), we use (\ref{eq:colehopf1}) with
\begin{equation}
\begin{aligned}
    \phi(x,t) = & \,\frac{1}{2\sqrt{t+1}}
    \Bigg[\exp\left({-\frac{x^{2}}{4\mu(t+1)}}\right)\left(\mathrm{erf}\left(\frac{t-x+1}{2\sqrt{\mu t}\sqrt{t+1}}\right)+\mathrm{erf}\left(\frac{t+x+1}{2\sqrt{\mu t}\sqrt{t+1}}\right)\right)\\
    &\qquad\qquad
    +\exp\left({-\frac{1}{4\mu}}\right)\sqrt{t+1}\left(\mathrm{erf}\left(\frac{x-1}{2\sqrt{\mu t}}\right)-\mathrm{erf}\left(\frac{x+1}{2\sqrt{\mu t}}\right)+2\right)\Bigg].
\end{aligned}
\label{eq:phiforsawtooth}
\end{equation}
We leave out the full solution as the formula is quite long.  We have plotted real-valued solutions with (\ref{eq:phiforsawtooth}) and (\ref{eq:colehopf1}) in figure~\ref{fig:Nwave}(a) for $\mu=1$ and $t=0.01$, $0.1$ and $0.5$, together with the $N$-wave solution (\ref{eq:nwave}).  Even for $t=0.5$, which is not a large time, the $N$-wave solution serves as a good approximation on this scale.

Phase portraits of the exact solution for this example are provided in figure~\ref{fig:Nwave}(b)-(f).  For very small times, we have checked that the pattern of the poles that emerge from the points of discontinuity, $x=\pm 1$, is very well predicted by the formula (\ref{eq:innerextended}), which was for the step-down initial condition.  This observation supports our conjecture that the local behaviour near discontinuities is essentially governed by the examples covered in section~\ref{Heaviside_initial_conditions}.

For the plots in figure~\ref{fig:Nwave}(b)-(c), we observe the four rays of poles emerging from $x=\pm 1$, two of which appear to collide near $x=\mathrm{i}$.  Like the example in section~\ref{sec:rectangular} with the top-hat initial condition, there is a vertical array of poles that emerges from where these paths meet at $t=0^+$.  The difference, however, is that in this example with the saw-tooth initial condition, each pair of poles collides before entering the vertical array.  We expect that the local asymptotic behaviour near each collision event is described by our analysis at the end of subsection~\ref{sec:stepdown}.  As such, the lowest of the two post-collision poles will briefly move downwards before changing direction moving up the imaginary axis, while the other post-collision pole will immediately move upwards.   Subsequently, the two poles settle into an equally spaced upward-moving vertical array, with a separation distance approximately given by $2\pi\mu t$.

Eventually, at around $t=0.1$ for this example, the final pair of poles will collide and enter the vertical array.  For larger times, for example $t=0.5$ in figure~\ref{fig:Nwave}(e), the vertical array breaks away from the rest of the structure, with each pole moving at a speed of $\mathcal{O}(1)$ as $t\rightarrow\infty$, while the closest poles in the remaining structure travel at a much slower speed, of $\mathcal{O}(\ln t)$ as $t\rightarrow\infty$.  For $t=5$ in figure~\ref{fig:Nwave}(f), the phase portrait is very similar to the $N$-wave solution (\ref{eq:nwave}) on this scale.

\section{Discussion}\label{sec:discussion}

In this study of Burgers' equation (\ref{eq:burgers0}), we have applied methods in matched asymptotic expansions to describe how infinitely many simple-pole singularities emerge from points of discontinuity (on the real line) in piecewise-continuous initial data and begin to propagate in the complex-$x$ plane.  We find that the closest singularities to the real axis move outward along a convex-shaped curve with a speed that is of $\mathcal{O}(t^{-1/2}\ln(1/t))$ as $t\rightarrow 0^+$, while singularities much further away move out along anti-Stokes lines at angles $\pi/4$ and $3\pi/4$ from the real axis with a speed that is of $\mathcal{O}(t^{-1/2})$.
For examples with more than one point of discontinuity, there are additional structures as some poles from one discontinuity travel along a path that collides with poles from another.  One of our strategies is to use the exact solution of Burgers' equation to generate numerical support for our asymptotic predictions, thereby providing confidence about the asymptotic methodology.
Further, the availability of the exact solution has allowed us to track the singularities over all times scales, helping to explain how the small-time trajectories are linked to known late-time behaviours, including interesting phenomena such as poles colliding with zeros and other poles, or arrays of poles from different directions weaving themselves into a new array.

While we have used some simple initial conditions that are piecewise continuous but not continuous, we include in Appendix~\ref{appendix:discderiv} an example for which the initial condition is continuous but not continuously differentiable.  We see that some details of the matched asymptotic expansions are more complicated, but the overall scalings are essentially the same as those for the examples in the main part of the paper.  Our conjecture is that for any initial condition in $C^n$ but not in $C^{n+1}$, for $n\geq 0$, an infinite number of simple-pole singularities emerge from the points at which the $(n+1)$th derivative is discontinuous in a convex-shaped configuration that extends out to rays at angles $\pi/4$ and $3\pi/4$.  Again, we expect the closest poles to the points of discontinuity will move with a speed that is of $\mathcal{O}(t^{-1/2}\ln(1/t))$ as $t\rightarrow 0^+$, while those that are much further away will move with a speed that is of $\mathcal{O}(t^{-1/2})$.

The starting point for a reflection on our results is to note that, in all of the studies of complex singularities of solutions of pdes mentioned in section~\ref{sec:intro}, the initial conditions used were analytic functions of the spatial variable. In fact, for many of these examples, the initial conditions were entire functions with no singularities in the plane at all.  On the other hand, when studying practical problems in applied mathematics, it is commonplace to stipulate initial conditions that are not analytic or, sometimes, not even continuous. As such, it is natural to enquire about how the initial conditions affect the manner in which complex singularities are created and how their position subsequently evolves in time.  Our study provides a step towards answering these questions.

To put our work into a broader perspective, it is worth comparing our results with the study of VandenHeuvel et al.~\cite{vandenheuvel2023burgers}, who study (\ref{eq:burgers}) with initial conditions that are analytic and have simple poles, using
\begin{equation}
u(x,0)=\frac{1}{1+x^2}
\label{eq:VdH}
\end{equation}
as their main example.  In that case, restricting ourselves to the upper half plane, an outer expansion of the form (\ref{eq:outergeneric}) breaks down near $x=\mathrm{i}$.  An inner problem for $\xi=(x-\mathrm{i})/t^{1/2}=\mathcal{O}(1)$ (see (\ref{eq:PhysDmodel})) has an exact solution that describes how infinitely many singularities are born at $x=\mathrm{i}$ and evolve like $s_n(t)\sim \mathrm{i}+\xi_n t^{1/2}$, with speed of $\mathcal{O}(t^{-1/2})$ where the $\xi_n$ are zeros of a certain parabolic cylinder function (see also \cite{lustri2023locating}).  Relatively speaking, the asymptotic scalings for Burgers' equation (\ref{eq:burgers}) with (\ref{eq:VdH}) are reasonably straightforward (involving power-laws only), as the initial condition already has a singularity that is a simple pole (remembering that all singularities of Burgers' equation must be simple poles via (\ref{eq:burgerpoleequation})).  On the other hand, for the initial condition
\begin{equation}
u(x,0)=\frac{1}{(1+x^2)^\beta}, \quad \beta>0,
\label{eq:Danielbetanotone}
\end{equation}
the asymptotics are more complicated for $\beta\neq 1$ (involving a cascade of logarithms; see Appendix E of \cite{vandenheuvel2023burgers}).  Of particular interest here is the regime $0<\beta<1$, for which it turns out there are many similarities with the analysis of the problems considered in this paper, including that there are infinitely many simple-pole singularities that emerge from $x=\mathrm{i}$ in a pattern that can be related to the Lambert-$W$ function.  The closest singularities to $x=\mathrm{i}$ initially propagate with a speed of $\mathcal{O}(t^{-1/2}\ln^{1/2}(1/t))$,
while the singularities that are further away move with speed $\mathcal{O}(t^{-1/2})$.
The reason for the similarities is that, for the initial condition (\ref{eq:Danielbetanotone}) with $0<\beta<1$, the outer expansion (\ref{eq:outergeneric}) breaks down near $x=\mathrm{i}$, where the subsequent inner problem for $\xi=(x-\mathrm{i})/t^{1/2}=\mathcal{O}(1)$ involves diffusion but not advection.  The treatment of this diffusion-dominated regime essentially follows the analysis for the examples we have treated here in this paper.

With this comparison in mind, we interpret the Heaviside example in section~\ref{Heaviside_initial_conditions} as a prototype for Burgers' equation with initial conditions that lead to regimes in which diffusion nominates advection, as it appears to be the cleanest possible example that can be studied using closed-form solutions and asymptotic methods.  More generally, for all of these diffusion-dominated scenarios, the application of matched asymptotics in the limit $t\rightarrow 0^+$ appears to involve three steps:
\begin{enumerate}
\item The first step is to write out a small-time expansion in powers of $t$ like (\ref{eq:outergeneric}) and note that it breaks down (becomes disordered) near a singularity of the leading-order term.  For the initial condition (\ref{eq:Danielbetanotone}) treated in Appendix E of \cite{vandenheuvel2023burgers}, the expansion (\ref{eq:outergeneric}) breaks down near $x=\mathrm{i}$; however, for the problems treated in this paper, the initial condition was not an analytic function, so in effect we applied expansions like (\ref{eq:outergeneric}) separately each side of the initial condition's point of discontinuity.

\item The second step is to rescale in an inner region using an appropriate similarity variable.  For diffusion-driven scenarios, the leading-order problem that comes out of a small-time expansion for this inner region will be a linear homogeneous second-order ode, as the nonlinear advection term will not form a part of the dominant balance.  The solution to this linear ode will be an entire function that grows exponentially in a direction away from the real axis of the similarity variable.  A correction term to this expansion will depend on the as-yet neglected advection term, and will grow even faster than the leading-order term in this direction, and so we will need to rescale again in regions for which this inner expansion breaks down.  This rescaling reveals that the speed of the closest singularities will generically be of  $\mathcal{O}(t^{-1/2}\ln^{1/2}(1/t))$ in the limit $t\rightarrow 0^+$.

\item The third step is to treat an inner-inner problem which, to leading order, will involve a second-order ode that includes the full balance between diffusion and advection.  We argue in Appendix~\ref{appendix:inner} that this ode will generically be of the form (\ref{eq:rotatedTWmodel}), which is a rescaled travelling wave problem.  See (\ref{eq:inner_regions7A}), for example (or equation (E.7) in~\cite{vandenheuvel2023burgers}).  The various possible solutions of (\ref{eq:rotatedTWmodel}) involve an array of simple poles that help explain the types of dipole-like and finger-like structure we see in all of our solutions (see figure~\ref{fig:genericsolns}).
\end{enumerate}

Returning to the motivating question of how singularities are born for Burgers' equation with various types of initial conditions, there are two other broad scenarios not covered in this paper (i.e., not involving inner problems that are diffusion dominated).  One involves borderline situations where, for step 2 above, there is a clean balance between nonlinear advection and linear diffusion.  This happens when an initial condition is an analytic function with a simple pole.  One of these scenarios is treated in detail in the bulk of VandenHeuvel et al.~\cite{vandenheuvel2023burgers} and mentioned in Appendix B of \cite{chapman2007}).  The important ode problem in these borderline situations
is (\ref{eq:PhysDmodel}), and the singularities all move initially with speed of $\mathcal{O}(t^{-1/2})$
(if the initial condition has a simple pole with residue $-2\mu$, then the small-time limit is simply that the pole moves with a speed of $\mathcal{O}(1)$ \cite{chudnovski1977,deconinck2007}).  For the second broad scenario, the inner region will be advection dominated, so the leading-order problem in this region will be a nonlinear first-order ode whose solutions will have branch points.  Further rescalings would be required to bring in the neglected diffusion, with inner-inner regions governed by an ode of the form $1+U_0U_0'=U_0''$ (cf. the last equation of page 18 of \cite{vandenheuvel2023burgers}).  Important examples include entire initial conditions such as $u(x,0)=-\sin x$  \cite{caflisch2015,derevianko2026,Kimura1995,sulem1983,trefethen2023,weideman2022dynamics} or $u(x,0)=4x^3-x/t_c$ \cite{bessis1984pole,bessis1990complex,senouf1997dynamics}.
While the small-time asymptotics for this advection-dominated scenario is touched upon very briefly in Appendix E.2 of \cite{vandenheuvel2023burgers}, where (\ref{eq:Danielbetanotone}) is considered with $\beta>1$, we shall revisit this case in future work.

\vspace{3ex}
\noindent{\bf Supporting materials:}
We have generated a series of video files and collected them on
\href{https://jacob-g.xyz/piecewise-continuous}{this website}
to supplement the images in this manuscript.  The videos provide a useful tool for understanding how pole fields evolve, especially during collision or near-collision events.


\vspace{3ex}
\noindent{\bf Acknowledgements:}
SWM is grateful for valuable discussions with John King.  JCG acknowledges the stipend and tuition fee offset provided by a QUT Postgraduate Research Award for the duration of his Master of Philosophy degree.  SWM and MCD acknowledge the support of the Australian Research Council Discovery Project DP250101095.


\appendix

\vspace{3ex}
\begin{Large}
\noindent {\bf Appendices}
\end{Large}

\vspace{-5ex}
\section{Inner problems for diffusion-dominated regimes}
\label{appendix:inner}

\subsection{Typical scalings}

In the process of applying the method of matched asymptotic expansions in the limit $t\rightarrow 0^+$, the appropriate inner problem (or ``inner-inner'' problem, depending on the details) for resolving the correct form of the singularities typically involves the scalings
\begin{equation}
u=\frac{\mu^{1/2}}{\epsilon(t)}U(X,t),
\quad
x=S(t)+\mu^{1/2}\epsilon(t)X,
\label{eq:genericinner}
\end{equation}
where $\epsilon(t)>0$ is some real-valued temporal scaling such that $\epsilon\rightarrow 0^+$ as $t\rightarrow 0^+$, $S(t)$ is a complex function of time that locates the inner region on this time-scale, and $X=\mathcal{O}(1)$ is the inner variable.  For each singularity of $U(X,t)$, which we could label $X=X_n$ for a natural number $n$, there is a singularity of the original solution of Burgers' equation $u(x,t)$ at $x=s_n(t)$, where $s_n(t)\sim S(t)+\mu^{1/2}\epsilon(t)X_n$ as $t\rightarrow 0^+$.

Applying the change of variables (\ref{eq:genericinner}), Burgers' equation (\ref{eq:burgers}) becomes
\begin{equation}
\epsilon^2 U_t-\epsilon\dot{\epsilon}(U+XU_X)-\frac{\epsilon\dot{S}}{\mu^{1/2}}U_X
+UU_X=U_{XX}.
\label{eq:Burgersinner}
\end{equation}
Boundary conditions will come from matching out to an outer region.  For any correct description of a singularity of Burgers' equation, we need the terms $UU_X$ and $U_{XX}$ to appear in our leading-order analysis.  Further treatment of (\ref{eq:Burgersinner}) depends on the relative size of $\epsilon(t)$ and $S(t)$.

For example, if $\epsilon\dot{\epsilon}=\mathcal{O}(1)$ with $|\dot{S}|\ll \dot{\epsilon}$, then we can choose $\epsilon=t^{1/2}$.  Thus, by writing $U\sim U_0(X)$ as $t\rightarrow 0^+$, we find
\begin{equation}
-\sfrac{1}{2}(U_0+XU_0')+U_0U_0'=U_0'',
\quad
U_0\sim \mbox{constant}/X \quad\mbox{as}\quad X\rightarrow -\mathrm{i}\infty.
\label{eq:PhysDmodel}
\end{equation}
This is equivalent to equations (19) and (21) in \cite{vandenheuvel2023burgers} and (1.2)-(1.3) in \cite{lustri2023locating}; the integrated version is equation (B.10) in \cite{chapman2007}.
As discussed in Appendix B.2 of \cite{vandenheuvel2023burgers}, we can solve (\ref{eq:PhysDmodel}) exactly in terms of parabolic cylinder functions (the solutions reduce to rational functions when the constant in the far-field condition in (\ref{eq:PhysDmodel}) is a natural number multiplied by $-2$).  Importantly, (\ref{eq:PhysDmodel}) arises from a scenario in which diffusion and advection are perfectly weighted.  In contrast, the examples below are all for diffusion-dominated regimes.

Another possibility for (\ref{eq:Burgersinner}) occurs if $\epsilon\dot{\epsilon}\ll 1$ with $|\dot{S}|\ll \dot{\epsilon}$.  In this case, by writing $U\sim U_0(X)$ as $t\rightarrow 0^+$, we have
$$
U_0U_0'=U_0''.
$$
The solution of this ode subject to $U_0\rightarrow u_+$ as $X\rightarrow\infty$ is
\begin{equation}
U_0=u_+\left(1-\frac{2}{1-\mathrm{e}^{-u_+X}}\right),
\quad u_+<0.
\label{eq:periodicarray}
\end{equation}
This function involves a vertical array of poles and zeros that align themselves up the imaginary $X$-axis.

\subsection{Important case for our study}

In this paper, we are interested in the case where $\epsilon(t)$ is smaller than $t^{1/2}$ in (\ref{eq:Burgersinner}), so that $\epsilon\dot{\epsilon}\ll 1$.  If we have $\epsilon|\dot{S}|/\mu^{1/2}=\mathcal{O}(1)$, then we arrive at an important leading-order balance for diffusion-dominated regimes.  For this purpose, we could write $\dot{S}(t)=|\dot{S}(t)|\mathrm{e}^{\mathrm{i}\alpha}$, where the angle $\alpha$ indicates the direction that the singularities are travelling in the original $x$-plane.  Then, by choosing $\epsilon=\mu^{1/2}/|\dot{S}|$ and writing $U\sim U_0(X)$ as $t\rightarrow 0^+$, the leading-order ode becomes
\begin{equation}
-\mathrm{e}^{\mathrm{i}\alpha}U_0'+U_0U_0'=U_0''.
\label{eq:rotatedTWmodel}
\end{equation}
This travelling-wave equation appears to be relevant in a variety of regimes for which diffusion is dominant over advection from an outer perspective.

The behaviour of the solution of (\ref{eq:rotatedTWmodel}) depends on the far-field conditions.  To make this clearer, we write $U_0=\mathrm{e}^{\mathrm{i}\alpha}\bar{U}_0(\bar{X})$, $X=\mathrm{e}^{-\mathrm{i}\alpha}\bar X$, so that
\begin{equation}
-\bar U_0'+ \bar U_0 \bar U_0'=\bar U_0'',
\label{eq:TWmodel}
\end{equation}
where now the dashes mean differentiation with respect to $\bar X$, and impose a far-field condition of the form
\begin{equation}
\bar U_0 \rightarrow u_+, \quad \bar U_0' \rightarrow 0,
\quad\mbox{as}\quad \bar X\rightarrow \infty.
\label{eq:TWmodelBC}
\end{equation}
Ultimately, this far-field condition will come from matching out to the outer region, suggesting that, with these rotated variables, $u_+$ is real-valued.  Together, (\ref{eq:TWmodel})-(\ref{eq:TWmodelBC}) comprise the very-well studied travelling wave problem for Burgers' equation.  We can integrate and solve exactly to give
$$
\bar U_0 = u_++\frac{2(1-u_+)}{1+K\mathrm{e}^{(1-u_+)\bar X}},
$$
where the constant $K$ allows us to fix the origin in $X$, given that (\ref{eq:TWmodel}) is invariant to translations in $\bar X$.  For example, if we suppose there is a singularity at $\bar X=0$, then we can set $K=-1$, in which case
\begin{equation}
\bar U_0 = u_++\frac{2(1-u_+)}{1-\mathrm{e}^{(1-u_+)\bar X}},
\quad u_+\neq 1.
\label{eq:twsoln}
\end{equation}
Then, in the neighbourhood of $\bar X=0$, we have $\bar U_0 \sim -2/\bar X+1$, which, in original variables, is $u\sim -2\mu/(x-S(t))+\dot{S}(t)$, agreeing with (\ref{eq:burgerpoleequation}).  There are infinitely many other simple-pole singularities of $\bar U_0$ located at
\begin{equation}
\bar X = \frac{2n\pi\mathrm{i}}{1-u_+}, \quad n\in\mathbb{Z},
\label{eq:locatepoles}
\end{equation}
all of which have the correct local behaviour.
In the special case that $u_+=1$, integrating (\ref{eq:TWmodel}) directly gives the exact solution
\begin{equation}
\bar U_0 = -\frac{2}{\bar X}+1,
\quad u_+= 1.
\label{eq:twsoln2}
\end{equation}
Clearly this solution also has the correct behaviour near $\bar X=0$.

\begin{figure}[!h]
\centering
\includegraphics[width=0.24\textwidth]{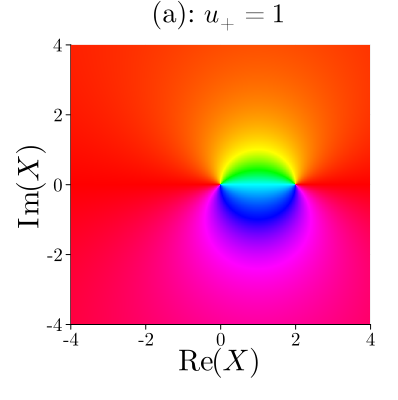}
\includegraphics[width=0.24\textwidth]{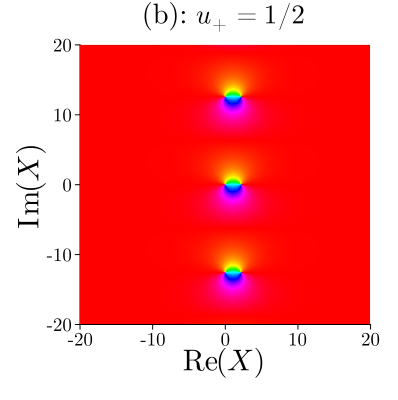}
\includegraphics[width=0.24\textwidth]{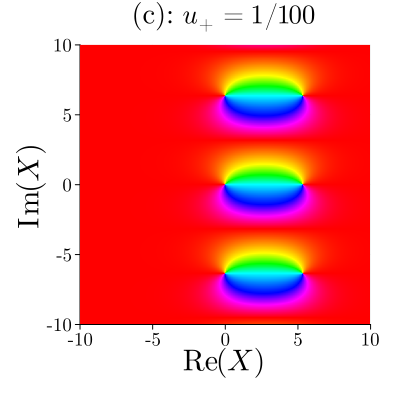}

\includegraphics[width=0.24\textwidth]{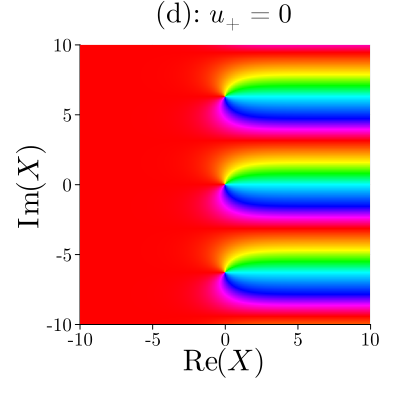}
\includegraphics[width=0.24\textwidth]{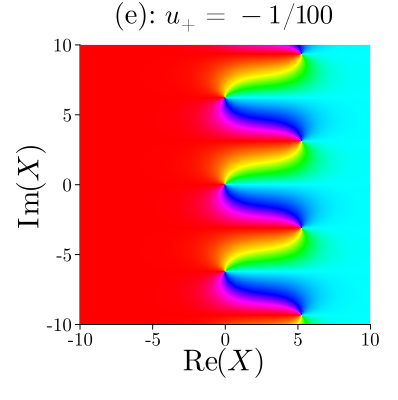}
\includegraphics[width=0.24\textwidth]{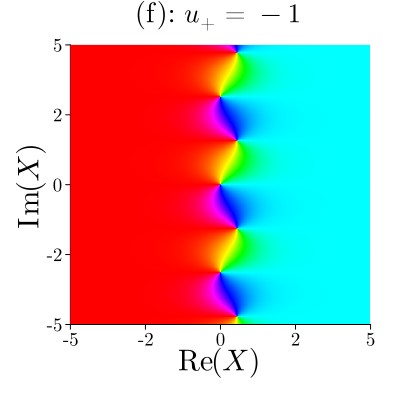}
\includegraphics[width=0.24\textwidth]{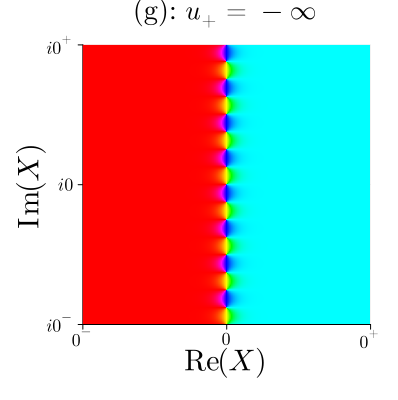}
\caption{Phase portraits of the generic solution (\ref{eq:twsoln}).
(a) The example $u_+=1$ involves a clean dipole-like structure.  (b)-(c) The examples $u_+=1/2$ and $u_+=1/100$, respectively, illustrate a vertical array of horizontal dipoles.  (d) The example with $u_+=0$ is the classical ``travelling wave'' solution consisting of an array of finger-like structures, each with a pole at the finger tip and a zero at the base at infinity.  (e)-(f) The examples $u_+=-1/100$ and $u_+=-1$, respectively, illustrate wave-like structures with poles at the ``crest'' and zeros at the ``troughs'' of the waves.  (g) The limiting solution of (\ref{eq:twsoln}) as $u_+\rightarrow -\infty$, which involves a vertical array of poles and zeros.  To represent this limit we have drawn (\ref{eq:periodicarray}) with $u_+=-5$.
}
\label{fig:genericsolns}
\end{figure}

It is worth summarising the properties of the ``travelling-wave'' solution (\ref{eq:twsoln}) or (\ref{eq:twsoln2}).  Note that if $u_+\leq 1$ then  $\bar U_0\rightarrow 2-u_+$ as $\bar X\rightarrow -\infty$, as well as the original condition that $U_0\rightarrow u_+$ as $\bar X\rightarrow +\infty$.  There are four qualitatively different cases to consider.  First, for $u_+=1$, the solution (\ref{eq:twsoln2}) is a clean dipole-like structure with a single simple pole at $\bar X=0$ and a simple zero at $\bar X=2$.  See figure~\ref{fig:genericsolns}(a).  Second, for $0<u_+<1$, (\ref{eq:twsoln}) has infinitely many ``dipoles'' in a vertical array up the imaginary $\bar X$ direction with simples poles at (\ref{eq:locatepoles})
and simple zeros at $\bar X=(1-u_+)^{-1}\ln((2-u_+)/u_+)+2n\pi\mathrm{i}/(1-u_+)$,
as indicated in figure~\ref{fig:genericsolns}(b) and (c).  As $u_+\rightarrow 1^-$, the zeros move to the left so that their real part approaches $2$, in line with the $u_+=1$ case, while as $u_+$ decreases the zeros move to the right until they are stretched out to infinity as $u_+\rightarrow 0^+$.  Indeed, for the third case, namely $u_+=0$, (\ref{eq:twsoln}) consists of an infinite array of finger-like structures, as in Figure~\ref{fig:genericsolns}(d).  In this special case, the solution
\begin{equation}
\bar U_0 = \frac{2}{1-\mathrm{e}^{\bar X}},
\label{eq:twU0}
\end{equation}
has poles at (\ref{eq:locatepoles}) but no zeros.  The fourth qualitative behaviour is for $u_+<1$, two examples of which are shown in Figures~\ref{fig:genericsolns}(e) and (f).  Here the poles at (\ref{eq:locatepoles}) and zeros at
$\bar X=(1-u_+)^{-1}\ln((u_+-2)/u_+)+(2n+1)\pi\mathrm{i}/(1-u_+)$ create a wave-like structure.  Note that, in the limit $u_+\rightarrow -\infty$, the solution (\ref{eq:twsoln}) approaches (\ref{eq:periodicarray}), as illustrated in Figure~\ref{fig:genericsolns}(g).

This catalogue of possible local solution behaviours covers all of the examples considered in this paper.  For example, for the step-down initial condition covered in subsection~\ref{sec:smalltime}, the inner problem (\ref{eq:inner_regions7A})-(\ref{eq:inner_regions7B}) is equivalent to (\ref{eq:TWmodel}) with the boundary conditions (\ref{eq:TWmodelBC}) forcing $u^+=0$, and the inner solution (\ref{eq:inner_regions9}) is equivalent to (\ref{eq:twU0}).  Thus, when we zoom in to the poles in figure~\ref{fig:stepdownphaseandasymptotic}(a), for example, the local finger-like behaviour is similar to that shown in figure~\ref{fig:genericsolns}(d), up to a rotation.

\section{An example with a discontinuous derivative}
\label{appendix:discderiv}

\subsection{Preamble}

\begin{figure}[!h]
    \centering
        \begin{subfigure}[t]{0.32\textwidth}
            \includegraphics[width=\linewidth]{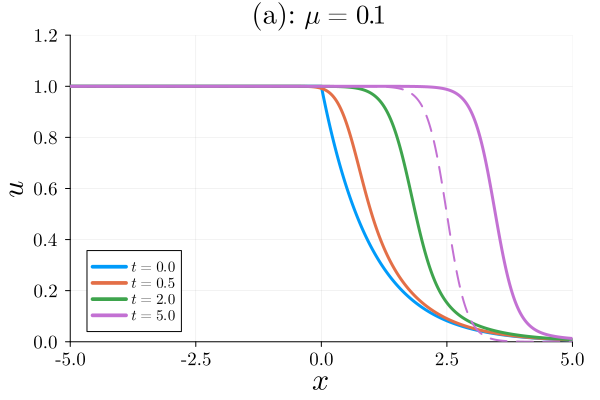}
        \end{subfigure}
        \begin{subfigure}[t]{0.32\textwidth}
            \includegraphics[width=\linewidth]{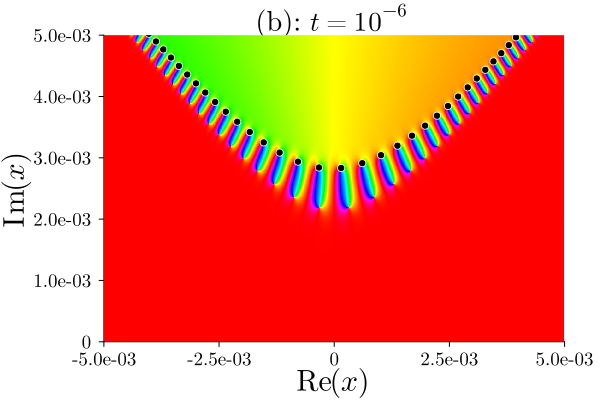}
        \end{subfigure}
        \begin{subfigure}[t]{0.32\textwidth}
            \includegraphics[width=\linewidth]{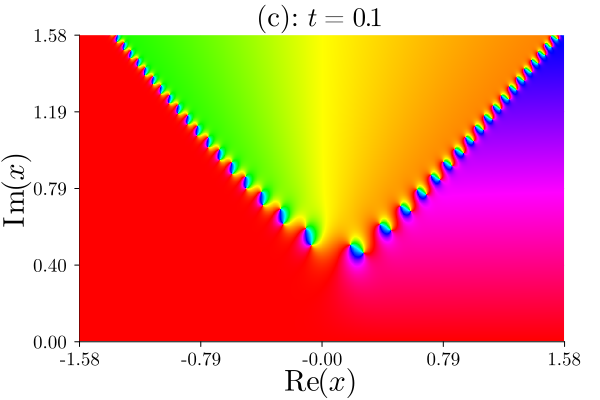}
        \end{subfigure}
        \begin{subfigure}[t]{0.32\textwidth}
            \includegraphics[width=\linewidth]{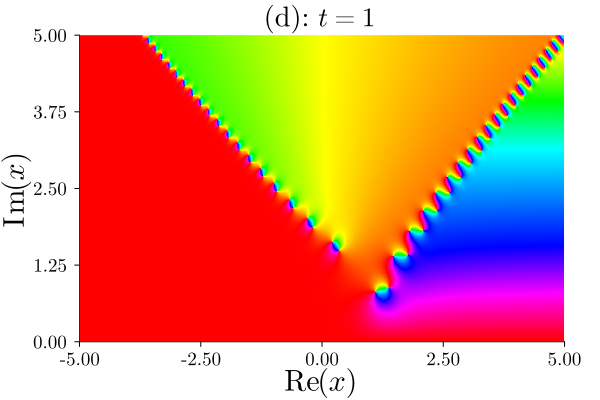}
        \end{subfigure}
        \begin{subfigure}[t]{0.32\textwidth}
            \includegraphics[width=\linewidth]{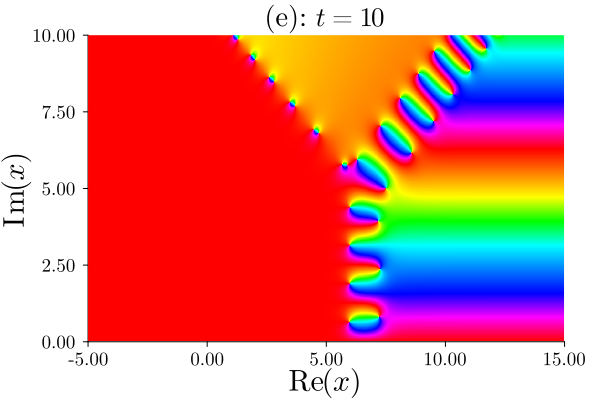}
        \end{subfigure}
        \begin{subfigure}[t]{0.32\textwidth}
            \includegraphics[width=\linewidth]{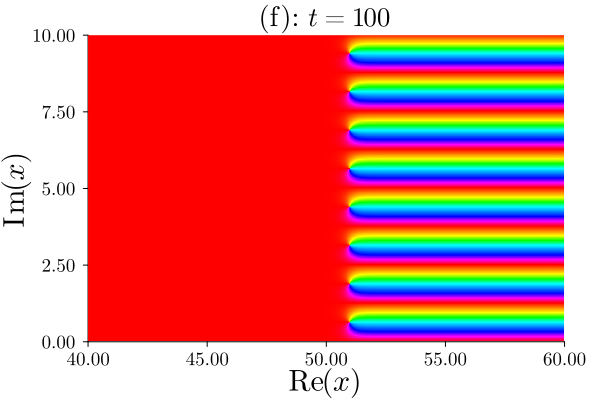}
        \end{subfigure}
        \caption{(a) Real-valued solution profiles of Burgers' equation (\ref{eq:burgers}), (\ref{eq:rampdown}) with $\mu=0.1$ for $t=0$, $0.5$, $2$ and $5$
        together with the travelling wave solution \eqref{eq:travwavesolution2} computed with $x_0=0$ (dashed line).  (b)-(f)
        Phase portraits of the exact solution of (\ref{eq:burgers}), (\ref{eq:rampdown}) for $\mu = 0.1$ at times $t=10^{-6}$, $0.1$, $1$, $5$, $10$ and $100$.
        Note that, while this last image in panel (f) is drawn for a finite time $t=100$, it is visually undistinguishable from the travelling wave solution \eqref{eq:travwavesolution} on this scale.
        }
        \label{fig:expo}
\end{figure}

Here we briefly outline how our analysis can be adapted for more complicated problems where the initial condition is continuous but not continuously differentiable.  The example we use is Burgers' equation (\ref{eq:burgers}) with
\begin{equation}
u_0(x)=
\left\{
\begin{array}{ll}
1, & x<0, \\
\mathrm{e}^{-x}, & x\geq 0,
\end{array}
\right.
\quad x \in \mathbb{R}.
\label{eq:rampdown}
\end{equation}
Leaving out the details, the exact solution of (\ref{eq:burgers}), (\ref{eq:rampdown}) is given by (\ref{eq:colehopf1}) with
\begin{equation}
\phi(x,t) =
        \sfrac{1}{2}
        \mathrm{e}^{-x^2/4\mu t}
        \Bigg[
        \mathrm{erfcx}\left(
        \frac{x-t}{2\sqrt{\mu t}}
        \right)
        +
        \mathrm{e}^{-1/2\mu}
        \sum_{n=0}^{\infty}
        \frac{\left(2\mu\right)^{-n}}{n!}
        \mathrm{erfcx}\left(
        \frac{-x+2\mu t n}{2\sqrt{\mu t}}
        \right)
        \Bigg],
\label{eq:expphi}
\end{equation}
where $\mathrm{erfcx}(z)=\mathrm{e}^{z^2}\mathrm{erfc}(z)$ is the scaled complementary error function.  Real-valued solutions of (\ref{eq:burgers}), (\ref{eq:rampdown}) are shown in figure~\ref{fig:expo}(a) for $\mu=0.1$.  We see here that, as with our first example in figure~\ref{fig:stepdowntotravwave}(a), the solution evolves towards  the travelling wave profile (\ref{eq:travwavesolution2}) for intermediate times.  Phase portraits are shown in figure~\ref{fig:expo}(b)-(f) for $\mu=0.1$, which we discuss below.

\subsection{Outer region, $|x|=\mathcal{O}(1)$}

While the initial condition (\ref{eq:rampdown}) is continuous, it is worthy of attention as its derivative is not.  Further, there is only one point at which the derivative is discontinuous, namely $x=0$; thus, this initial condition serves as an obvious choice within this class.  We show below that complex-plane singularities will be born at $x=0$ at $t=0^+$ in a similar way to our first example in subsection~\ref{sec:smalltime}.  The key difference here is that we need more care in deriving the appropriate temporal scalings involved.

A starting point is to consider an outer expansion for $x>0$ of the form $u\sim u_0+tu_1+\ldots$, leading to
\begin{equation}
u\sim \mathrm{e}^{-x}+t\left(\mathrm{e}^{-2x}+\mu\mathrm{e}^{-x}\right)+\ldots
\quad\mbox{as}\quad t\rightarrow 0^+,
\quad x>0.
\label{eq:expouter}
\end{equation}
Writing this outer expression (\ref{eq:expouter}) in terms of the inner variable $\xi=x/t^{1/2}$ and expanding in the limit $t\rightarrow 0^+$ gives
\begin{equation}
u\sim 1-t^{1/2}\xi+t\left(\sfrac{1}{2}\xi^2+\mu+1\right)
+t^{3/2}\left(-\sfrac{1}{6}\xi^3-(\mu+2)\xi\right)+\ldots,
\quad \xi>0.
\label{eq:expouter2}
\end{equation}
For $x<0$, the outer region in terms of $\xi$ is simply
\begin{equation}
u\sim 1,
\quad \xi<0.
\label{eq:expouter3}
\end{equation}

\subsection{Inner region}

This way of expressing the outer region in (\ref{eq:expouter2}) and (\ref{eq:expouter3}) motivates an inner region with scalings (\ref{eq:similarity}), where $f$ satisfies (\ref{eq:asymptoticsol1}).
It turns out we need to include the first four terms in the expansion for $f$,
\begin{equation}
f\sim f_0(\xi)+t^{1/2}f_1(\xi)+tf_2(\xi)+t^{3/2}f_3(\xi)
\quad\mbox{as}\quad t\rightarrow 0+,
\label{eq:expoexpandf}
\end{equation}
(cf.~(\ref{eq:asymptoticsol2}), where only the first two terms $f_0$ and $f_1$ were required), although the leading-order term $f_0$, which satisfies
$$
-\sfrac{1}{2}\xi f_0'=\mu f_0'',
\quad f_0\rightarrow 1 \quad\mbox{as}\quad \xi\rightarrow\pm\infty,
$$
is simply the trivial solution $f_0=1$.  With this leading-order term solution, the problems for $f_1$, $f_2$ and $f_3$ reduce to
\begin{eqnarray}
-\sfrac{1}{2}\xi f_1'+\sfrac{1}{2}f_1 & = & \mu f_1'',
\label{eq:expof1}\\
-\sfrac{1}{2}\xi f_2'+f_2+f_1' & = & \mu f_2'',
\label{eq:expof2}\\
-\sfrac{1}{2}\xi f_3'+\sfrac{3}{2}f_3+f_1f_1'+f_2' & = & \mu f_3'',
\label{eq:expof3}
\end{eqnarray}
together with far-field conditions
\begin{eqnarray}
f_1\rightarrow 0 \quad\mbox{as}\quad \xi\rightarrow -\infty,
& \quad &
f_1\rightarrow -\xi \quad\mbox{as}\quad \xi\rightarrow \infty,
\label{eq:expof1far}
\\
f_2\rightarrow 0 \quad\mbox{as}\quad \xi\rightarrow -\infty,
& \quad &
f_2\rightarrow \sfrac{1}{2}\xi^2+\mu+1 \quad\mbox{as}\quad \xi\rightarrow \infty,
\label{eq:expof2far}
\\
f_3\rightarrow 0 \quad\mbox{as}\quad \xi\rightarrow -\infty,
& \quad &
f_3\rightarrow -\sfrac{1}{6}\xi^3-(\mu+2)\xi \quad\mbox{as}\quad \xi\rightarrow \infty,
\label{eq:expof3far}
\end{eqnarray}
that come from (\ref{eq:expouter2}) and (\ref{eq:expouter3}).  These problems (\ref{eq:expof1})-(\ref{eq:expof3}) with (\ref{eq:expof1far})-(\ref{eq:expof3far}) are all linear.

We can easily solve for $f_1$ directly to give
\begin{equation}
f_1=-\xi-\sqrt{\frac{\mu}{\pi}}\mathrm{e}^{-\xi^2/4\mu}
+\sfrac{1}{2}\xi\,\mathrm{erfc}\left(\frac{\xi}{2\sqrt{\mu}}\right).
\label{eq:expof1soln}
\end{equation}
Given the problem for $f_1$ is linear and homogeneous, the solution (\ref{eq:expof1soln}) is an entire function and therefore cannot provide any approximation for the location of the poles (just like $f_0$ given by (\ref{eq:asymptoticsolf0}) in our original problem in subsection~\ref{sec:smalltime}).  Instead, $f_1$ grows exponentially in the imaginary direction as
\begin{equation}
f_1\sim -\frac{2\mu^{3/2}}{\pi^{1/2}}
\frac{1}{\xi^2}\mathrm{e}^{-\xi^2/4\mu}
\quad\mbox{as}\quad \xi\rightarrow\mathrm{i}\infty,
\label{eq:expof1growth}
\end{equation}
and thus $f_1$ has an essential singularity at infinity.

For our analysis, it is not necessary to solve (\ref{eq:expof2}) exactly for $f_2$; instead, we can simply focus on far-field behaviour to find
\begin{equation}
f_2\sim -\frac{\mu^{1/2}}{\pi^{1/2}}
\frac{1}{\xi}\mathrm{e}^{-\xi^2/4\mu}
\quad\mbox{as}\quad \xi\rightarrow\mathrm{i}\infty.
\label{eq:expof2growth}
\end{equation}
We see the exponential growth here in the imaginary direction for $f_2$ (see (\ref{eq:expof2growth})) is at the same rate as in $f_1$ (see (\ref{eq:expof1growth})), although the algebraic prefactor is different.  Further, we see the original nonlinear term $t^{1/2}ff_\xi$ in (\ref{eq:asymptoticsol1}) does not play a role in the equations for $f_1$ or $f_2$, (\ref{eq:expof1})-(\ref{eq:expof2}).  For these reasons, we need to go to the next term $f_3$ by considering (\ref{eq:expof3}).

Focusing on the far-field behaviour only, we find
\begin{equation}
f_3\sim -\frac{2\mu^{3}}{\pi}
\frac{1}{\xi^5}\mathrm{e}^{-\xi^2/2\mu}
\quad\mbox{as}\quad \xi\rightarrow\mathrm{i}\infty.
\label{eq:expof3growth}
\end{equation}
This time we see the stronger exponential growth as $\xi\rightarrow \mathrm{i}\infty$ ($\mathrm{e}^{-\xi^2/2\mu}$, as opposed to $\mathrm{e}^{-\xi^2/4\mu}$), which is entirely due to the term $f_1f_1'$ in (\ref{eq:expof3}).  This contribution comes from the original nonlinear advective term $t^{1/2}ff_\xi$ in (\ref{eq:asymptoticsol1}), which had been neglected up to this point.  Thus, in the expansion (\ref{eq:expoexpandf}), the term $f_3$ is the first to feel the effects of the nonlinearity in Burgers' equation, $uu_x$, explaining why we need to go to this order for this problem.

\subsection{Inner-inner region}

To determine the region in which (\ref{eq:expoexpandf}) breaks down, we need to balance $t^{1/2}f_1$ (the first term that is singular at infinity) with $t^{3/2}f_3$ (the first term whose far-field behaviour is driven by the nonlinearity in Burgers' equation), giving
$$
\frac{\pi^{1/2}}{\mu^{3/2}}\xi^3\mathrm{e}^{\xi^2/4\mu}=\mathcal{O}(t).
$$
Solving this equation asymptotically, we find
\begin{eqnarray*}
\frac{\xi^2}{4\mu} & \sim & -\ln(1/t)-\ln(4\sqrt{\pi})+2m\pi\mathrm{i}
-\sfrac{3}{2}\,\mathrm{Log}\left(
-\ln(1/t)-\ln(4\sqrt{\pi})+2m\pi\mathrm{i}
\right) \\
& \sim & -\ln(1/t)-\ln(4\sqrt{\pi})+(2m-\sfrac{3}{2})\pi\mathrm{i}
-\sfrac{3}{2}\,\mathrm{Log}\left(
\ln(1/t)+\ln(4\sqrt{\pi})-(2m-\sfrac{3}{2})\pi\mathrm{i}
\right),
\end{eqnarray*}
as $t\rightarrow 0^+$, suggesting the inner-inner region is defined by
\begin{equation}
\frac{\xi^2}{4\mu}=-\ln(1/t)-\ln(4\sqrt{\pi})+(2m-\sfrac{3}{2})\pi\mathrm{i}
-\sfrac{3}{2}\,\mathrm{Log}\left(
\ln(1/t)+\ln(4\sqrt{\pi})-(2m-\sfrac{3}{2})\pi\mathrm{i}
\right)+\mathrm{i}X,
\quad X=\mathcal{O}(1).
\label{eq:expoxi}
\end{equation}
Using this new variable $X$, together (\ref{eq:expof1growth}) and (\ref{eq:expof3growth}) give
\begin{equation*}
t^{1/2}f_1+t^{3/2}f_3
\sim \frac{\mu^{1/2}}{t^{1/2}}
\left(
\ln(1/t)+\ln(4\sqrt{\pi})-(2m-\sfrac{3}{2})\pi\mathrm{i}
\right)^{1/2}
\left(-2\mathrm{i}\mathrm{e}^{-\mathrm{i}X}
-2\mathrm{i}\mathrm{e}^{-2\mathrm{i}X}
\right)
\end{equation*}
as $t\rightarrow 0^+$.  This equation provides the scalings and far-field conditions for this inner-inner region, suggesting we define
$$
f=\frac{\mu^{1/2}}{t^{1/2}}
\left(
\ln(1/t)+\ln(4\sqrt{\pi})-(2m-\sfrac{3}{2})\pi\mathrm{i}
\right)^{1/2}F(X,t).
$$
We write $F\sim F_0(X)$ as $t\rightarrow 0^+$, so that, after some algebra, to leading order the pde (\ref{eq:asymptoticsol1}) gives
$$
-\mathrm{i}F_0'+F_0F_0'=F_0'',
\quad
F_0\sim -2\mathrm{i}\mathrm{e}^{-\mathrm{i}X}
-2\mathrm{i}\mathrm{e}^{-2\mathrm{i}X}
\quad\mbox{as}\quad X\rightarrow -\mathrm{i}\infty.
$$
This inner-inner problem is the same as (\ref{eq:inner_regions7A})-(\ref{eq:inner_regions7B}), with the solution (\ref{eq:inner_regions9}), namely
$$
F_0=\frac{2\mathrm{i}}{1-\mathrm{e}^{\mathrm{i}X}},
$$
except this variable $X$ is defined by (\ref{eq:expoxi}).
There is a pole at $X=0$ and others at $X=2k\pi$, for $k\in\mathbb{Z}$.  The counter $k$ can be absorbed into $m$ above, so an asymptotic approximation for the poles $x=s_m(t)$ is
\begin{equation}
s_m(t) \sim 2\mu^{1/2}\mathrm{i}t^{1/2}\left(
\ln(1/t)+\ln(4\sqrt{\pi})-(2m-\sfrac{3}{2})\pi\mathrm{i}
\right)^{1/2}
\left[
1+\frac{\sfrac{3}{4}\,\mathrm{Log}
\left(
\ln(1/t)+\ln(4\sqrt{\pi})-(2m-\sfrac{3}{2})\pi\mathrm{i}
\right)
}{\ln(1/t)+\ln(4\sqrt{\pi})-(2m-\sfrac{3}{2})\pi\mathrm{i}}
\right]
\label{eq:expsm}
\end{equation}
as $t\rightarrow 0^+$.

\subsection{Phase portraits}

We illustrate the early-time behaviour for this example in figure~\ref{fig:expo}(b), which shows a phase portrait calculated from the exact solution (\ref{eq:colehopf1}) with (\ref{eq:expphi}) at $t=10^{-6}$.  Also included in this image are black dots that indicate the location of the poles according to the approximation (\ref{eq:expsm}).  At this small time, we see the poles arranged along a convex-shaped curve, with the asymptotic result (\ref{eq:expsm}) doing an excellent job of approximating the pole locations on this scale.

At moderate times, we see in figure~\ref{fig:expo}(c)-(e) how the pole structure begins to propagate to the right while the poles and zeros closest to the real axis begin to merge and construct the late-time formation.  The large-time examples in figure~\ref{fig:expo}(e)-(f) indicate the transition to the travelling wave configuration.

In summary, the example in this Appendix, namely (\ref{eq:burgers}) with (\ref{eq:rampdown}), is chosen to illustrate how simple-pole singularities are born and initially propagate when the initial condition is in $C^0$ but not $C^1$.  While more care is needed to resolve the matched asymptotics for this example than the original example in section~\ref{sec:smalltime}, the overall structure is essentially the same.

\bibliographystyle{plain}
\bibliography{Gentner_references}

\end{document}